\documentclass[11pt,a4paper]{article}
\pdfoutput=1
\usepackage{jheppub}
\usepackage{slashed}
\usepackage{physics}
\usepackage{tensor}
\makeatletter
\def\@fpheader{\relax}
\makeatother

\usepackage{empheq}
\usepackage{ulem}

\usepackage[czech,english]{babel}
\usepackage{graphicx}
\usepackage{amsmath,amsfonts,amssymb}
\usepackage{url}
\usepackage{mathtools}

\usepackage[makeroom]{cancel}
\usepackage{leftidx}
\usepackage{xcolor}
\numberwithin{equation}{section}
\usepackage{adjustbox}
\usepackage{appendix}
\usepackage{tikz}
\tikzstyle{process} = [rectangle, minimum width=3cm, minimum height=1cm, text centered, draw=black, fill=orange!30]
\tikzstyle{arrow} = [thick,->,>=stealth]

\usepackage{comment}

\preprint{LCTP-23-09}

\title{The Attractor Flow for AdS$_5$ Black Holes in $\mathcal{N}=2$ Gauged Supergravity}

\author[a]{Marina David}

\author[b]{,~Nizar Ezroura}

\author[b,c]{,~ and Finn Larsen}

\emailAdd{marina.david@kuleuven.be}

\emailAdd{nezroura@umich.edu}

\emailAdd{larsenf@umich.edu}

\affiliation[a]{Instituut voor Theoretische Fysica, KU Leuven Celestijnenlaan 200D, B-3001 Leuven, Belgium}

\affiliation[b]{Leinweber Center for Theoretical Physics, University of Michigan, Ann Arbor, MI 48109, U.S.A.}

\affiliation[c]{Department of Physics and Stanford Institute for Theoretical Physics,\\Stanford University, Palo Alto, CA 94305, USA.}

\abstract{We study the flow equations for BPS black holes in $\mathcal{N} = 2$ five-dimensional gauged supergravity coupled to any number of vector multiplets via FI couplings. We develop the Noether-Wald procedure in this context and exhibit the conserved charges as explicit integrals of motion, in the sense that they can be computed at any radius on the rotating spacetime. The boundary conditions needed to solve the first order differential equations are discussed in great detail. We extremize the entropy function that controls the near horizon geometry and give explicit formulae for all geometric variables at their supersymmetric extrema. We have also considered a complexification of the near-horizon variables that elucidates some features of the theory from the near-horizon perspective.

}


\arxivnumber{}

\begin{document}

\maketitle

\section{Introduction}

The radial dependence of physical fields in a black hole background relates the field configuration far from the black hole to the region near the black hole horizon. It is important in holography
because the radial evolution is identified with the renormalization group flow in the dual quantum field theory, so it determines the low energy QFT observables in terms of UV data. For supersymmetric black holes in asymptotically flat spacetimes, these ideas are realized beautifully by the so-called attractor flow. Unfortunately, the analogous construction for BPS black holes in asymptotically AdS spacetimes, where holography is more precise, is less developed. The goal of this article is to 
give an explicit and detailed account of the attractor flow for BPS black holes in AdS$_5$.  

All extremal black holes, whether supersymmetric or not, enjoy an attractor {\it mechanism}, in that the end point of the radial flow is a horizon region with enhanced $SO(2,1)$ symmetry. The field configuration in this 
AdS$_2$ region is determined by the entropy function formalism, an extremization principle that was studied in many contexts \cite{Sen:2005wa, Sen:2005iz, David:2006yn, Sen:2007qy, Sen:2008yk, Sen:2008vm, Sen:2009vz, Astefanesei:2006dd, Castro:2007sd}, including its applications to subleading corrections of the black hole entropy \cite{Kraus:2005vz, Banerjee:2009af, Banerjee:2010qc}. The attractor {\it flow} refers to the entire evolution from the asymptotic space to the event horizon. The long throat characterizing the final approach to the horizon gives a geometric reason to intuit that only a restricted set of endpoints are possible. In this case the flow is \textit{attracted} to specific points in configuration space. 

The study of supersymmetric attractor flows was initiated in ${\cal N}=2$ ungauged supergravity \cite{Ferrara:1995ih, Ferrara:1996dd}. In this context an attractor mechanism was realized. The horizon values of scalars in ${\cal N}=2$ vector multiplets is independent of their freely chosen asymptotic values. The attractor flow for asymptotically flat black holes was later generalized to different 
amounts of supersymmetry in various dimensions \cite{Ferrara:1996um, Ferrara:2006em, Larsen:2006xm, Ceresole:2007wx, LopesCardoso:2007qid, Perz:2008kh, Galli:2010mg, Ceresole:2009iy, Andrianopoli:2007gt, Andrianopoli:2009je}. 
In this work we inquire about the analogous radial flow in gauged supergravity, the setting for asymptotically AdS black holes. 
This research direction is very well motivated by holography, but it is technically more involved and much less developed.
Moreover, the reference to an \textit{attractor} is a misnomer in gauged supergravity: in general no \textit{initial} data is introduced at the asymptotic AdS boundary, so nothing is lost when 
the horizon is reached. However, the attractor terminology is so ingrained by now that we
keep it, even though it can be misleading.  
Indeed, versions of attractor flows and attractor mechanisms in gauged supergravity previously appeared in \cite{Ceresole:2001wi, Hosseini:2017mds, DallAgata:2010ejj, Cacciatori:2009iz, Halmagyi:2014qza, Hristov:2018spe}.

In the canonical set-up \cite{Ferrara:1995ih, Ferrara:1996dd}, ungauged ${\cal N}=2$ supergravity coupled to vector multiplets, scalar fields are the only variables needed to characterize the attractor flow. Classical black hole solutions involve vector fields as well but, in stationary black hole backgrounds, their radial dependence is entirely determined  by the conserved electric charges, augmented by magnetic charges in $D=4$. The scalars can take any values
in the asymptotically flat space as they are moduli that parametrize the vacuum. 
However, for given charges, their radial flow is governed by an effective black hole potential that guides them to an attractor value that depends only on the charges. 

BPS black holes in AdS are fundamentally different because the scalar fields in gauged supergravity are subject to a potential that depends on the couplings of the theory, so generally the scalars are not moduli that parametrize vacua. When scalars do not depend on free 
parameters at infinity, there is no scope for an attractor mechanism that imposes
horizon values independent of such parameters. There is an exception when FI-couplings are fine-tuned to create flat directions in the scalar potential. In this case the scalar fields do obey an attractor mechanism.

This work focuses on black holes with nonzero electrical charge and rotation in 5D asymptotically AdS spacetimes \cite{Gutowski:2004ez, Gutowski:2004yv, Kunduri:2006ek, Chong:2005hr,Cvetic:2005zi} whose solution has been well studied in various contexts. Several of the technical challenges we encounter in our setup arise because all known BPS black holes in AdS$_5$ have non-vanishing angular momentum. In our implementation of the attractor flow, the angular momentum $J$ is a conserved quantity like any other: it can be computed as a flux integral over any topological sphere surrounding the black hole. 
The attractor flow from asymptotically AdS to the horizon
corresponds to decreasing radial coordinate. 
Because of the rotation, the spatial sphere at any given
radius is squashed and co-rotating. 

There has been much previous work on the attractor mechanism in gauged supergravity, including \cite{Ceresole:2001wi, Hosseini:2017mds,Halmagyi:2014qza, Huebscher:2008wob,Bellucci:2008cb,Cacciatori:2009iz,Cassani:2009na,Hristov:2010eu,DallAgata:2010ejj,Hristov:2010ri, Hristov:2014eba, Bobev:2020jlb, Hristov:2018spe,Hosseini:2018usu, Cabo-Bizet:2017xdr, Hristov:2016vbm, Hristov:2014eza, Kachru:2011ps,Chimento:2015rra,Astesiano:2021hro} and references therein. As a guide, we highlight some of the
features we focus on: 

\begin{itemize}
\item 
We construct the entire attractor flow of gauged supergravity, ie. the radial solution that interpolates between the AdS$_5$ vacuum far from the black hole and the horizon vacuum of AdS$_2$-type. We do this directly in 5D, without the dimensional reduction to 4D pursued by some researchers \cite{Hosseini:2017mds, Hristov:2018spe}. 

\item
We construct {\it conserved charges} in terms of flux integrals. They are 
not specifically defined asymptotically far from the black hole, nor in the horizon region. Rather, they can be evaluated on any member of a nested family of surfaces that interpolate between the horizon and a homologous asymptotic surface. This is essential for the attractor flow. 

Technically, we construct these conserved charges using the covariant Noether-Wald procedure. This is similar to what is done in \cite{Suryanarayana:2007rk}, although their focus is on the near-horizon region. 
One can also take the approach of a generalized Komar integral \cite{Rodriguez:2021hks, Kastor:2008xb, Kastor:2009wy, Magnon:1985sc, Ortin:2021ade, Cano:2023dyg}, where a specific gauge must be chosen to deal with diverging asymptotics in AdS spacetimes.

\item
Our work incorporates {\it general special geometry}, including symplectic invariance. As such, we incorporate general cubic prepotentials. We provide a reader friendly guide to the widely studied STU and minimal SUGRA models that correspond to special cases. For a recent study on the attractor mechanism in the STU model, see \cite{Ntokos:2021duk}.

\item
We carry out the {\it entropy extremization} procedure in complete detail, reconstructing all details of the horizon structure. This elucidates the relation between real and complex fields. 
\end{itemize}

It may be useful to also note some aspects of the attractor mechanism for gauged supergravity that are interesting but {\it not} developed in this article:
\begin{itemize}
    \item 
    We focus on electric black holes in {\it only five} spacetime dimensions. 
    \item
    We study {\it supersymmetric} attractor flows but other extremal flows are interesting as well.
   \item We specialize to black holes with {\it equal angular momenta}, so our ansatz for the geometry preserves an $S^2$ throughout the flow. More general AdS$_5$ black holes with unequal angular momenta, studied in \cite{Chong:2005hr, Kunduri:2006ek}, depend on a complex structure on $S^1\times S^3$ that evolves radially and is defined only in Euclidean signature \cite{Closset:2013vra, Closset:2014uda}. 
\item
We study AdS$_5$ supergravity with FI-gauging, so only $U(1)$ gauge groups appear. This class of theories is technically simpler, because it is entirely specified by a linear superpotential, and there is no need for the moment map. Moreover, in this theory all scalar fields are neutral \cite{Markeviciute:2018yal}.
 \item
We do not connect to standard aspects of the AdS$_5$/CFT$_4$ correspondence, such as holographic renormalization, time-dependent sources, and so on. This direction was recently studied in \cite{Ntokos:2021duk}. 
\end{itemize}
We hope to expand on some of these omissions in future work.

This article is organized as follows.  In section~\ref{section: effective 2d lagrangian}, we introduce the $\mathcal{N}=2$ gauged supergravity action, the 5D form of the black hole solution, and its dimensional reduction to both 2D and 1D. section~\ref{section: noether-wald surface charge} is devoted to an analysis of Noether-Wald surface charges, and the subtleties of gauge invariant conserved charges for actions with Chern-Simons terms. section~\ref{section: the flow equations} is the longest and most detailed. We derive the first order differential equations imposed by preserved supersymmetry, in the context of our ansatz. We study the boundary conditions needed to solve the equations perturbatively, 
from both the horizon and the asymptotic AdS point of view. With both these perspectives, we 
recover the known black hole solutions by establishing truncation of the pertubative expansion. 
In section~\ref{section: entropy extremization} we develop the entropy extremization formalism 
and compute all near horizon aspects of the black hole, including its entropy. We also construct a complexification of the near-horizon variables that elucidates some aspects of the solution.
We conclude in section~\ref{section: discussion} with a discussion of open problems concerning the attractor flow in gauged supergravity and related topics. A series of appendices are devoted to technical details, conventions and notations regarding differential forms in Appendix~\ref{appendix: conventions and notations}, real special geometry in \ref{appendix subsection: real special geometry}, and the
supersymmetry conditions in Appendix~\ref{appendix: supersymmetry}.

\section{The Effective 2D Lagrangian} \label{section: effective 2d lagrangian}

In this section we introduce the action of $\mathcal{N}=2$ 5D supergravity with coupling to $n_V$ vector multiplets and gauging by Fayet–Iliopoulos couplings, as well as its dimensional reduction to a 2D theory. This also serves to define conventions and notation. 
For additional details on real special geometry and supersymmetry we refer to Appendices~\ref{appendix subsection: real special geometry} and~\ref{appendix: supersymmetry}, respectively.

\subsection{The 5D theory}

We study five dimensional $\mathcal{N}=2$ gauged supergravity with bosonic action
\begin{align} \label{eq: general 5D action}
\begin{split}
    S=\frac{1}{16 \pi G_{5}} &\int_{\mathcal{M}}\mathcal{L}_{5} + \frac{1}{8\pi G_{5}} \int_{\partial \mathcal{M}} d^4 x \sqrt{|h|}\Tr K~,
\end{split}
\end{align}
where the 5D Lagrangian density is given by
\begin{align} \label{eq: 5D Lagrangian}
    \mathcal{L}_{5} &= (-\mathcal{R}_{5}-2V) \star_{5} 1 - G_{I J} F^{I}_{5} \wedge \star_{5} F^{J}_{5}+G_{I J} d X^{I} \wedge \star_{5} d X^{J} - \frac{1}{6} c_{IJK} F^{I}_{5} \wedge F^{J}_{5} \wedge A^{K}_{5} \, .
\end{align}
We have included the subscript $5$ to emphasize that we are in five dimensions and the five dimensional Hodge dual is given by $\star_{5}$. The Gibbons-Hawking-York boundary term must be included to have a well-defined variation of the action \eqref{eq: general 5D action} and is given by the trace of the second fundamental form $K$ which is integrated over the induced metric $h$ on the boundary. Other conventions and notations regarding differential forms and the Hodge dual are in Appendix~\ref{appendix: conventions and notations}.

The field content includes the field strengths $F^{I}_{5}=dA^{I}_{5}$ where $I=1,\dots, n$ and the scalars $X^{I}$, correspond to $n-1$ physical scalars, constrained via the following relation
\begin{align} \label{eq: constraint on X}
    \frac{1}{6}c_{IJK}X^{I}X^{J}X^{K} = 1~.
\end{align}
The scalar potential is given by
\begin{align} 
    V &= - c^{I J K} \xi_{I} \xi_{J} X_{K}=- \xi_{I} \xi_{J}\left(X^{I} X^{J}-\frac{1}{2} G^{I J}\right)~,
\end{align}
where $\xi_{I}$ are the real Fayet–Iliopoulos parameters. The scalars with lowered index
\begin{align}
    X_{I} &= 2G_{IJ}X^{J}~,
\end{align}
obey the analogous constraint
\begin{align}
\label{eq: X cube}
    \frac{1}{6}c^{IJK}X_{I}X_{J}X_{K} = 1~,
\end{align}
when closure relation \eqref{eq:cclosure up} is satisfied. For further details on definitions, conventions and identities, we refer the reader to  Appendix~\ref{appendix subsection: real special geometry}.

Alternatively, the scalar potential can be expressed as
\begin{align} \label{eq: V pot W}
    V=-\left(\frac{2}{3}W^{2}-\frac{1}{2} G^{I J} D_{I} W D_{J} W\right)~,
\end{align}
where the superpotential $W$ is
\begin{align} \label{eq: W def}
    W=\xi_{I} X^{I}~,
\end{align}
and the K\"{a}hler covariant derivative $D_I$ takes the constraint 
\eqref{eq: X cube} into acount. Using this form of the potential $V$, 
the condition for a supersymmetric minimum becomes
\begin{equation}
    D_I W = \xi_{I} - \frac{1}{3}X_{I}(\xi \cdot X)  \underset{\rm min}{=} 0 ~. 
\end{equation}
According to this equation, the asymptotic values of the scalars $X_{I,\infty}$ must be parallel to $\xi_I$, in the sense of real special geometry vectors, and  the constraint \eqref{eq: X cube} determines the proportionality constant between the two:
\begin{equation}
\label{eq: X infty}
    X_{I,\infty} = \left(\frac{1}{6}c^{JKL}\xi_J \xi_K \xi_L \right)^{-1/3} \xi_I  ~. 
\end{equation}
The value of the potential $V$ at the minimum
must be related to the $\text{AdS}_5$ length scale $\ell$ and the
cosmological constant in the usual manner
\begin{align} \label{eq: def of V}
    V_{\infty} &= - c^{I J K} \xi_{I} \xi_{J} X_{K,\infty} \equiv - 6\ell^{-2}~.
\end{align}
This gives the constraint 
\begin{align} \label{eq: xi cubed constraint}
    \frac{1}{6}c^{IJK} \xi_I \xi_J \xi_K = \ell^{-3}~,
\end{align}
on the FI-parameters $\xi_I$ and the simple relation for the asymptotic values of the scalars
\begin{equation}
\label{eq: xinf xi}
X_{I,\infty} = \ell \xi_{I}~.
\end{equation}
Thus the $n_V+1$ independent FI-parameters $\xi_I$ determine the asymptotic values $X^I_\infty$ of the $n_V$ scalars, as well as the AdS$_5$ scale $\ell$. 

For contrast, recall that in ungauged supergravity, the scalar fields are moduli as they experience no potential. Then their asymptotic values $X_{I,\infty}$ far from the black hole are set arbitrarily by boundary conditions, which is related to the fact that the spacetime is asymptotically flat. The fact that the value of the scalars $X_I$ at the {\it horizon} is independent of the asymptotic values $X_{I,\infty}$ is the attractor mechanism for BPS black holes in ungauged supergravity. 

As we have seen, the present context is very different in that the asymptotic values of the scalars
are set by the theory through the FI-parameters $\xi_I$, rather than by boundary conditions. This is a generic feature of gauged supergravity, theories with asymptotically AdS vacuum. It precludes an attractor mechanism that is analogous to the one in asymptotically flat space. We will discuss this point in more depth in section~\ref{section: the flow equations} when we study the linear flow equations derived from supersymmetry. 

The equations of motion $\mathcal{E}_{\Phi}$, where $\Phi$ is any field in the theory corresponding to the Lagrangian density \eqref{eq: 5D Lagrangian}, are the Einstein equation
 \begin{align}
 \begin{split} \label{eq: 5d EE}
      \mathcal{E}_{g} &= R_{AB} - \tfrac{1}{2}g_{AB} R + G_{IJ} \left( F^{I}_{5,AC}F^{J,C}_{5,B} - \tfrac{1}{4} g_{AB} F^{I}_{5,CD} F^{J,CD}_{5}\right) \\ & \quad - G_{IJ}\left(\nabla_{A}X^{I}\nabla_{B}X^{J} - \tfrac{1}{2} g_{AB}\nabla_{C}X^{I}\nabla^{C}X^{J}  \right) - g_{AB} V = 0\, ,
\end{split}
\end{align}
and the matter equations for the Maxwell field $A^{I}_{5}$ and the constrained scalars $X^{I}$
\begin{align}
\begin{split} \label{eq: 5d EoM F}
    \mathcal{E}_{A} &= d\left(G_{IJ} \star F^{J}_{5} \right) + \tfrac{1}{4}c_{IJK} F^{J}_{5} \wedge F^{K}_{5} = 0,
\end{split}
    \\
\begin{split} \label{eq: 5d EoM X}
    \mathcal{E}_{X^I} &= -d\star dX_I + \tfrac{1}{3}X_I X^J d\star dX_J+2c^{JKL} \xi_K \xi_L \left(\tfrac{2}{3}X_I X_J - c_{IJM}X^M\right) \star 1  
    + \left(X_J X^L c_{IKL} \right. \\&\quad  \left.  - \tfrac{1}{2}c_{IJK}-\tfrac{2}{3}X_I X_J X_K  
    + \tfrac{1}{6}X_I c_{JKN}X^N\right) (F^J _5 \wedge \star F^K _5 - dX^J \wedge \star dX^K) = 0~. 
\end{split}
\end{align}

\subsection{The effective 2D theory} \label{subsection: the effective 2D theory}

We do not study all solutions to the 5D theory \eqref{eq: general 5D action}, just stationary black holes. Then it is sufficient
to consider a reduction to 2D --- and eventually to 1D. 
We impose  the metric ansatz\footnote{In our conventions the metric has a mostly negative signature.}
\begin{align} \label{eq: 5D metric ansatz}
    ds^2_5 = ds^2_2 - e^{-U_1} d\Omega^2 _2 - e^{-U_2} (\sigma_3 + a^{0})^2~,
\end{align}
with $ds_2 ^2$ a general 2D metric and the 1-form ansatz for the gauge potential
\begin{align} \label{eq: 5D potential ansatz}
    A^I_{5} &= a^I + b^I (\sigma_3 + a^0)~ \, .
\end{align}
In our conventions, the left invariant 1-forms
\begin{equation}
\label{eq: sigmas123}
\begin{aligned}
\sigma_{1} &=\sin \phi \, d \theta-\cos \phi \sin \theta \, d \psi~, \\
\sigma_{2} &=\cos \phi \, d \theta+\sin \phi \sin \theta \, d \psi~, \\
\sigma_{3} &=d \phi+\cos \theta \, d \psi~,
\end{aligned}
\end{equation}
parametrize $SU(2)$ with
\begin{align}
\label{eq: theta phi psi ranges}
    0 \leq \theta \leq \pi~, \qquad 0 \leq \phi \leq 4 \pi~, \qquad 0 \leq \psi \leq 2 \pi \, .
\end{align}
The ansatz \eqref{eq: 5D metric ansatz} suggests the vielbein
\begin{equation}
    \begin{aligned}
        e^{0} &= e^{0}_{\mu}dx^{\mu}~, & \quad
        e^{1} &= e^{1}_{\mu}dx^{\mu}~,  & \quad
        e^{2} &= e^{-\frac{1}{2}U_{1}}\sigma_1~, & \quad \\
        e^{3} &= e^{-\frac{1}{2}U_{1}}\sigma_2~, & \quad
        e^{4} &= e^{-\frac{1}{2}U_{2}}(\sigma_{3} + a^{0})~.
    \end{aligned}
\end{equation}
We use Greek indices to denote the curved coordinates $t$ and $R$ in 2D. For extremal near-horizon geometries, the 2D coordinates describe the AdS$_2$ throat of the solution. The dimensional reduction via \eqref{eq: 5D metric ansatz} and \eqref{eq: 5D potential ansatz} of the 5D Lagrangian \eqref{eq: 5D Lagrangian} introduces the scalar fields $U_1, U_2$ and $b^I$, along with the 1-forms $a^{0}, a^I$. All these fields depend only on the 2D coordinates.

The effective 2D Lagrangian density that follows from \eqref{eq: 5D Lagrangian} is given by
\begin{align} \label{eq: 2D bulk Lagrangian}
    \begin{split}
        \mathcal{L}_2 &= \frac{\pi}{G_{5}}e^{-U_{1}-\frac{1}{2}U_{2}}
        \Big\{(-{\cal R}_{2}  + 2e^{U_{1}} -\tfrac{1}{2}e^{2U_{1}-U_{2}}) \star 1 - \tfrac{1}{2}dU_{1}\wedge \star d\left( U_{1} + 2U_{2} \right)
       \\ & \left. \quad\quad -\tfrac{1}{2}e^{-U_{2}}da^{0} \wedge \star da^{0}- 2V 
        - G_{IJ}\Big((d a^{I} + b^{I} da^{0}) \wedge \star (d a^{J} + b^{J} da^{0}) + e^{2U_{1}}b^{I}b^{J} \star 1
        \right. \\ &  \quad\quad
        + e^{U_{2}}db^{I} \wedge \star db^{J} - dX^{I}\wedge \star dX^{J}\Big) + \tfrac{1}{3}e^{U_{1}+\frac{1}{2}U_{2}} c_{IJK}\left(\tfrac{3}{2}b^{I}b^{J}da^{K} + b^{I}b^{J}b^{K}da^{0}\right)\Big\}  \\
        & \quad\quad + 
        \frac{\pi}{G_{5}}d \left((e^{-U_{1}-\frac{1}{2}U_{2}}\star d (2U_{1}+U_{2}))-\tfrac{1}{6}b^{I}b^{J}a^{K}\right) \, .
        \end{split}
\end{align}
We denote the Ricci scalar of the reduced 2D metric $\mathcal{R}_{2}$ and the Hodge dual is now in 2D. 
The overall exponential factor $e^{-U_{1}-\frac{1}{2}U_{2}}$ comes from the 5D metric on a deformed $S^3$.
The first line in \eqref{eq: 2D bulk Lagrangian} is due to the reduction of the 5D Ricci scalar, which introduces additional kinetic and potential terms associated to the scalars $U_1$ and $U_2$, as well as for the 1-form $a^0$ in the beginning of the second line. 
The terms preceded by $G_{IJ}$ are the reduction of the Maxwell field which yield kinetic terms for the 1-forms $a^0, a^I$ and the scalars $b^I$, and the reduction of the kinetic term of $X^I$. The remainder of the third line of \eqref{eq: 2D bulk Lagrangian} is the Chern-Simons term. Finally, in the last line, there is a total derivative that is inconsequential for the equations of motion but is required in order that $\mathcal{L}_2$ \eqref{eq: 2D bulk Lagrangian} is the dimensional reduction of the
5D Lagrangian \eqref{eq: 5D Lagrangian}. The latter does not include the Gibbons-Hawking-York boundary term, the extrinsic curvature that appears separately in \eqref{eq: general 5D action}.

Boundary terms present an important subtlety that we will return to repeatedly in our study.
The Chern-Simons term in the 5D Lagrangian \eqref{eq: 5D Lagrangian} is not manifestly gauge invariant, but it transforms to a total derivative under a gauge variation. Gauge invariance could be restored by introducing a total derivative in the action. Such a term does not change the equations of motion but the resulting theory is not covariant in 5D, so there is a tension 
between important principles. The bulk part of the 2D Lagrangian \eqref{eq: 2D bulk Lagrangian} is not only covariant, it is
also manifestly gauge invariant: $a^I$ appears only as the field strength $da^I$. Manifest gauge invariance also applies to
$a^0$ which encodes 5D rotational invariance. These are benefits of reducing to 2D. 

From the dimensionally reduced Lagrangian density \eqref{eq: 2D bulk Lagrangian}, we can derive the equations of motion for the fields $U_{1}, U_{2}, a^{0}, a^{I}$ and $b^{I}$. The solutions to these 2D equations of motion are solutions of the 5D theory. 
The field equations for the 2D scalar fields are given by
\begin{align}
     \begin{split}
        \mathcal{E}_{U_1} &= d(e^{-U_{1}-\frac{1}{2}U_{2}}\star (dU_{1}+dU_{2})) + e^{-U_{1}-\frac{1}{2}U_{2}} \{  
        \left(\mathcal{R}_{2} + 2V - \tfrac{1}{2}e^{2U_{1}-U_{2}}\right)\star 1 \\& \quad + \tfrac{1}{2}dU_{1} \wedge \star (dU_{1} + 2 dU_{2})  + \tfrac{1}{2}e^{-U_{2}}da^{0} \wedge \star da^{0} + G_{IJ}\left((b^{I} da^{0} + d a^{I}) \wedge \star (b^{J} da^{0} + d a^{J})  \right. \\& \left. \quad - e^{2U_{1}}b^{I}b^{J} \star 1 + e^{U_{2}}db^{I} \wedge \star db^{J} -
        dX^{I}\wedge \star dX^{J} \right)\} = 0\, ,
    \end{split}
    \\
   \begin{split}
        \mathcal{E}_{U_2} &= d(e^{-U_{1}-\frac{1}{2}U_{2}}\star dU_{1}) + \tfrac{1}{2}e^{-U_{1}-\frac{1}{2}U_{2}}\{\left(\mathcal{R}_{2} + 2V - 2e^{U_{1}} + \tfrac{3}{2}e^{2U_{1}-U_{2}} \right)\star 1  \\& \quad 
        + \tfrac{1}{2}dU_{1} \wedge \star (dU_{1} + 2 dU_{2}) + \tfrac{3}{2}e^{-U_{2}}da^{0} \wedge \star da^{0} + G_{IJ}\left((b^{I} da^{0} + d a^{I}) \wedge \star (b^{J} da^{0} + d a^{J}) 
         \right.\\&  \left. \quad 
        + e^{2U_{1}}b^{I}b^{J} \star 1-  e^{U_{2}}db^{I} \wedge \star db^{J} -  dX^{I}\wedge \star dX^{J} \right)\} = 0\, ,
    \end{split}
    \\
    \begin{split}
        \mathcal{E}_{b^I} &= 2d(e^{-U_{1}+\frac{1}{2}U_{2}}G_{IJ} \star db^{J}) -2e^{-U_{1}-\frac{1}{2}U_{2}} G_{IJ} da^{0} \wedge \star (b^{J}da^{0}+da^{J})-2G_{IJ}e^{U_{1}-\frac{1}{2}U_{2}}b^{J} \star 1 \\&\quad + c_{IJK}b^{J}da^{K} + c_{IJK}b^{J}b^{K}da^{0} =0\, ,
    \end{split}
\end{align}
and the 1-forms satisfy
\begin{align} \label{eq: 2D EoM}
    \begin{split}
        \mathcal{E}_{a^0} &= - d (e^{-U_{1}-\frac{3}{2}U_{2}}\star da^{0}) 
        -  2 d (G_{IJ} b^{I}e^{-U_{1}-\frac{1}{2}U_{2}}\star (b^{J}da^{0} + da^{J})) + \tfrac{1}{3}c_{IJK} d(b^{I}b^{J}b^{K})=0~, 
        \\
        \mathcal{E}_{a^I} &= - 2 d \left( e^{-U_{1}-\frac{1}{2}U_{2}}G_{IJ} \star \left(b^{J} da^{0} + da^{J} \right)\right) + \tfrac{1}{2} c_{IJK} d\left(b^{J}b^{K}\right) = 0~.
    \end{split}
\end{align}

\subsection{An effective 1D theory} \label{subsection: the effective 1D theory}

We conclude the section by reducing the 2D reduced Lagrangian \eqref{eq: 2D bulk Lagrangian} to a one-dimensional radial effective theory where all of the functions that appear in the effective Lagrangian \eqref{eq: 2D bulk Lagrangian} are set to be exclusively radial functions, with respect to the radial coordinate $R$. 
In this additional reduction we pick a diagonal gauge for the 2d line element of \eqref{eq: 5D metric ansatz}:
\begin{equation}
\label{eq:2d line el}
    ds_2 ^2 = e^{2\rho} dt^2 - e^{2\sigma} dR^2 ~. 
\end{equation}
The operators $d$ and $\star$ acting on the fields in the Lagrangian \eqref{eq: 2D bulk Lagrangian} simplify with this ansatz. 
For example, the 2D Ricci scalar becomes
\begin{align}
     \mathcal{R}_2 &= 2e^{-\rho-\sigma}\partial_R(e^{-\sigma} \partial_R e^{\rho}) \, .
\end{align}
Second derivatives are awkward so it is advantageous to rewrite this term as
\begin{align}
\begin{split} \label{eq: Ricci 2 decomposition}
    e^{\rho+\sigma-U_{1}-\frac{1}{2}U_{2}} \mathcal{R}_2 &= 2 \partial_{R}\left(e^{\rho-\sigma-U_{1}-\frac{1}{2}U_{2}}\partial_{R}\rho\right) + e^{\rho+\sigma-U_{1}-\frac{1}{2}U_{2}}(e^{-2\sigma}\partial_{R}\rho) \partial_{R}\left(2U_{1}+U_{2}\right) \, ~.
\end{split}
\end{align}
The first term is a total derivative, an additional boundary term. To examine the total boundary contribution, we consider a constant radial slice at infinity. As we are now reducing to 1D, the boundary terms must be evaluated at the bounds for the time coordinate. This is trivial since there is no explicit time dependence. 
After dimensional reduction, the Gibbons-Hawking-York term in \eqref{eq: general 5D action}
corresponds to the total derivative 
\begin{align}
    \begin{split} \label{eq: GHY term}
        \mathcal{L}_{\text{GHY}} &= 
        \frac{2\pi}{G_{5}} 
        \partial_R \left( e^{\rho-\sigma-U_{1}-\frac{1}{2}U_{2}}\partial_{R}(\rho-U_{1}-\tfrac{1}{2}U_{2})\right) \, .
    \end{split}
\end{align}
The total derivative term in \eqref{eq: Ricci 2 decomposition}, the Gibbons-Hawking-York term \eqref{eq: GHY term}, and the boundary terms in the last line of \eqref{eq: 2D bulk Lagrangian} after dimensional reduction to 1D, precisely cancel. 
This leaves only the contribution arising from the Chern-Simons term
\begin{align}
\begin{split} \label{eq: 1D boundary}
    \mathcal{L}_{\text{bdry}} &=
    -\frac{1}{6}\frac{\pi}{G_{5}} d \left( c_{IJK} b^{I} b^{J} a^{K}_{t}\right)  \,.
\end{split}
\end{align}
This remaining boundary term in \eqref{eq: 1D boundary} is crucial as it will affect the conserved charges we seek to compute. We will comment on this in depth in the subsequent section~\ref{section: noether-wald surface charge}. In summary, the 1D Lagrangian density takes the form
\begin{equation}
\label{eq: 1D bulk Lagrangian}
    \begin{split}
        \mathcal{L}_{1}  &= \frac{\pi}{G_5} e^{\rho +\sigma-U_{1}-\frac{1}{2}U_{2}}\left[-e^{-2\sigma}(\partial_{R}\rho) \partial_{R}\left(2U_{1}+U_{2}\right) - \tfrac{1}{2}e^{-2\sigma} (\partial_R U_1) (\partial_R U_1 + 2 \partial_R U_2) \right. \\ 
        &\quad \left. - G_{IJ} e^{-2\sigma} \left( \partial_R X^I \partial_R X^J - e^{U_2} \partial_R b^I \partial_R b^J \right) + \tfrac{1}{2}e^{-U_2 - 2\rho -2\sigma} (\partial_R a^0 _t) ^2 \right. \\ 
        &\quad \left. + G_{IJ} e^{-2\rho - 2\sigma} (\partial_R a^I_t + b^I \partial_R a^0 _t)(\partial_R a^J _t + b^J \partial_R a^0 _t) -2V+2e^{U_1} - \tfrac{1}{2}e^{2U_1 - U_2} \right. \\ 
        &\quad \left. - G_{IJ} e^{2U_1} b^I b^J \right]  + \frac{\pi}{G_5} \frac{1}{3}c_{IJK} \left[-\frac{3}{2} b^I b^J \partial_R a^K _t -  b^I b^J b^K \partial_R a^0 _t \right]\,.
    \end{split}
\end{equation}
Having established the effective Lagrangian in 2D \eqref{eq: 2D bulk Lagrangian} and 1D \eqref{eq: 1D bulk Lagrangian}, we proceed in the next subsection with construction of the Noether-Wald surface charges in our theory.

\section{Noether-Wald surface charges} \label{section: noether-wald surface charge}

In this section, we review the Noether-Wald procedure for computing the conserved charge due to a general symmetry \cite{Wald:1993nt, Iyer:1994ys}. We specifically consider an isometry generated by a Killing vector and a gauge symmetry in the presence of Chern-Simon terms. In each case, we express
the conserved charge as a flux integral that is the
same for any surface surrounding the black hole.

\subsection{The Noether-Wald surface charge: general formulae} \label{subsection: Noether-Wald surface charge}

We consider a theory in $D$ dimensions described by a Lagrangian $\mathcal{L}$ that is presented as a $D$-form. The Lagrangian depends on fields $\Phi_{i}$ that include both the metric $g_{\mu\nu}$ and matter fields, as well as the derivatives of these fields. 

A symmetry $\zeta$ is such that the variation of $\mathcal{L}$ with respect to $\zeta$ 
is a closed form (locally), i.e. $d$ acting on a $D-1$ form $\mathcal{J}_\zeta$:
\begin{align} 
    \mathcal{L} ~~ \underset{\zeta}{\to} ~~\mathcal{L} + \delta \mathcal{L} = \mathcal{L} + d\mathcal{J}_\zeta~.
    \label{eq:symmvar}
\end{align}
The variation of the Lagrangian due to \textit{any} change in the fields is given by\footnote{In practice, when we solve for the Einstein equations, we will consider a variation of the metric and will not directly use \eqref{eq:varLag}.} 
\begin{align}
    \begin{split}
    \delta \mathcal{L}
    &=
    \delta \Phi_{i} \frac{\partial \mathcal{L}}{\partial \Phi_{i}} + (\partial_{\mu}\delta \Phi_{i})\frac{\delta \mathcal{L}}{\delta \partial_{\mu}\Phi_{i}}
    \\&=
    \delta \Phi_{i} 
    \left[ \frac{\partial \mathcal{L}}{\partial \Phi_{i}} - \partial_{\mu}\left(\frac{\delta \mathcal{L}}{\delta \partial_{\mu}\Phi_{i}}\right) \right] +  \partial_{\mu}\left(\delta \Phi_{i}\frac{\delta \mathcal{L}}{\delta \partial_{\mu}\Phi_{i}}\right)~.
    \label{eq:varLag}
    \end{split}
\end{align}
The usual variational principle determines the equations of motion $\mathcal{E}_{\Phi}$ as the vanishing of the expression in the square bracket. The remaining term, by definition, is the total derivative of the presymplectic potential 
\begin{align}
    \Theta^{\mu} &\equiv \delta \Phi_{i}\frac{\delta \mathcal{L}}{\delta \partial_{\mu}\Phi_{i}}~. 
    \label{eq:presymplecticdef}
\end{align}
In our informal notation, the left hand side of this equation is indistinguishable from a vector. However, the Lagrangian is a $D$-form and the $\delta$-type ``derivative" removes an entire 1-form. Therefore, the presymplectic potential $\Theta$ becomes
a $D-1$ form, with indices obtained by contracting the volume form with the vector that is normal to the boundary. A more precise version of \eqref{eq:varLag} reads
\begin{align} \label{eq: variation of L}
    \delta \mathcal{L}= \delta \Phi_{i}
    \left[ \frac{\partial \mathcal{L}}{\partial \Phi_{i}} - \partial_{\mu}\left(\frac{\delta \mathcal{L}}{\delta \partial_{\mu}\Phi_{i}}\right) \right]  + d \Theta[\Phi_{i}, \delta \Phi_{i}]~.
\end{align}
Comparing this formula for a general variation with its analogue \eqref{eq:symmvar} for a symmetry establishes $d\mathcal{J}_\zeta= d\Theta$ and so 
the $D-1$ form
\begin{align} \label{eq: current general formula}
    J_{\zeta} &= \mathcal{J}_{\zeta}  - \Theta[\Phi_{i}, \delta \Phi_{i}]~
\end{align}
is closed when the equations of motion $\mathcal{E}_{\Phi}$ are imposed. This identifies the familiar conserved Noether current associated to the symmetry $\zeta$. The corresponding Noether charge is
\begin{align}
    Q_{\zeta, \text{Noether}} = \int_\Sigma  J_{\zeta}~,
\end{align}
where $\Sigma$ is a Cauchy surface on the background manifold. Conservation amounts to this charge being the same on all Cauchy surfaces. Conceptually, the total charge is the same at all times. That is the point of conservation in a truly dynamical setting, but it is not terribly interesting in a stationary black hole spacetime which is, by definition, independent of time. 

For black holes it is important that, given the closed $(D-1)$ form $J_{\zeta}$, there exists a 
$(D-2)$-form $Q_\zeta$ such that
\begin{align}
\label{eq: JdQ}
     J_{\zeta} \cong d  Q_\zeta~. 
\end{align}
The $Q_\zeta$ is the Noether-Wald {\it surface} charge. It amounts to a conserved {\it flux} in the sense of Gauss' law: integration of the flux over any surface enclosing the source gives the same result. 

The surface charge $Q_\zeta$ is more subtle than the conserved charge integrated over an entire Cauchy surface.
The semi-equality $\cong$ reminds us that generally the closed form $J_\zeta$ is only $d$ of something locally so, in general, the charge $Q_\zeta$, is only defined up to $d$ of some $D-3$ form. Therefore, it does not necessarily satisfy Gauss' law.

One way around this is to evaluate the surface charge at infinity. For example, the presence of a Chern-Simons term can be interpreted physically as a charge density that obstructs flux conservation but this contribution is subleading at infinity and will not contribute to $Q_{\zeta,\text{Noether}}$.

Alternatively, following \cite{Kastor:2008xb, Kastor:2009wy, Ortin:2021ade, Cano:2023dyg, Cassani:2023vsa}, we can modify our definition of the surface charge by adding a $D-2$ form to $Q_{\zeta}$. This new surface charge satisfies a Gauss law and can be integrated at any given surface $\Sigma$. 

A third approach \cite{Sen:2007qy}, is the one taken in this paper. It is to compute the surface charges in a dimensionally reduced 2D theory. 

Moreover, we integrate by parts such that in the process of dimensional reduction to 2D, we ensure gauge invariance. We will carry this procedure out in section~\ref{subsection: Chern-Simons Terms}. Therefore, in this case, $Q_{\zeta}$ will satisfy a Gauss law.

The procedure for computing the conserved charges is extremely general. In the following, we make the abstract procedure explicit for two particular symmetries: isometries generated by a spacetime Killing vector $\xi$ and gauge symmetries $\lambda$ in the presence of Chern-Simons terms.

\subsection{Killing vector fields}
\label{subsection: killing vector fields}

A Killing vector $\xi$ generates a spacetime isometry.
It transforms the Lagrangian as 
\begin{align}
    \delta_{\xi}\mathcal{L} 
    &=
    L_{\xi}\mathcal{ L}~. 
    \label{eq:delL KV}
\end{align}
Here $L_\xi$ is the Lie derivative along the Killing vector $\xi$. \\
The Lie derivative acting on a general form $\omega$ is given by Cartan's magic formula
\begin{align}
    L_{\xi}  \omega = d (i_{\xi}  \omega) + i_{\xi}d  \omega~.
\end{align}
Since $\mathcal{ L}$ is a $D$-form it must be closed
$d\mathcal{ L}=0$ and then the Lie derivative becomes
\begin{align} \label{eq: variation of L version 2}
    \delta_{\xi}\mathcal{L} 
    &=
    L_{\xi}\mathcal{L}
    =
    d(i_{\xi}\mathcal{L}) + i_{\xi}(d\mathcal{L}) = d(\xi \cdot \mathcal{L})~,
\end{align}
where $\cdot$ denotes the contraction of $\xi$ with the first index of $\mathcal{ L}$. Comparing \eqref{eq: variation of L version 2} with 
\eqref{eq:symmvar}, we identify
\begin{align} \label{eq: d of xi L}
   {\cal J}_\xi  &= \xi \cdot \mathcal{L}~,
\end{align}
up to a closed form that is unimportant in our application.
Thus, for a Killing vector $\xi$, 
the Noether current 
\eqref{eq: current general formula}
becomes
\begin{align} \label{eq: J for killing vector}
     J_{\xi} = \xi \cdot \mathcal{ L} -  \Theta[\Phi,\mathcal{ L}_{\xi}\Phi]~.
\end{align}
The computations show that 
this current $(D-1)$ form is closed on-shell. In other words, it is conserved when the equations of motion are satisfied.

\subsection{Incorporating gauge invariance}
\label{subsection: gauge invariance}

We now consider a \textit{gauge invariant} Lagrangian and compute the conserved current as defined in \eqref{eq: current general formula} for the conserved charges of the theory, whether derived from spacetime isometries or gauge invariance. 

The relevant gauge invariant Lagrangian is the one defined in \eqref{eq: general 5D action} \textit{without} the Chern-Simons term. In other words, we consider the Lagrangian density
\begin{align} \label{eq: 5d Lagrangian pot and kin terms}
    \begin{split}
        \mathcal{L}_{5,\text{pot}} & = -\frac{1}{16 \pi G_{5}} \sqrt{g_5}\left(\mathcal{R}_{5}+2V\right)\, ,
        \\
        \mathcal{L}_{5,\text{kin}} &= \frac{1}{16 \pi G_{5}} \sqrt{g_5} \left(- \frac{1}{2}G_{I J} F^{I}_{5,AB}F^{J,AB}_{5}+G_{I J} \nabla^{A} X^{I} \nabla_{A} X^{J} \right)~.
    \end{split}
\end{align}
We use early capital Latin indices $A, B, \ldots$ to denote 5D coordinates. The Lagrangian 
$\mathcal{L}_{5,\text{kin}}+\mathcal{L}_{5,\text{pot}}$ 
is manifestly gauge invariant
\begin{align}
\label{eq: delta L  pot + kin alpha}
    \delta_{\alpha} (\mathcal{L}_{5,\text{pot}} + \mathcal{L}_{5,\text{kin}}) &= 0~,
    %\\
   %\delta_{\xi} (\mathcal{L}_{5,\text{pot}} + \mathcal{L}_{5,\text{kin}}) &= \nabla_{A}\left(\xi^{A}(\mathcal{L}_{5,\text{pot}} + \mathcal{L}_{5,\text{kin}})\right)~.
\end{align}
As detailed in the previous subsection, there is a conserved charge for any Killing vector that generates a spacetime isometry. According to \eqref{eq: variation of L version 2},
the Lagrangian 
$\mathcal{L}_{5,\text{kin}}+\mathcal{L}_{5,\text{pot}}$ 
transforms as
\begin{align}
\label{eq: delta L  pot + kin xi}
   \delta_{\xi} (\mathcal{L}_{5,\text{pot}} + \mathcal{L}_{5,\text{kin}}) &= \nabla_{A}\left(\xi^{A}(\mathcal{L}_{5,\text{pot}} + \mathcal{L}_{5,\text{kin}})\right)~.
\end{align}
The presymplectic potential \eqref{eq:presymplecticdef} for $\mathcal{L}_{5,\text{pot}}$ is
\begin{align}
\begin{split}\label{eq: Theta pot}
    \Theta^{A, 5}_{\xi, \text{pot}}&= 
    \frac{1}{16\pi G_{5}}\sqrt{g_5} \left(\nabla_{B}\nabla^{A}\xi^{B}+\nabla_{B}\nabla^{B}\xi^{A} -2 \nabla^{A}\nabla_{B}\xi^{B} \right)
    \\&=\frac{1}{16\pi G_{5}}\sqrt{g_5} \left(\nabla_{B}\left(\nabla^{B}\xi^{A}-\nabla^{A} \xi^{B} \right) + 2R^{AB}\xi_{B}\right) \,,
\end{split}
\end{align}
where in the second line, we have used the commutator relation for two covariant derivatives. In addition, the presymplectic potential for the kinetic terms $\mathcal{L}_{5,\text{kin}}$ is
\begin{align} \label{eq: Theta kin}
    \Theta^{A,5}_{\alpha, \xi, \text{kin}} &= \frac{1}{16\pi G_{5}}\sqrt{g_5} G_{IJ} \left( 2F^{I,AB}_{5}\left(\xi^{C} F^{J}_{5,CB}+\nabla_{B}\left(\xi^{C} A^{J}_{5,C} + \alpha^{J}_{5} \right)\right) \right) \, ,
\end{align}
where we have used the variation
\begin{align}
    \delta A^{I}_{A,5} &= \delta_{\xi}A^{I}_{5,A} + \delta_{\alpha}A^{I}_{5,A} = \xi^{B} F^{I}_{5,BA}+\nabla_{A}\left(\xi^{B} A^{I}_{5,B}\right) + \nabla_{A}\alpha^{I}\,.
\end{align}
Inserting the variations \eqref{eq: delta L  pot + kin alpha} and \eqref{eq: delta L  pot + kin xi} and the presympletic potentials given in \eqref{eq: Theta pot} and \eqref{eq: Theta kin} into the current density \eqref{eq: current general formula}, we find
\begin{align}
\begin{split}
    J^{A}_{\alpha, \xi} &= 
    \frac{1}{16\pi G_{5}} \sqrt{g_5} \Big[\left.- \nabla_{B}\left(\nabla^{B}\xi^{A}-\nabla^{A} \xi^{B} \right)- 2\nabla_{B}\left(G_{IJ}F^{I,AB}(\xi^{C}A^{J}_{C}+\alpha^{J})\right) 
    \right. \\& \left. \quad\quad\quad\quad\quad\quad\quad\quad\quad\quad\quad\quad\quad\quad\quad\quad\quad\quad -2\xi_{B}\mathcal{E}_{g}^{B} - 2\mathcal{E}_{J,A_{5}}^{A}(\xi^{C}A^{J}_{C}+\alpha^{J})\right. \Big] ~,
\end{split}
\end{align}
where the second line is proportional to the equations of motion $\mathcal{E}_{g}^{B}$ and $\mathcal{E}_{J,A_{5}}^{A}$ and vanish on-shell giving 
\begin{equation} \label{eq: current on shell without CS}
\begin{aligned}
    J^{A}_{\alpha, \xi} &= - \frac{1}{16\pi G_{5}} \sqrt{g_5} \nabla_{B}\Big[\left(\nabla^{B}\xi^{A}-\nabla^{A} \xi^{B} \right) +  2\left(G_{IJ}F^{I,AB}(\xi^{C}A^{J}_{C}+\alpha^{J})\right) \Big].
\end{aligned}
\end{equation}
The Noether-Wald surface charges of the theory can now be read off from the current \eqref{eq: current on shell without CS}. To find the conserved charges, we integrate over a surface $\Sigma$ enclosing the source and we find
\begin{align} \label{eq: 5D charges}
    \begin{split}
        Q_{\alpha} &=  - \frac{1}{8\pi G_{5}} \int_{\Sigma} d\Sigma_{AB} \sqrt{g_5} G_{IJ}F^{I,AB}\alpha^{J} \, ,
        \\
        Q_{\xi} &=  - \frac{1}{16\pi G_{5}} \int_{\Sigma} d\Sigma_{AB} \sqrt{g_5} \left[ \left(\nabla^{B}\xi^{A}-\nabla^{A} \xi^{B} \right) + 2G_{IJ}F^{I,AB}\xi^{C}A^{J}_{C}\right]
         \, .
    \end{split}
\end{align}

\subsection{Chern-Simons Terms}
\label{subsection: Chern-Simons Terms}

The charge $Q_{\xi}$ that corresponds to angular momentum depends explicitly on the gauge field $A^J$ whereas the electric charges $Q_\alpha$ depend on the field strength. When Chern-Simons terms are taken into account, $Q_\alpha$ also depends on the gauge field $A^J$. This gauge dependence renders the value of the charges ambiguous.

To address the situation, we dimensionally reduce the theory \eqref{eq: general 5D action} to 2D, as was done in subsection~\ref{subsection: the effective 2D theory} and express the resulting action as a covariant theory in 2D \cite{Sen:2007qy}. As part of the process, we must ensure that the field strength does not have a nonzero flux through the squashed sphere. This can be achieved by adding total derivatives before the dimensional reduction to remove the derivatives acting on the gauge potentials and gauge fields that act nontrivally through the squashed sphere.

We now show how this can be done. Let us consider the Lagrangian \eqref{eq: 5d Lagrangian pot and kin terms} along with the five-dimensional Chern-Simons term of the form

\begin{align} \label{eq: 5d Lagrangian CS term}
  \mathcal{L}_{5,\text{CS}} = - \frac{1}{16\pi G_{5}} \frac{1}{6} c_{IJK}  F^{I}_{5} \wedge F^{J}_{5} \wedge A^{K}_{5}~.
\end{align}
We are interested in transforming \eqref{eq: 5d Lagrangian CS term} by the inclusion of total derivatives such that the potential term associated to the electric charge is manifestly gauge invariant. Note this procedure is not covariant in 5D and therefore we explicitly break covariance along the way. However, because of the dimensional reduction, the 2D Lagrangian still remains covariant.

We consider the ansatz in \eqref{eq: 5D metric ansatz} such that the potential and gauge fields are of the form
\begin{align} \label{eq: A and F reduced}
    \begin{split}
        A^{I}_{5} &= A^{I}_{5,A} dx^{A} = A^{I}_{5,\mu}dx^{\mu} + A^{I}_{5,a}dx^{a}~,
        \\
        F^{I}_{5} &= \frac{1}{2} F^{I}_{5,AB} \, dx^{A} \wedge dx^{B} =  \frac{1}{2}F^{I}_{5,\mu\nu}dx^{\mu} \wedge dx^{\nu} +  F^{I}_{5,\mu a} dx^{\mu} \wedge dx^{a} + \frac{1}{2} F^{I}_{5,ab} dx^{a} \wedge dx^{b}~,
    \end{split}
\end{align}
where lowercase Latin indices denote the indices on the compact space and as before, the Greek indices correspond to the 2D space. Expanding out the Chern-Simons term in component form using \eqref{eq: A and F reduced}, there are two types of terms, having the following structure of indices: $F^{I}_{\mu \nu} F^{J}_{bc} A^{K}_{a}$ and $F^{I}_{\mu a} F^{J}_{bc} A^{K}_{\nu}$. Only for the second expression we must transfer the derivative such that in the process of dimensional reduction, we find it to be gauge invariant in the 2D theory. This means the integration by parts of this term takes the form
\begin{align} \label{eq: total derivative term CS}
    c_{IJK}\epsilon^{\mu abc \nu}F^{I}_{\mu a} F^{J}_{bc} A^{K}_{\nu} = 2 c_{IJK}\epsilon^{\mu abc \nu}\left(\partial_{\mu}(A^{K}_{\nu}A^{I}_{a}F^{J}_{bc}) - (\partial_{\mu}A^{K}_{\nu})(A^{I}_{a}F^{J}_{bc})\right)~,
\end{align}
and the presymplectic potential is found to be
\begin{align}
\begin{split}
    \Theta^{A,5}_{\alpha, \xi, \text{CS}} &= \frac{1}{16\pi G_{5}}\sqrt{g_5} \left[\frac{1}{6}c_{IJK}\left( \epsilon^{ABCDE} F^{I}_{BC}A^{J}_{D}\left(\xi^{F} F^{K}_{FE}+\nabla_{E}(\xi^{F} A^{K}_{F}) + \nabla_{E}\alpha^{K}\right) \right) \right] \\ & \quad - \frac{1}{8\pi G_{5}}\sqrt{g_5}\left[c_{IJK}\epsilon^{A abc \nu}A^{I}_{a} F^{J}_{b c} \nabla_{\nu} \alpha^{K}\right]\, ,
\end{split}
\end{align}
where the last term is the contribution of \eqref{eq: total derivative term CS} coming from adding a total derivative. To investigate the current and the Noether-Wald surface charges, we proceed to dimensionally reduce over the squashed $S^3$ where covariance over the 2D spacetime is still maintained. The 5D rotational isometries in $\varphi$ and $\psi$ take on a different role in the 2D perspective. Moreover, we find that they become 2D gauge transformations of $a^{0}$ and $a^{I}$ coming from the dimensionally reduced potential $A^I$ \eqref{eq: 5D potential ansatz}.

\subsection{The 2D conserved charges} \label{subsection: the 2D charges}

The 2D Lagrangian \eqref{eq: 2D bulk Lagrangian} inherits some symmetries from the 5D theory \eqref{eq: 5D Lagrangian}, including gauge symmetry associated with the 5D gauge potential $A^{I}$ and rotational isometries associated to the Killing vectors $\partial_{\phi}$ and $\partial_{\psi}$. In the 2D theory, all symmetries become gauge symmetries and have associated charges. 
We denote the 2D charge originally coming from the 5D rotational isometries $J$ and the 2D charges originally coming from the 5D gauge transformations $Q_{I}$. These 2D gauge transformations are associated to $a^{0}$ and $a^{I}$ as they come from the dimensionally reduced potential $A^I$ \eqref{eq: 5D potential ansatz}. Therefore, we consider the following symmetries
\begin{align} \label{eq: lambda and chi symmetries}
    \delta_{\lambda} a^{0} = d\lambda, \qquad \delta_{\chi} a^{I} = d\chi^{I}~,
\end{align}
with total corresponding conserved current
\begin{align} \label{eq: current}
    J_{\lambda, \chi} &= J_{\lambda} + J_{\chi} = \sum_{i=\lambda, \chi}\left(\mathcal{J}_{i} - \Theta_{i} \right), 
\end{align}
where $J_{\lambda}$ and $J_{\chi}$ are the currents corresponding to $\lambda$ and $\chi$, respectively, and the second equality is given by \eqref{eq: current general formula}. The effective 
2D Lagrangian \eqref{eq: 2D bulk Lagrangian} is manifestly gauge invariant
and the 
variations 
with respect to each symmetry \eqref{eq: lambda and chi symmetries} yield
\begin{align}
    \delta_{\lambda} \mathcal{L}_{2} &= d\mathcal{J}_{\lambda} = 0~, \quad \delta_{\chi} \mathcal{L}_{2} = d\mathcal{J}_{\chi} = -\frac{\pi}{G_{5}}\frac{1}{6}c_{IJK}d\left(b^{I}b^{J}d\chi^{K}\right)~.
  \end{align}
The presymplectic potentials given in \eqref{eq: variation of L} become
\begin{align}
    \Theta_{\lambda} &= - \frac{\pi}{G_{5}}e^{-U_{1}-\frac{1}{2}U_{2}} \left[e^{-U_{2}}\star da^{0} +  2G_{IJ} b^{I} \star\left(b^{J}da^{0} + da^{J}\right)\right]d\lambda + \frac{\pi}{G_{5}}\frac{1}{3}c_{IJK}  b^{I}b^{J}b^{K}d\lambda~,
    \\
    \Theta_{\chi} &= - \frac{\pi}{G_{5}}e^{-U_{1}-\frac{1}{2}U_{2}}\left[2G_{IJ} d \chi^I \wedge \star \left(b^{J}da^{0} + da^{J}\right) \right] + \frac{\pi}{G_{5}}\frac{1}{3}c_{IJK}  b^{I}b^{J}d \chi^{K}~.
\end{align}
We used the symmetries \eqref{eq: lambda and chi symmetries} and included the additional total derivative term \eqref{eq: total derivative term CS}. Using 
the equations of motion \eqref{eq: 2D EoM},
the on-shell current \eqref{eq: current}  can be recast in the form of \eqref{eq: JdQ}:
\begin{align}
    \begin{split} \label{eq: J for lambda and chi}
        J_{\lambda,\chi} &\cong
        \frac{\pi}{G_{5}}d\left[\lambda \left(e^{-U_{1}-\frac{1}{2}U_{2}} \left[e^{-U_{2}}\star da^{0} +  2G_{IJ} b^{I} \star \left(b^{J}da^{0} + da^{J}\right) \right] - \frac{1}{3}c_{IJK}b^{I}b^{J}b^{K}\right) \right.
        \\& \left. \qquad \qquad + \chi^{I} \left( 2e^{-U_{1}-\frac{1}{2}U_{2}}G_{IJ} \star \left(b^{J}da^{0} + da^{J}\right)  - \frac{1}{2}c_{IJK}b^{J}b^{K} \right)\right].
    \end{split}
\end{align}
The conserved charges $J$ and $Q_{I}$ can be directly read off from  \eqref{eq: J for lambda and chi} and we find

\begin{align} \label{eq: 2D charges}
    \begin{split}
       J &= \frac{\pi}{G_{5}}\left[e^{-U_{1}-\frac{1}{2}U_{2}}\left[e^{-U_{2}}\star da^{0} +  2G_{IJ} b^{I} \star \left(b^{J}da^{0} + da^{J}\right)\right] - \frac{1}{3}c_{IJK} b^{I}b^{J}b^{K}\right]~,
        \\
        Q_{I} &= \frac{\pi}{G_{5}}\left[2e^{-U_{1}-\frac{1}{2}U_{2}} G_{IJ} \star \left(b^{J}da^{0} + da^{J} \right) - \frac{1}{2} c_{IJK}b^{J}b^{K} \right]~.
    \end{split}
\end{align}
From now on, we use the rescaled charges
\begin{align}
\label{eq: JQ rescale}
    \widetilde{J} &\equiv \frac{4G_{5}}{\pi}J~, \qquad
    \widetilde{Q} \equiv \frac{4G_{5}}{\pi}Q_{I}~.
\end{align}
The charges and the current are indeed conserved including the charge associated to $a^{I}$ since we demanded gauge invariance at the level of the Lagrangian in \eqref{eq: 2D bulk Lagrangian}. This added a total derivative that shifted the charge but did not affect the equations of motion. Moreover, the charges computed in 2D are proportional to those computed in 5D. In the 1D reduction \eqref{eq: 1D bulk Lagrangian}, the charges take the following form
\begin{align} \label{eq: 2D charges explicit}
    \begin{split}
        \widetilde{J}_{} &= 4\left(e^{-U_{1}-\frac{3}{2}U_{2}-\rho-\sigma}
       \left(\partial_{R}a^{0}_{t} +  2G_{IJ} e^{U_2} b^{I} \left(
       \partial_{R}a^{J}_{t} + b^{J} \partial_{R}a^{0}_{t}\right)\right) - \frac{1}{3}c_{IJK} b^{I}b^{J}b^{K}\right)~,
        \\
        \widetilde{Q}_{I} &= 4\left(2e^{-U_{1}-\frac{1}{2}U_{2} -\rho-\sigma} G_{IJ} \left(
        \partial_{R}a^{J}_{t} + b^{J}\partial_{R}a^{0}_{t} \right) - \frac{1}{2} c_{IJK}b^{J}b^{K} \right)~.
    \end{split}
\end{align}
These formulae are essential for the radial flow in the black hole background. A {\it very} rough reading is that each of the conserved charges $\widetilde{J}$ and $\widetilde{Q}_I$ are radial derivatives of their conjugate potentials $a^0_t$, $a^I_t$, as in elementary electrodynamics. With this na\"{\i}ve starting point, the overall factors depending on $U_{1}, U_{2}, \rho$ and $\sigma$ serve to take on the non-flat spacetime into account and $G_{IJ}$ incorporates special geometry as required by symmetry. All remaining terms depend on $b^I$ and take rotation into account in a manner that combines kinematics (rotation ``looks" like a force) and electrodynamics (electric and magnetic fields mix in a moving frame). These effects defy simple physical interpretations. 

From our point of view, the formulae \eqref{eq: 2D charges explicit} for the charges 
$\widetilde{J}$ and $\widetilde{Q}_{I}$ are  complicated functions of various fields, each of which are themselves nontrivial functions of the radial coordinate. Our construction shows that symmetry
guarantees that these {\it combinations} must be independent of the radial position, within the framework of our {\it ansatz}. 

In the following section we study the conditions that {\it supersymmetric} AdS$_5$ black holes must satisfy. The vanishing of the supersymmetry variations of the theory for a subset of the supersymmetries always imposes first-order radial differential equations on the joint geometry/matter configuration. We refer to these first order equations as {\it flow equations}. They are very constraining but, 
as usual for first order equations imposed by supersymmetry, they are not sufficient to determine 
the solution. The raison d'\^{e}tre of this entire section is that the additional data needed, sometimes referred to as the integrability conditions, is furnished by the conserved charges.

\section{The flow equations} \label{section: the flow equations}

In this section we derive the first order flow equations for AdS$_5$ black holes. They follow from preservation of supersymmetry, complemented by conservation of the charges. We study the flow equations using two perturbative expansions: one starting at the near-horizon and one starting at the asymptotic boundary. Enforcing the conservation of charges at both the horizon and at the asymptotic boundary allows us to make contact between the two expansions.

\subsection{Supersymmetry conditions}
\label{subsection: Supersymmetric variations and the (1+4) split}

We study bosonic backgrounds that preserve some supersymmetry \cite{gunaydin1984geometry, Gauntlett:2002nw, Gutowski:2004yv}. Thus there exists a supersymmetric spinor $\epsilon^\alpha$ for which the gravitino and the dilatino variations vanish. This condition amounts to 
\begin{align}
\label{eq: gaugino var}
    0 &= \left[ G_{IJ}  \left( \frac{1}{2} \gamma^{AB} F^J_{~AB}   - \gamma^A \nabla_A X^J \right) \epsilon^\alpha - \xi_I \epsilon^{\alpha \beta} \epsilon^\beta \right] 
\partial_i X^I  ~, \\ 
\label{eq: gravitino var}
 0 &= \left[ (\partial_A - \frac{1}{4} \omega_A^{BC} \gamma_{BC}) + \frac{1}{24} (\gamma_A^{~BC} - 4\delta^{~B}_A \gamma^C) X_I F^I_{BC} \right] 
 \epsilon^\alpha  +\frac{1}{6}  \xi_I (3A^I_A - X^I\gamma_A) \epsilon^{\alpha \beta} \epsilon^\beta  ~,
\end{align}
where 
$\epsilon^\alpha \ (\alpha=1,2)$ are symplectic Majorana spinors. In the following, we recast these variations as radial flow equations.

For the analysis of supersymmetry, it is
convenient to split the 5D spacetime geometry into $(1+4)$ dimensions
as 
\begin{align}
\label{eq:14ds5}
    ds_5^2 &= f^2 (dt + w \sigma_3 )^2 - f^{-1} ds_4 ^2 ~, \\
    \label{eq:14ds4}
    ds_4 ^2 &= g_m ^{-1} dR^2 + \frac{1}{4}R^2 (\sigma_1 ^2 + \sigma_2 ^2 + g_m \sigma_3 ^2)  \, .
\end{align}
This form highlights the 4D base space $ds_4 ^2$ which is
automatically K\"ahler.
This can be shown by picking the flat vielbein
\begin{equation}
\begin{split}
\label{eq: 14 vierbein}
    e^1 &= g_m ^{-1/2} dR ~, \quad
    e^2 = \frac{1}{2}R \sigma_1 ~, \quad
    e^3 = \frac{1}{2}R \sigma_2 ~, \quad
    e^4 = \frac{1}{2}R g_m ^{1/2}\sigma_3 ~,
\end{split}    
\end{equation}
which gives the manifestly closed K\"ahler 2-form 
\begin{equation} 
\label{eq: J Kahler def}
J^{(1)} = \epsilon (e^1 \wedge e^4 - e^2 \wedge e^3)~.
\end{equation}
The symbol $\epsilon=\pm 1$ denotes the orientation of the base manifold. It should not be confused with the supersymmetry parameter $\epsilon^\alpha$.

The $(1+4)$ split of the 5D gauge potential $A^I$ defined in \eqref{eq: 5D potential ansatz} can be expressed as
\begin{equation}
\begin{split}
\label{eq:AIdict}
    A^I &= fY^I (dt + w \sigma_3) + u^I \sigma_3 ~.
\end{split}
\end{equation}
In the rest of the paper, as well as in Appendix~\ref{appendix: supersymmetry}, we use lowercase Latin letters for the four spatial indices.

\subsection{Dictionary between the $(1+4)$ and the $(2+3)$ splits}
We can relate the $(1+4)$ split introduced in the previous subsection to simplify the supersymmetric variations \eqref{eq: gaugino var} and \eqref{eq: gravitino 
var}, to the $(2+3)$ split \eqref{eq: 5D metric ansatz} and \eqref{eq: 5D potential ansatz} that was used earlier to perform the reduction from 5D to 2D. The 5D geometry 
\eqref{eq: 5D metric ansatz} with the diagonal gauge
\eqref{eq:2d line el} for the 2D line element is
\begin{equation}
    \label{eq: 5D metric ansatz full}
    ds_5 ^2 = e^{2\rho}dt^2 - e^{2\sigma}dR^2 - e^{-U_1}(\sigma_1 ^2 + \sigma_2 ^2) - e^{-U_2} (\sigma_3 + a^0)^2 ~.
\end{equation}
By identifying the metric components of \eqref{eq:14ds5} and \eqref{eq: 5D metric ansatz full}, we find the 
dictionary of variables in the $(2+3)$ split of the 5D line element $ds_5^2$, expressed in terms of the variables in the $(1+4)$ split
\begin{equation}
    \begin{aligned} \label{eq:geometrydict}
        e^{-U_1} &= \frac{1}{4}R^2 f^{-1}~, & \quad
        e^{-U_2} &= \frac{1}{4}R^2 g_m f^{-1}-f^2 w^2~, & \quad b^I &= fY^I w + u^I ~, \\
        e^{2\sigma} &= f^{-1} g_m ^{-1} ~ , & \quad
        a^0 _t &= \frac{-f^2 w}{\tfrac{1}{4}R^2 g_m f^{-1}-f^2 w^2} ~, & \quad e^{2\rho} &= f^2 - \frac{f^4w^2}{(\frac{1}{4}R^2 g_m f^{-1}-f^2 w^2)^2 } ~.
    \end{aligned}
\end{equation}
In this section we primarily use the $(1+4)$ variables $fX^I$, $u^I$, $w$ and $g_m$, along with the conserved 
charges $Q_I$ and $J$. 

As noted in the previous subsection, the 4D base of the $(1+4)$ split \eqref{eq:14ds5} is automatically K\"ahler. In the variables of the $(2+3)$ split in \eqref{eq:geometrydict}, the K\"ahler condition amounts to the relation
\begin{equation}
    \label{eq:Kahlercond}
    e^{\sigma + \rho - U_2/2} = \frac{1}{2}R~,
\end{equation}
between $\rho$, $\sigma$, and $U_2$. This is 
explained further in Appendix~\ref{appendix subsection: Kahler condition}.

\subsection{The attractor flow equations}
The preserved supersymmetries are defined by the projections on the spinors $\epsilon^\alpha$ 
\begin{align}
\label{eq:proj0}
    \gamma^0 \epsilon^\alpha &= \epsilon^\alpha ~, \\
    \label{eq:projJ}
    \frac{1}{4}J^{(1)}_{mn} \gamma^{mn} \epsilon^\alpha &= -\epsilon^{\alpha \beta} \epsilon^\beta ~.
\end{align}
The $J^{(1)}_{mn}$ are components of the K\"{a}hler form $J^{(1)}$ \eqref{eq: J Kahler def}, and the spatial gamma matrices $\gamma^m$ satisfy the usual Clifford algebra. The details on the simplification of the equations \eqref{eq: gaugino var} and \eqref{eq: gravitino var} are presented in Appendix~\ref{appendix: supersymmetry}. The result is the following differential conditions on the variables $fX^I$, $u^I$, $w$ and $g_m$

\begin{align}
\label{eq:susyrelation1}
    0 &= G_{IJ}\left(\partial_R(fY^I) -\partial_R (fX^I)\right)\partial_i X^J ~,
    \\
\label{eq:susyrelation2}
    0 &=\left(\partial_{R^2} + \frac{1}{R^2}\right)u^I - \frac{1}{2}\epsilon f^{-1}c^{IJK} X_J \xi_K ~,
    \\
\label{eq:susyrelation3}
    0 &= \left(\partial_{R^2} - \frac{1}{R^2}\right)w + \frac{1}{2}f^{-1}X_I \left(\partial_{R^2}-\frac{1}{R^2}\right)u^I ~,
    \\
\label{eq:susyrelation4}
    0&= - \epsilon R^2 (\partial_{R^2} g_m) + 2\epsilon(1-g_m) + 2\xi_I u^I~. 
\end{align}
The variation \eqref{eq:susyrelation1} allows for the electric potential $fY^I$ in 
\eqref{eq:AIdict} to be identified with the scalar field $fX^I$, and thus
\begin{equation} \label{eq:AI}
    A^I = fX^I (dt + w \sigma_3) + u^I \sigma_3  ~. 
\end{equation}
This sets the variables $a^I$, $a^0$ and $b^I$ in the  decomposition \eqref{eq: 5D potential ansatz} to
\begin{equation}
\label{eq:potential field 23 to 14}
   a^I + b^I a^0 = fX^I dt \ , \ b^I = fX^I w + u^I ~,
\end{equation}
with $a^0 = a^0 _t dt$ given in \eqref{eq:geometrydict}.

In the limit of ungauged supergravity, $fX^{I}$ is a harmonic function. Then, the identification of $X^I=Y^I$ means that the corresponding electric potential is also a harmonic function. We may expect the same functional dependence in the case of gauged supergravity.\footnote{There are $n_V+1$ potentials $Y^I$ and $n_V$ scalars $X^I$ so there is freedom to adjust a single integration constant that we do not exploit. It is unclear to us if this freedom is physically significant.}

In the context of a black hole, solutions to the supersymmetry conditions (\ref{eq:susyrelation1}-\ref{eq:susyrelation4})
are specified in part by the conserved charges of the theory. The charges in $\widetilde{J}$ and $\widetilde{Q}_I$ in \eqref{eq: 2D charges explicit} are expressed in terms of 
the $(2+3)$ variables $U_1, U_2, b^I, a^0 _t, a^I _t $ but
we can recast them in terms of $(1+4)$ variables $f, u^I, w, g_m$
using the dictionary for the geometry \eqref{eq:geometrydict} and the potential \eqref{eq:potential field 23 to 14}.
We can also remove most of the derivatives in the 
equations \eqref{eq: 2D charges explicit} for the charges $\widetilde{J}$ and $\widetilde{Q}_I$ using the radial equations for the variables $u^I$, $w$ and $g_m$ (\ref{eq:susyrelation2}-\ref{eq:susyrelation4}). 
Our final expressions of the charges, which we use for the remainder of the section are given by
\begin{align} \label{eq: 2D charges in 1+4}
    \begin{split}
        \widetilde{J} &= \widetilde{Q}_I u^I + \frac{2}{3}c_{IJK}u^I u^J u^K -R^2 g_m \left(f^{-1}X\cdot u + 2w -\frac{1}{2}\epsilon R^2 f^{-3} \xi \cdot fX\right) \\& \quad +2R^2 w (1+\epsilon \xi \cdot u) ~, 
        \\
        \widetilde{Q}_{I} &= -2c_{IJK}u^J u^K -2\epsilon w R^2 \xi_I -g_m R^4 \partial_{R^2} (f^{-1}X_I)~.
    \end{split}
\end{align}
The second of these equations is a first order differential equation for $f^{-1}X_I$. Together with the three radial differential equations (\ref{eq:susyrelation2}-\ref{eq:susyrelation4}) for the variables $u^I$, $g_m$ and $w$, we find
the four equations
\begin{subequations}
\begin{align}
    \left(\partial_{R^2} + \frac{1}{R^2}\right)u^{I} &= \frac{1}{2}\epsilon c^{IJK} (f^{-1}X_J)\xi_K ~, \label{eq: flow eq uI}\\ 
    \left(\partial_{R^2} + \frac{2}{R^2}\right)g_m &= \frac{2}{R^2}+\frac{2}{R^2}\epsilon \xi_I u^I ~\, , \label{eq: flow eq gm}
    \\
    \left(\partial_{R^2} - \frac{1}{R^2}\right)w &= -\frac{1}{2}f^{-1}X_I \left(\partial_{R^2} - \frac{1}{R^2}\right)u^I~, \label{eq: flow eq w}
    \\
    R^4 \partial_{R^2}(f^{-1}X_I ) &=-\frac{1}{g_m}\left(\widetilde{Q}_I + 2 c_{IJK} u^J u^K + 2  \epsilon w R^2 \xi_I\right) ~. \label{eq: flow eq fX}
\end{align}
\end{subequations}
We refer to the set of first order differential equations (\ref{eq: flow eq uI}-\ref{eq: flow eq fX}) as the attractor flow equations for black hole solutions to the theory \eqref{eq: general 5D action}.

\subsection{Solution of the attractor flow equations} \label{subsec:solutiontoattractoreq}

The attractor flow equations (\ref{eq: flow eq uI}-\ref{eq: flow eq fX}) are first order differential equations. 
In this subsection we discuss the boundary conditions needed to specify their solutions completely. This turns out to be surprisingly subtle. 
We then solve the equations using perturbative expansions.

\subsubsection{Boundary conditions}

The attractor flow equations (\ref{eq: flow eq uI}-\ref{eq: flow eq fX}) 
are first order differential equations with $\xi_I$ and $\tilde{Q}_I$ as given parameters. As such the superficial expectation is that the specification of all the unknown functions $u^I, g_m, w, f^{-1}X_I$ at any coordinate $R^2$ yields the corresponding derivatives at that position. Further iterations should then be sufficient to reconstruct the entire radial dependence, at least in principle. We seek to implement this strategy starting from either asymptotically AdS$_5$, or from the horizon. We consider each in turn. 

For a solution to be asymptotically AdS$_5$, the metric ansatz \eqref{eq:14ds5} 
requires the leading order behavior $f \to R^{0}$, 
$g_{m} \to R^{2}$, and 
$w \to R^2$ as $R \to \infty$. With these boundary conditions for $f$, $g_m$, and $w$, 
\eqref{eq: flow eq uI} and \eqref{eq: flow eq fX} yield
$u^I \to c^{IJK} \xi_J \xi_K R^2$ and $X_I \to \xi_I \cdot R^0$ for the matter fields as $R \to \infty$.

Alternatively, we can impose boundary conditions at the horizon of the black hole. 
There, the near-horizon geometry has a manifest AdS$_2$ factor of the form
\begin{align}
    ds^2_{2} = R^4 dt^2 - \frac{dR^2}{R^2}~,
\end{align}
and so $g_{m} \to R^{0}$, $f \to R^{2}$, and $w \to R^{-2}$. With these leading asymptotics, the attractor flow equations \eqref{eq: flow eq uI} and \eqref{eq: flow eq fX} determine $u^I \to R^0$ and $X_I \to R^{0}$ near the horizon. 

Staying with our superficial expectation, we would start from either asymptotically AdS$_5$ or from the AdS$_2$ horizon. 
Mathematically, it could be a concern that the differential equations are coupled and non-linear, because then the expansions might fail to converge. This situation is most likely incompatible with a black hole solution that is regular throughout the entire flow from asymptotically AdS$_5$ to the AdS$_2$ horizon, or {\it vice versa}. Nonlinearity does not appear to pose a conceptual challenge.

In order to study the unknown functions $u^{I}, g_{m}, w$ and $f^{-1}X_{I}$ around a regular point, we multiply by an appropriate factor of $R^2$. Near the horizon we consider $R^2 u^{I}, R^2 g_{m}, R^2 w$, and $R^2 f^{-1}X_{I}$. At infinity, we expand the functions $R^{-2} u^{I}, R^{-2} g_{m}, R^{-2} w$ and $R^{-2} f^{-1}X_{I}$. After
such rescalings the left hand sides of each of the attractor flow equations (\ref{eq: flow eq uI}-\ref{eq: flow eq fX}) will take the form
\begin{align}
    \begin{split} \label{eq: rescaling flow equation formula}
        \left(\partial_{R^2} + \frac{\alpha+\beta}{R^2}\right)P = R^{-2\beta}\left(\partial_{R^2} + \frac{\alpha}{R^2}\right)(R^{2\beta}P),
    \end{split}
\end{align}
for some field $P$ and some integers $\alpha$ and $\beta$ that can either be positive or negative. The challenge we will encounter repeatedly is that, when $P\sim R^{-2(\alpha+\beta)}$, this expression vanishes. We refer to this situation as a {\it zero-mode} of the perturbative expansion. What it means is that an attractor flow equation does not reveal a derivative, contrary to expectation. Instead, it yields a constraint between the unknown functions on the right hand side of the equation in question. This constraint will be nonlinear and, in general, difficult to implement. In other words, the initial value problem, at both the horizon and at asymptotic infinity, turns out to be unexpectedly complicated. 

\begin{table}[t]
\centering
\begin{tabular}{l|l|l}
        & $R \to \infty$   & $R\to 0$ \\ \hline
$g_{m}$ & $R^2$   & $R^{0}$       \\ \hline
$f$     & $R^{0}$ & $R^{2}$       \\ \hline
$w$     & $R^2$   & $R^{-2}$      \\ \hline
$u^{I}$ & $R^{2}$ & $ R^{0}$     \\ \hline
$X_{I}$ & $R^{0}$ & $R^{0}$       \\ 
\end{tabular}
\caption{Asymptotics of the various functions}
\label{table: asymptotics}
\end{table}

In the following subsections we develop this general theme explicitly, first starting from the horizon, and then from asymptotic infinity. We subsequently merge the two perturbative expansions to seek a global understanding.

\subsubsection{Perturbative solution starting from the horizon}

To satisfy the regularity conditions at the horizon, we take $\beta=1$ in \eqref{eq: rescaling flow equation formula} 
and rewrite the attractor flow equations (\ref{eq: flow eq uI}-\ref{eq: flow eq fX}) as
\begin{subequations}
\begin{align}
     \partial_{R^2}(R^2 u^I) &= \frac{1}{2}\epsilon c^{IJK} (R^2 f^{-1}X_J)\xi_K ~, \label{eq: flow eq uI rescaled}\\ 
    \left(\partial_{R^2} + \frac{1}{R^2}\right)(R^2 g_m) &= 2+\frac{2}{R^2}\epsilon \xi_I (R^2 u^I) ~, \label{eq: flow eq gm rescaled}
    \\
    \left(\partial_{R^2} - \frac{2}{R^2}\right)(R^2 w) &= -\frac{1}{2}R^{-2}(R^2f^{-1}X_I) \left(\partial_{R^2} - \frac{2}{R^2}\right)(R^2 u^I)~, \label{eq: flow eq w rescaled} \\ 
    \left(\partial_{R^2}-\frac{1}{R^2}\right)(R^2 f^{-1}X_I ) &= -R^{-2} g_m^{-1}\left(\widetilde{Q}_I + 2 R^{-4} c_{IJK} (R^2 u^J)(R^2 u^K) + 2  \epsilon (R^2 w) \xi_I\right)~. \label{eq: flow eq fX rescaled}
\end{align}
\end{subequations}
We then expand the unknown functions near the horizon. Since the radial dependence is of the form $R^{2n}$ where $n$ is some integer, the expansion can be written as
\begin{subequations}
\begin{align}
    \label{eq:u expansion R}
    R^2 u^{I} &= \sum_{n=1}^{\infty} u_{(n)}^I R^{2n}~,
    \\
    \label{eq:g expansion R}
    R^2 g_{m} &=  \sum_{n=1}^{\infty} g_{m,(n)} R^{2n}~,
    \\
    \label{eq:w expansion R}
    R^2 w &=  \sum_{n=0}^{\infty} w_{(n)} R^{2n}~,
    \\
    \label{eq:x expansion R}
    R^2 f^{-1} X_{I} &= \sum_{n=0}^{\infty} x_{I,(n)}R^{2n}~.
\end{align}
\end{subequations}
With the asymptotic structure of Table~\ref{table: asymptotics}, the expansions for $R^2 u^I$ and $R^2 g_m$ do not start with a constant term. Moreover, with the horizon expansions above, the differential operators on the left hand sides of \eqref{eq: flow eq w rescaled} and \eqref{eq: flow eq fX rescaled} are such that the coefficients $w_{(2)}$ and $x_{(1)}$ drop out. These coefficients are the zero-modes that make the initial value problem more complicated. There are no analogous zero-modes for $u^{I}$ and $g_{m}$. 

To study the structure of the attractor equations (\ref{eq: flow eq uI rescaled}-\ref{eq: flow eq fX rescaled}), we temporarily treat the unknown scalar field $f^{-1}X_I$ as a given function of the radial coordinate $R$. Then the linear flow equation \eqref{eq: flow eq uI rescaled}, which is sourced by $f^{-1} X_I$, yields all the coefficients $u^{I}_{(n)}$ in terms of $x_{I,(n)}$. At this point we know both of the functions $f^{-1}X_I$ and $u^I$ and then the attractor flow equation \eqref{eq: flow eq gm rescaled} similarly yields the series coefficients $g_{m,(n)}$ in terms of $x_{I,(n)}$. Given all of $f^{-1}X_I$, $u^I$, and $g_m$, it would seem straightforward to exploit \eqref{eq: flow eq w rescaled} and find all the coefficients $w_{(n)}$ in terms of $x_{I,(n)}$. This mostly works, but the zero-mode $w_{(2)}$ can {\it not} be determined this way. That is the obstacle where, as advertized, the derivatives are such that an expansion coefficient simply drops out. 

The final flow equation \eqref{eq: flow eq fX rescaled}, due to the conserved charge $Q_I$ \eqref{eq: 2D charges in 1+4}, is crucial for the complete story. Assuming for a moment that the zero mode $w_{(2)}$ is given as an initial condition, along with the entire function $f^{-1}X_I$, this equation determines the expansion parameters $x_{I,(n)}$ in terms of the $x_{I,(n)}$ themselves, and so the entire system would appear to be solved. However, this last equation also has a zero mode $x_{I,(1)}$ which cannot be determined by the iterative procedure. In short, a careful analysis must be considered and accordingly, this is what we proceed to do now: we solve the expansion coefficients order by order, following the procedure we have outlined.

First, inserting the expansions \eqref{eq:u expansion R} and \eqref{eq:x expansion R} into 
the flow equations \eqref{eq: flow eq uI rescaled}, we find 
\begin{align}
    \begin{split}
        \sum_{n=1}^{\infty} n u_{(n)} ^I R^{2n-2} &= \frac{\epsilon}{2}\sum_{n=0}^{\infty} c^{IJK} \xi_{K} x_{J,(n)}R^{2n}~.
    \end{split}
\end{align}
Comparing each order in $R^2$ leads to a relation between the coefficients $u^{I}_{(n)}$ and $x_{I,(n)}$:
\begin{align} \label{eq: un coefficients}
    \begin{split}
        u_{(n)}^{I} &= \frac{\epsilon}{2n}c^{IJK}\xi_{K} x_{J,(n-1)}~, \quad n\geq1~.
    \end{split}
\end{align}
We insert this result in the flow equation \eqref{eq: flow eq gm rescaled} for the expansion of 
$g_{m}$ \eqref{eq:g expansion R} and find
\begin{align}
    \begin{split}
        \sum_{n=1}^{\infty} (n+1) g_{m,(n)} R^{2n-2}  &= 2+ \sum_{n=1}^{\infty} \frac{1}{n}c^{IJK}\xi_{I}\xi_{J} x_{K,(n-1)}R^{2n-2}~\,. 
    \end{split}
\end{align}
Thus all expansion coeficients of $g_{m}$ can be expressed in terms of the $x_{I,(n)}$:
\begin{align}
\label{eq: gm coefficients}
\begin{split}
    g_{m,(n)} =\frac{1}{n+1}\left(2\delta_{n,1} + \frac{1}{n}c^{IJK}\xi_{I}\xi_{J}x_{K,(n-1)}\right) , \quad n\geq 1 \, .
\end{split}
\end{align}
The steps we have taken so far leads us to express the functions $u^{I}$ and $g_{m}$ solely in terms of $f^{-1}X_I$. This was expected from the general discussion. 

We then start with \eqref{eq: flow eq w rescaled} and use the expansions (\ref{eq:u expansion R}-\ref{eq:x expansion R}) to obtain
\begin{equation}
    \begin{split}
\sum_{n=0}^\infty (n-2) w_{(n)} R^{2n-2} = -\frac{1}{2} \sum_{n=0} ^\infty \left(\sum_{k=0}^n (n-1-k) x_{I,(k)}  u^I _{(n+1-k)}\right)R^{2n-2}~.
    \end{split}
\end{equation}
Comparing powers of $R^2$ we find
\begin{align}
    (n-2)w_{(n)} = -\frac{\epsilon}{4}\sum_{k=0}^n \frac{n-1-k}{n+1-k}c^{IJK}\xi_I x_{J,(k)} x_{K,(n-k)}\label{eq: wn relation 1}~.
\end{align}
We see explicitly that the differential equation 
\eqref{eq: flow eq w rescaled}
fails to express the zero mode $w_{(2)}$ in the expansion \eqref{eq:w expansion R} in terms of other data. However, 
the right hand side of \eqref{eq: wn relation 1}
still reveals important information at $n=2$ since it imposes 
a constraint on the $x_{I,(n)}$ expansion coefficients

\begin{equation}
\label{eq: x0 xi x2 perp}
\begin{split}
    0 &= \frac{\epsilon}{3}c^{IJK} x_{I,(0)} x_{J,(2)}\xi_K ~. 
\end{split}
\end{equation}
Thus determination of $w_{(2)}$ is replaced by a constraint on the functions $f^{-1}X_I$ which we have considered given so far. 

To make further progress it remains to study the constants of motion due to the 
conservation \eqref{eq: 2D charges in 1+4} of electric charge and angular momentum. The electric charge 
\eqref{eq: flow eq fX rescaled} yields 
\begin{equation}
\label{eq: wn relation charge}
    \begin{split}
        &-\sum_{n=1} ^\infty \sum_{k=1}^n (n-1-k) g_{m,(k)} x_{I,(n-k)} R^{2n-2} \\& \quad\quad\quad = \widetilde{Q}_I + 2c_{IJK}\sum_{n=1} ^\infty\sum_{k=1}^n u^J _{(k)} u^K _{(n+1-k)} R^{2n-2} 
        + 2\epsilon \xi_I\sum_{n=0} ^\infty w_{(n)} R^{2n}~. 
    \end{split}
\end{equation}
Comparing each power of $R^2$ we find
the sequence of relations
\begin{equation}
\label{eq: wn relation charge coeffs}
    -\sum_{k=1} ^{n+1} (n-k)  g_{m,(k)}  x_{I,(n-k+1)} = \widetilde{Q}_I \delta_{n,0} + 2c_{IJK} \sum_{k=1} ^{n+1}  u^J _{(k)}  u^K _{(n+2-k)} + 2\epsilon \xi_I  w_{(n)} ~,
\end{equation}
where it is understood in this relation that the coefficients
$g_{m,(k)}$, $u^I _{(k)}$, and $w_{(k\neq 2)}$ depend on the $x_{I,(k)}$ through \eqref{eq: un coefficients}, \eqref{eq: gm coefficients}, and \eqref{eq: wn relation 1}. Thus
the relations \eqref{eq: wn relation charge coeffs} constrain the 
given functions $f^{-1}X_I$ significantly. Unfortunately, 
these constraints are nonlinear and difficult to solve. 

For the constant order in the $R^2$ expansion, we take $n=0$ in \eqref{eq: wn relation charge coeffs} and find the electric charge 
\begin{align}
\label{eq:Qx1}
\begin{split}
    \widetilde{Q}_{I} &= x_{I,(0)}\left(1+\tfrac{1}{2}c^{JKL}x_{J,(0)}\xi_K \xi_L\right) - \tfrac{1}{2}c_{IJK}c^{JML}c^{KNP}\xi_{M}\xi_{N}x_{L,(0)}x_{P,(0)} \\& \quad + \tfrac{1}{4} \xi_{I}c^{LMN}x_{L,(0)}x_{M,(0)}\xi_{N}~.
\end{split}
\end{align}
The charges $\widetilde{Q}_{I}$ depend only on 
the scalar fields at the horizon $x_{I,(0)}$ and 
the FI-parameters $\xi_{I}$. In fact, if positivity conditions are imposed on the $x_{I,(0)}$, the $x_{I,(0)}$ in \eqref{eq:Qx1} can be inverted in terms of the $\widetilde{Q}_I$, allowing to replace the pair of inputs $(\xi_I, x_{I,(0)})$ by the pair $(\xi_I, \widetilde{Q}_I)$. This fits nicely with the understanding of the physical inputs and charges that go in defining the radial flow at every radial hypersurface. 

To rewrite \eqref{eq:Qx1} in a more canonical form, 
we simplify the second term involving a triple product of $c_{IJK}$ by contracting \eqref{eq:cclosure up} with $\xi_{M}\xi_{Q}x_{L,(0)}x_{P,(0)}$. This gives
\begin{equation}
 \label{eq:Qxix}
    \widetilde{Q}_I = x_{I,(0)} -\frac{1}{2}\xi_I \left(\frac{1}{2}c^{JKL}x_{J,(0)}x_{K,(0)}\xi_L\right) + \frac{1}{2}c_{IJK}\left(\frac{1}{2}c^{JNO}\xi_N \xi_O\right) c^{KLM}x_{L,(0)}x_{M,(0)}~.
\end{equation}
This expression makes contact with the form of the charge given in \cite{Gutowski:2004yv}.\footnote{The equation agrees with $Q_I$ given in (3.53) of \cite{Gutowski:2004yv} with the following map between notations:
\begin{align}
    q_I &= \frac{1}{3}x_{I,(0)}, \ \bar{X}_I = \frac{1}{3}\ell \xi_I ~, \ \bar{X}^I = \frac{1}{2}\ell^2 c^{IJK}\xi_J \xi_K ~, 
    \ Q_\text{there} = \frac{\pi}{4G}\widetilde{Q}_\text{here}~.
\end{align}}
The improvement in our work is that we introduce the charge independently of the radial coordinate so it can be computed at any hypersurface we choose, which --- in this case --- is the black hole horizon.

Before analyzing the consequences of electric charge conservation \eqref{eq: wn relation charge coeffs}
for $n\geq 1$, we consider the analogous equations due to conservation of the black hole angular momentum $J$ \eqref{eq: 2D charges in 1+4}. As the first line of \eqref{eq: 2D charges in 1+4} involves the scalar field $X^{I}$, we must recast it in terms of the scalar field with a lowered index, as our expansion \eqref{eq:x expansion R} dictates. Utilizing \eqref{eq: XI relation 1}, we have
\begin{equation}
\begin{split}
    f^{-3} fX^I &= \frac{1}{2}c^{IJK} (f^{-1} X_J) (f^{-1} X_K)~.
\end{split}
\end{equation}
Introducing the near-horizon expansions (\ref{eq:u expansion R}-\ref{eq:x expansion R}) we find
\begin{equation}
\begin{split}
\label{eq: J coeffs n}
    \widetilde{J} \delta_{n,0} &= \widetilde{Q}_I u^I _{(n+1)} + \frac{2}{3}c_{IJK} \sum^{n+1} _{k=1} \sum_{\ell = 1}^k u^I _{(\ell)} u^J _{(k+1-\ell)} u^K _{(n+2-k)}  -\sum_{k=0} ^n \sum_{\ell=0} ^{n-k} g_{m,(k+1)} x_{I,(\ell)} u^I _{(n+1-\ell-k)}\\
    & \quad -2\sum_{k=0}^n g_{m,(k+1)} w_{(n-k)} +\frac{\epsilon}{4}c^{IJK}\xi_I  \sum_{k=0}^n \sum_{\ell=0} ^{n-k} g_{m,(k+1)}x_{J,(\ell)} x_{K,(n-k-\ell)} \\& \quad +2w_{(n)} +2\epsilon \xi_I \sum_{k=0} ^n w_{(k)} u_{(n-k+1)} ^I ~.
\end{split}
\end{equation}
As before, it is understood that $g_{m,(k)}$, $u^I _{(k)}$, and $w_{(k)}$ depends on
the $x_{I,(k)}$ according to \eqref{eq: un coefficients}, \eqref{eq: gm coefficients} and \eqref{eq: wn relation 1}, and here we also need the explicit form of $\widetilde{Q}_I$ \eqref{eq:Qxix}. Thus angular momentum conservation gives another infinite set of relations between 
the $x_{I,(k)}$. Unfortunately, they are even more nonlinear than their analogues for 
conservation of electric charge. 

For $n=0$ \eqref{eq: J coeffs n} gives the angular momentum expressed in terms of $x_{I(0)}$ and $\xi_I$
\begin{equation}
    \begin{split} \label{eq:Jxix}
    \widetilde{J} &= \frac{\epsilon}{4}c^{IJK}
    x_{I,(0)}x_{J,(0)} \xi_K -\frac{\epsilon}{4}c^{IJK}c^{LMN}\xi_I \xi_N \xi_K x_{J,(0)}x_{L,(0)}x_{M,(0)} \\
    &\quad + \epsilon c_{IJK}c^{ILM}c^{JNO}c^{KPQ}\left(\tfrac{1}{8} x_{L,(0)}x_{P,(0)}x_{Q,(0)}\xi_N \xi_O \xi_M + \tfrac{1}{12} x_{N,(0)}x_{P,(0)}\xi_M \xi_O \xi_Q \right) ~,
    \end{split}
\end{equation} 
where we have used the value of $\widetilde{Q}_I$ given in \eqref{eq:Qxix}. To make contact with the form of the angular momentum in \cite{Gutowski:2004yv}, we rewrite the formula 
as\footnote{This agrees with the angular momentum reported in (3.50) of \cite{Gutowski:2004yv} 
with the following map between conventions:
\begin{align}
    q_I &= \frac{1}{3}x_{I,(0)}~, \ \bar{X}_I = \frac{1}{3}\ell \xi_I ~, \ J_\text{there} = \frac{\pi}{4G}\widetilde{J}_\text{here}~.
\end{align}}
\begin{equation}
\label{eq: J final nh}
    \widetilde{J} = \frac{\epsilon}{4}c^{IJK}
    x_{I,(0)}  x_{J,(0)} \xi_K +  \frac{1}{36}\left(c^{IJK} \xi_I \xi_J \xi_K \right) \left(c^{LMN}  x_{L,(0)} x_{M,(0)} x_{N,(0)} \right) ~.
\end{equation}
Again, we are able to express the final result for the conserved charge entirely in terms of near horizon data.
Moreover, since $\widetilde{Q}_I$ and $\widetilde{J}$ depend on the same integration constants $x_{I,(0)}$, 
the charges are indeed not independent of each other. 

We now turn to the $n=1$ component of \eqref{eq: wn relation charge coeffs}, i.e., electric charge conservation at order $R^2$ 
away from the horizon. It amounts to
\begin{equation}
    g_{m,(2)}x_{I,(0)} = 4c_{IJK}u^J _{(1)} u^K _{(2)} + 2\epsilon \xi_I w_{(1)} ~. 
\end{equation}
The absence of $x_{I,(1)}$ in this equation is due to the zero-mode in 
\eqref{eq:x expansion R}. However, a constraint on the $x_{I,(1)}$ will follow, in analogy with the zero-mode $w_{(2)}$ giving the condition \eqref{eq: x0 xi x2 perp}. 
The values of $g_{m,(2)}$, $u^I _{(1)}$, $u^I_{(2)}$, and $w_{(1)}$ from 
\eqref{eq: un coefficients}, \eqref{eq: gm coefficients} and \eqref{eq: wn relation 1} give 
the vector relation
\begin{equation}
    \label{eqn: x1 constraint vec}
    \left[\frac{1}{6}c^{JKL}\xi_K \xi_L x_{I,(0)} - \frac{1}{2}c_{IKM}c^{JLM}c^{KNP} \xi_L \xi_N x_{P,(0)} + \frac{1}{2}c^{JKL}\xi_K x_{L,(0)} \xi_I \right] x_{J,(1)} = 0 ~.
\end{equation}
We see that $x_{I,(1)}$ is constrained even though it is a zero-mode of the differential operator. 
Using the cubic condition on the $c_{IJK}$ \eqref{eq: cijk symmetrization}, we can show that
the matrix in square brackets has the null vector 
\begin{equation}
\label{eq: x1 direction}
x_{I,(1)} = \ell \xi_I ~. 
\end{equation}
It is unique, at least for generic structure constants $c_{IJK}$ and generic charges, which are
parametrized by $x_{I,(0)}$. The constraint \eqref{eqn: x1 constraint vec} does not determine the
overall normalization. However, the scale of the radial coordinate $R^2$ is
arbitrary from the near horizon point of view, 
so the choice \eqref{eq: x1 direction} involves no loss of generality. 

The $n=2$ component of \eqref{eq: wn relation charge coeffs} gives another vector-valued relation
\begin{equation}
\label{eq: w2 nh expansion}
    T_I ^J x_{J,(2)} = 2\epsilon \left(w_{(2)} + \frac{\epsilon}{2\ell}\right) \xi_I  ~,   
\end{equation}
where we have simplified using \eqref{eq: x1 direction} and
\begin{equation}
\label{eq: TIJ matrix} 
    T_I ^J = -\left(1+\frac{1}{2}c^{KLM}\xi_K \xi_L x_{M,(0)}\right)\delta_I ^J + \frac{1}{12}c^{JKL}\xi_K \xi_L x_{I,(0)} - \frac{1}{3}c_{IKL}c^{KMJ}c^{LNP}\xi_M \xi_N x_{P,(0)}  ~.
\end{equation}
The matrix $T_I^J$ is invertible so, given the inputs $\xi_I, x_{I,(0)}, w_{(2)}$, the coefficient $x_{J,(2)}$ is completely determined by \eqref{eq: w2 nh expansion}. However, the value of $x_{J,(2)}$
computed this way fails to satisfy the previously established constraint \eqref{eq: x0 xi x2 perp}. 
This apparent contradiction can be avoided only if 
\begin{equation}
\label{eq: xI2zero}
x_{I,(2)} =0~. 
\end{equation}
Because the left hand side of \eqref{eq: w2 nh expansion} vanishes, 
the right side requires
\begin{equation}
w_{(2)} = - \frac{\epsilon}{2\ell}~.
\end{equation}

At this point, we can finally consider generic components of the electric charge 
conservation \eqref{eq: wn relation charge coeffs}, i.e. the 
infinite set of equations $n\geq 3$.
The coefficients $w_{(n\geq 3)}$ can be eliminated using \eqref{eq: wn relation 1} for $w_{(n)}$. The $u^I _{(n)}$ and $g_{m,(n)}$ are similarly traded for $x_{I,(n)}$, this time using \eqref{eq: un coefficients} and \eqref{eq: gm coefficients}.
For all $n\geq 3$ this gives
\begin{equation}
\label{eq: xn recurrence}
\begin{split}
    &-\sum_{k=0} ^n \frac{n-1-k}{k+2}\left(2\delta_{0,k} + \frac{1}{k+1}c^{JKL}\xi_J \xi_K x_{L,(k)} \right)x_{I,(n-k)} \\
    &= \frac{1}{2}c_{IJK}c^{JLM}c^{KNP} \xi_L \xi_N \sum_{k=0} ^n \frac{1}{(k+1)(n-k+1)}x_{M,(k)}x_{P,(n-k)}\\
    &\quad - \frac{1}{2(n-2)}\xi_I \sum_{k=0}^n \frac{n-1-k}{n+1-k} c^{JKL}\xi_J x_{K,(k)}x_{L,(n-k)}~.
\end{split}
\end{equation}
This messy expression can be reorganized 
as a recurrence relation giving $x_{I,(n)}$ in terms of 
the preceding $x_{I,(0\leq k \leq n-1)}$:
\begin{equation}
\begin{split}
\label{eq: nh x recursion}
    &\left[-\frac{n-1}{2}\left(2 + c^{KLM}\xi_K \xi_L x_{M,(0)} \right)\delta^J _I + \frac{1}{(n+1)(n+2)}c^{JKL}\xi_K \xi_L x_{I,(0)} \right. \\
    &\left. -\frac{1}{n+1}c_{IKL}c^{KMJ}c^{LPQ}\xi_M \xi_P x_{Q,(0)} -\frac{1}{(n-2)(n+1)}c^{JKL}\xi_K x_{L,(0)} \xi_I \right]x_{J,(n)}\\
    &=\sum_{k=1} ^{n-1} \left[\frac{(n-1-k)}{(k+1)(k+2)}c^{JKL}\xi_J \xi_K x_{L,(k)} x_{I,(n-k)}+ \frac{1}{2}c_{IJK}c^{JLM}c^{KNP} \xi_L \xi_N x_{M,(k)}x_{P,(n-k)}\right. \\
    &\left. \quad - \frac{1}{2(n-2)}\xi_I \frac{n-1-k}{n+1-k} c^{JKL}\xi_J x_{K,(k)}x_{L,(n-k)}\right]~. 
\end{split}
\end{equation}
The left hand side can be inverted, at least for some specific models of $c_{IJK}$, such as the STU model ($c_{IJK}=|\epsilon_{IJK}|$ for $I,J,K$ running from 1 to 3). In such cases the recurrence 
relation  \eqref{eq: nh x recursion} determines all higher-order $x_{I,(n\geq 3)}$ in terms of the coefficients $x_{I,(0)}$, $x_{I,(1)}$ and $x_{I,(2)}$ as well as the FI-parameters $\xi_I$. 
In fact, the constraints \eqref{eq: x1 direction} and \eqref{eq: xI2zero} from low $n$ will be sufficient 
to show that the series {\it truncates} at $n=2$. This is discussed in subsection
\ref{subsec:combi}. 

At this point we have exhausted the information that comes from the 
conservation of electric charge $\widetilde{Q}_I$ charge in \eqref{eq: wn relation charge coeffs}. We did not yet study 
the $\widetilde{J}$ conservation relations \eqref{eq: J coeffs n}. As noted already, 
the constant order $n=0$ determines the angular momentum from a near horizon perspective. 
We have worked out the first few orders $n\geq 1$ and found either redundant relations, involving already known coefficients such as $x_{I,(0)}$, $\xi_I$ and $x_{I,(1)}$, or relations that tie together higher order $x_{I,(k\geq 2)}$ with lower-order ones. 
We do not foresee any further constraints due to the $\widetilde{J}$ relations. 

In summary, starting from the near horizon region, we have exploited
supersymmetry and found the entire black hole solution. 
The fields $u^I, g_m, w$ and $f^{-1} X_I$ are reported in \eqref{eq: un coefficients}, \eqref{eq: gm coefficients}, \eqref{eq: wn relation 1} and \eqref{eq: w2 nh expansion}. Additionally, we computed the electric charges $\widetilde{Q}_{I}$ and the angular momenta in terms of the horizon values of
the scalars $x_{I,(0)}$, and the subleading coefficients
$x_{I,(1)}$ which, according to \eqref{eq: x1 direction}, coincide with the FI-parameters $\xi_I$.

\subsubsection{Perturbative solution starting from asymptotic AdS}

We now adapt the approach from the previous subsection 
and expand the unknown functions $u^I$, $g_m$, $w$ and $f^{-1}X_I$ at large $R$, near the asymptotic AdS$_5$ boundary. 

Given the asymptotic behaviors listed in Table~\ref{table: asymptotics},  regularity requires taking $\beta=-1$ in \eqref{eq: rescaling flow equation formula}. We then recast the flow equations (\ref{eq: flow eq uI}--\ref{eq: flow eq fX}) as
\begin{subequations}
\begin{align}
    \left(\partial_{R^2} + \frac{2}{R^2}\right)(R^{-2}u^{I}) &= \frac{1}{2}\epsilon c^{IJK} (R^{-2}f^{-1}X_J)\xi_K ~, \label{eq: flow eq uI infty}\\ 
    \left(\partial_{R^2} + \frac{3}{R^2}\right)(R^{-2}g_m) &= \frac{2}{R^4}+\frac{2}{R^2}\epsilon \xi_I (R^{-2}u^I) ~. \label{eq: flow eq gm infty}
    \\
    \partial_{R^2}(R^{-2}w) &= -\frac{1}{2}R^{2}(R^{-2}f^{-1}X_I) \partial_{R^2}(R^{-2}u^I)~, \label{eq: flow eq w infty}
    \\
    \begin{split}
    \left(\partial_{R^2}+\frac{1}{R^2}\right)(R^{-2}f^{-1}X_I ) &=-\frac{R^{-8}}{(R^{-2}g_m)}\left(\widetilde{Q}_I + 2 c_{IJK} R^{4}(R^{-2}u^J) (R^{-2}u^K) \right. \\& \qquad\qquad\qquad\qquad\qquad\qquad\qquad \left. + 2  \epsilon R^{4} (R^{-2}w) \xi_I\right) ~\, . \label{eq: flow eq fX infty} \end{split}
\end{align}
\end{subequations}
We define the perturbative expansions at infinity as
\begin{subequations}
\begin{align}
    \label{eq:u expansion R infinity}
    R^{-2} u^{I} &= \sum_{n=0}^{\infty} \bar u_{(n)}^I R^{-2n}~,
    \\
    \label{eq:g expansion R infinity}
    R^{-2} g_{m} &=  \sum_{n=0}^{\infty} \bar g_{m,(n)} R^{-2n}~ \, ,
    \\
    \label{eq:w expansion R infinity}
    R^{-2} w &=  \sum_{n=0}^{\infty} \bar w_{(n)} R^{-2n}~,
    \\
    \label{eq:x expansion R infinity}
    R^{-2} f^{-1} X_{I} &=  \sum_{n=1}^{\infty} \bar x_{I,(n)}R^{-2n}~.
\end{align}
\end{subequations}
The bar distinguishes the expansion coefficients at the asymptotically AdS$_5$ boundary from their analogues at the horizon. 

As before, we initially specify the entire series $\bar x_{I,(n)}$. 
Additionally, examination of (\ref{eq: flow eq uI infty}--\ref{eq: flow eq fX infty}) shows that $\bar u_{(2)} ^I$, $\bar g_{m,(3)}$, $\bar w_{(0)}$, and $\bar x_{I,(1)}$ do not appear on the left hand sides of the equations. These are the zero modes that we also regard as inputs, at least provisionally. Among the zero-modes, we can determine $\bar x_{I,(1)}$ from the outset because they give the asymptotic values of the scalars 
\begin{align} \label{eq: x1 infinity solution}
 \bar x_{I,(1)} = \ell \xi_I \, ,
\end{align}
as we found in \eqref{eq: xinf xi}, by extremizing the potential of gauged supergravity.

We now proceed to solve for the expansion coefficients of each variable, order by order. Starting with the $u^I$ flow equation \eqref{eq: flow eq uI}, and using the expansions \eqref{eq:u expansion R infinity} and \eqref{eq:x expansion R infinity}, we find
\begin{align}
\begin{split}
    \sum_{n=0}^{\infty} (2-n)\bar u_{(n)}^I R^{-2n-2} &=\frac{1}{2}\epsilon c^{IJK} \xi_J  \sum_{n=1}^{\infty} \bar x_{K,(n+1)}R^{-2n-2}~ \, .
\end{split}
\end{align}
Comparing inverse powers of $R^2$, we find $\bar u_{(n)}^I$ for $n\neq 2$:
\begin{align} \label{eq: un eq}
    \begin{split}
        (2-n) \bar u_{(n)}^I &= \frac{1}{2}\epsilon c^{IJK} \xi_J  \bar x_{K,(n+1)}~, \quad n \geq 0 \, .
    \end{split}
\end{align}
The zero mode $\bar u_{(2)}^I$ drops out of the equation. Instead, we find a vectorial constraint on $x_{I,(3)}$
\begin{equation} 
\label{eq: x3 cond}
c^{IJK} \xi_J \bar x_{K,(3)} = 0 ~\, .
\end{equation}
It has the obvious solution
\begin{align} \label{eq: x3 zero}
    \bar x_{I,(3)} = 0 ~,
\end{align}
for all values of $I$. This solution is unique if the matrix $c^{IJK} \xi_J$ is non-singular. In \eqref{eq: xi inversion}, we show that it is indeed invertible
\begin{equation}
   (c^{IJK} \xi_J)^{-1} = \frac{1}{2}\ell^3 \left(c_{IJK}c^{JLM} \xi_L \xi_M - \xi_I \xi_K \right)~\,.
\end{equation}
Next, we consider the $g_m$ flow equation \eqref{eq: flow eq gm}. The expansions \eqref{eq:u expansion R infinity} and \eqref{eq:g expansion R infinity} give
\begin{align}
    \sum_{n=0}^{\infty} (3-n) \bar g_{m,(n)} R^{-2n-2} &= 2R^{-4} + 2\epsilon\xi_{I} \sum_{n=0}^{\infty} \bar u_{(n)}^I R^{-2n-2}\, .
\end{align}
The expansion coefficients $\bar g_{m,(n)}$ --- with the exception of the zero mode $g_{m,(3)}$ --- can be expressed in terms of $x_{I,(n)}$ and the zero mode $u^{I}_{(2)}$ as 
\begin{equation} \label{eq: gm coefs infty}
(3-n) \bar g_{m,(n)}  = 
\left\{
    \begin{array}{cc}
         2\delta_{1,n} + \frac{1}{2-n}c^{IJK}\xi_{I}\xi_{J} \bar x_{K,(n+1)} &  n \neq 2,
         \\
         2\epsilon\xi_{I} \bar u_{(2)}^I &  ~n = 2 \, .
    \end{array}
\right.
\end{equation}
In compensation for not determining $\bar g_{m,(3)}$, we find the constraint $\xi_I \bar u_{(3)}^I=0$.
Rewriting this constraint using \eqref{eq: un eq} gives a projection on $x_{I,(4)}$
\begin{equation}
\label{eq: x4 cond}
c^{IJK} \xi_I \xi_J \bar x_{K,(4)}=0 ~.
\end{equation}
This constraint is a real special geometry scalar, unlike the vector-valued condition \eqref{eq: x3 cond}.
It will nevertheless prove useful when simplifying results at large $R$.

We now turn to the nonlinear flow equation for $w$ \eqref{eq: flow eq w}. After using the expansions \eqref{eq:u expansion R infinity}, \eqref{eq:w expansion R infinity} and \eqref{eq:x expansion R infinity}, we find 
\begin{align}
    \sum_{n=1} ^\infty (n-1) \bar w_{(n-1)} R^{-2n} &= -\frac{1}{2}\sum_{n=0}^\infty \sum_{k=1}^n (n-k) \bar x_{I,(k)} \bar u^I _{(n-k)} R^{-2n} ~, 
\end{align}
and so
\begin{equation}
\label{eq: wn flow equation}
    (n-1) \bar w_{(n-1)} = -\frac{1}{2}\sum_{k=1}^{n} (n-k) \bar x_{I,(k)} \bar u^I_{(n-k)}
   ~ , \qquad n\geq 1 \, . 
\end{equation}
For $n=1$, the left hand side vanishes, so the zero-mode $\bar w_{(0)}$ is undetermined. 
The right hand side also vanishes for $n=1$ so in this case 
the equation with a zero-mode offers no additional information. 
We omit the $n=1$ case and rewrite \eqref{eq: wn flow equation} to
\begin{align} \label{eq: wn inf for n geq 1}
    \bar w_{(n)} &= -\frac{1}{2n}\sum_{k=1}^{n} (n+1-k) \bar x_{I,(k)} \bar u^I _{(n+1-k)} ~, \qquad n\geq 1 ~. 
\end{align}
We have refrained from eliminating $\bar u^I _{(k)}$ in favor of $\bar{x}_{I,(n)}$ via \eqref{eq: un eq} because, generally, the equation involves the zero mode $\bar{u}^{I}_{(2)}$ which cannot be removed this way. 

The final flow equation \eqref{eq: flow eq fX infty} was derived by combining supersymmetry with
conservation of electric charge. 
Using the expansions (\ref{eq:u expansion R infinity}--\ref{eq:x expansion R infinity}), we find
\begin{equation}
\begin{split}
    & R^{-4}\sum_{n=0}^\infty  \sum_{k=0}^n \bar g_{m,(k)} (n-k) \bar x_{I,(n+1-k)} R^{-2n} 
    \\&= R^{-4}\widetilde{Q}_I\delta_{n,2} + 2c_{IJK}R^{-4}\sum_{n=0}^\infty \sum_{k=0}^n \bar u^J _{(k)} \bar u^K_{(n-k)} R^{-2n} + 2\epsilon \xi_I R^{-4} \sum_{n=0} ^\infty \bar w_{(n)} R^{-2n}  ~. 
\end{split}
\end{equation}
For all $n\geq 0$, this gives
\begin{equation}
\label{eq: wn infty}
    \sum_{k=0} ^n \bar g_{m,(k)} (n-k) \bar x_{I,(n+1-k)} = \widetilde{Q}_I \delta_{n,2} + 2c_{IJK}\sum_{k=0}^n \bar u^J _{(k)} \bar u^K _{(n-k)} + 2\epsilon \xi_I \bar w_{(n)} ~\,.
\end{equation}
Again, we have chosen to maintain \eqref{eq: wn infty} as implicit functions of $\bar{x}_{I,(n)}$ due to the presence of the zero modes. 

It is worth examining the first few orders of \eqref{eq: wn infty} in detail. The $n=0$ component of \eqref{eq: wn infty} gives
\begin{equation} \label{eq:w0 value}
\begin{split}
    \xi_I \bar w_{(0)} = -\epsilon c_{IJK} \bar u^J _{(0)} \bar u^K _{(0)} = -\frac{\epsilon}{16} c_{IJK}c^{JLM}c^{KNP} \bar x_{L,(1)} \xi_M \bar x_{N,(1)} \xi_P = -\frac{\epsilon}{2\ell}\xi_I ~,
\end{split}
\end{equation}
where we have used \eqref{eq: x1 infinity solution}. Thus, it
provides the value of the zero mode $w_{(0)}$
\begin{align}
    \bar w_{(0)} = -\frac{\epsilon}{2\ell} \, .
\end{align}
At the next order, the $n=1$ component of \eqref{eq: wn infty} is redundant as it just confirms the value for $\bar{w}_{(1)}$ already obtained from \eqref{eq: wn inf for n geq 1}.

The $n=2$ component of \eqref{eq: wn infty} is particularly important, because it relates the electric charge to the expansion parameters at infinity
\begin{equation}
\label{eq: Q infty}
\begin{split}
    \widetilde{Q}_I &=  \bar x_{I,(2)} -\frac{1}{2}\xi_I \left(\frac{1}{2}c^{JKL} \bar x_{J,(2)} \bar x_{K,(2)}\xi_L\right) + \frac{1}{2}c_{IJK}\left(\frac{1}{2}c^{JNO}\xi_N \xi_O\right) c^{KLM} \bar x_{L,(2)} \bar x_{M,(2)} \\&\quad + \epsilon \ell \left( \xi_I  \xi_{J} - c_{IJK} c^{KLM} \xi_{L} \xi_{M} \right) \bar u^J _{(2)} \, ,
\end{split}
\end{equation}
where we imposed \eqref{eq: x1 infinity solution} and \eqref{eq: x3 zero} and recast the charge in a form 
similar to \eqref{eq:Qxix}. Since the electric charge is conserved, the expression  \eqref{eq: Q infty} for $\widetilde{Q}_I$, written in terms of the expansion parameters at infinity,
must be equal to its analogue \eqref{eq:Qxix} obtained from expansion near the horizon.

We have established that $\widetilde{Q}_I$ at infinity has been determined with the only inputs necessary being the coefficients $\bar{x}_{I,(2)}$, $\xi_I$ and $\bar{u}^I _{(2)}$. We now move on to the components of \eqref{eq: wn infty} for $n \geq 3$, to establish the recursion relation for the coefficients at infinity. 

For $n=3$, \eqref{eq: wn infty} simplifies after eliminating the $g_{m}$, $u^I$ and $w$ coefficients using \eqref{eq: un eq}, \eqref{eq: gm coefs infty} and \eqref{eq: wn inf for n geq 1} to
\begin{equation}
\label{eq: x4 infty}
    4\ell^{-2} \bar{x}_{I,(4)} + 2\epsilon\left( \bar{x}_{I,(2)} \xi_J + \frac{1}{3}\xi_I \bar{x}_{J,(2)} - c_{IJK}c^{KLM}\xi_L \bar{x}_{m,(2)}\right)\bar{u}^J _{(2)} = 0 ~.
\end{equation}
This relation indicates that $\bar{x}_{I,(4)}$ is described only with the help of $\bar{x}_{I,(2)}$, $\xi_I$ and $\bar{u}^I _{(2)}$.

Furthermore, we note that for $n\geq 4$, the simplification of \eqref{eq: wn infty} yields a generalization of \eqref{eq: x4 infty}
\begin{align}
\label{eq: x induction infty general} 
    \begin{split}
        &\frac{(n-1)^2}{n-2}\ell^{-2} \bar{x}_{I,(n+1)}
        \\&= -(n-1) \left(1+\tfrac{1}{2}(c \cdot \xi \xi \bar{x}_{(2)})\right) \bar{x}_{I,(n)} -2\epsilon (n-2) \xi_J \bar{u}^J _{(2)} \bar{x}_{I,(n-1)} - (n-3) \bar{g}_{m,(3)} \bar{x}_{I,(n-2)} \\
        &\quad  - \sum_{k=4}^n \frac{n-k}{(k-2)(k-3)} (c \cdot \xi \xi \bar{x}_{(k+1)}) \bar{x}_{I,(n+1-k)} + 4c_{IJK} \bar{u}^J _{(2)} \bar{u}^K _{(n-2)}
        \\&\quad + 2c_{IJK} c^{JLM}c^{KNP} \xi_L \xi_N \sideset{}{'}\sum_{\substack{k=1}} ^{n-1}  \frac{\bar{x}_{M,(k+1)} \bar{x}_{P,(n-k+1)}}{4(k-2)(n-k-2)} +\xi_I \sideset{}{'}\sum_{\substack{k=2}} ^{n}  \frac{n+1-k}{n-1-k} \frac{(c \cdot \xi \bar{x}_{(k)}\bar{x}_{(n-k+2)})}{2n}\,,
    \end{split}
\end{align}
where the apostrophes on the summation symbols indicate that we exclude the terms in the sum with vanishing denominators. We have imposed the value of $\bar x_{I,(1)}$ as given in \eqref{eq: x1 infinity solution} and products of the form $c \cdot xyz$ indicate special geometry contractions under $c^{JKL}$ of the form $c^{JKL}x_J y_K z_L$. The expression \eqref{eq: x induction infty general} has been expanded to distinguish contributions coming from $\bar{x}_{I,(n+1)}$, given by  the left hand side of \eqref{eq: x induction infty general}, and $\bar{x}_{I,(2\leq k \leq n)}$, given by the right hand side of the equality in \eqref{eq: x induction infty general}. It becomes clear that a given $\bar{x}_{I,(n+1)}$ depends only on the expansion parameters $\bar{x}_{I,(1)}, \bar{x}_{I,(2)}, \dots, \bar{x}_{I,(n)}$. By recursion, i.e. applying \eqref{eq: x induction infty general} repeatedly, 
we can now determine all $\bar{x}_{I,(4\leq k \leq n)}$ in terms of of $\bar{x}_{I,(2)}$, $\xi_I$, $\bar{u}^I _{(2)}$ and $\bar{g}_{m,(3)}$. 

Lastly, we analyse the conservation of angular momentum $\widetilde{J}$ by
expanding the first equation in \eqref{eq: 2D charges in 1+4} at infinity, 
making sure to rescale the functions $u^I$, $g_m$, $w$ and $f^{-1} X_I$ appropriately 
\begin{equation}
\label{eq: J expansion infty}
\begin{split}
    \widetilde{J} &= R^2 Q_I (R^{-2}u^I) + \tfrac{2}{3}R^6 c_{IJK}(R^{-2} u^I)(R^{-2} u^J)(R^{-2} u^K) -R^8(R^{-2} g_m) \left(2R^{-2}(R^{-2}w)  \right. \\
    &\quad \left. + (R^{-2}f^{-1}X)\cdot (R^{-2}u) -\tfrac{1}{2}\epsilon R^{-2} f^{-3} \xi \cdot fX)\right)  +2R^4 (R^{-2}w) (1+\epsilon R^2\xi \cdot (R^{-2}u)) ~.
\end{split}
\end{equation}
This can be expanded like \eqref{eq: J coeffs n} in terms of the expansions at infinity (\ref{eq:u expansion R infinity}-\ref{eq:g expansion R infinity}), leading to a relation for the component of $R^{-2n}$. At constant order ($n=0$), we obtain
\begin{equation}
\label{eq: J final inf}
\begin{split}
    \widetilde{J} &= \frac{\epsilon}{4}c^{IJK}
    \bar{x}_{I,(2)}\bar{x}_{J,(2)} \xi_K +  \frac{1}{36}\left(c^{IJK} \xi_I \xi_J \xi_K \right) \left(c^{LMN} \bar{x}_{L,(2)} \bar{x}_{M,(2)} \bar{x}_{N,(2)} \right) + \frac{\epsilon}{\ell}\bar{g}_{m,(3)} \\
& \quad + \ell \left(\frac{1}{2}\xi_I c^{JKL}\xi_J \xi_K \bar{x}_{L,(2)} + \xi_I + \frac{5}{3}\ell^{-3}\bar{x}_{I,(2)} \right)\bar{u}^I _{(2)} ~,
\end{split}
\end{equation}
where we have imposed \eqref{eq: x1 infinity solution} and \eqref{eq: x3 zero}. 
This expression at most depends on the inputs $\bar{x}_{I,(2)}$, $\xi_I$, $\bar{g}_{m,(3)}$ and $\bar{u}^I _{(2)}$.
All higher order powers of the $\widetilde{J}$ relation at infinity in \eqref{eq: J expansion infty} are
redundant because they yield  relations between the coefficients that have already been established.

In summary, we have studied the first order equations 
(\ref{eq: flow eq uI infty}-\ref{eq: flow eq fX infty}) due to supersymmetry
and conservation of electric charge, by expanding perturbatively near infinity. 
Given the asymptotic values of the scalars fields \eqref{eq: x1 infinity solution}, as well as
the conserved charges $\widetilde{Q}_I$, $\widetilde{J}$ defined by fall-off conditions at infinity, 
the simplest outcome would have been for supersymmetry to determine the entire black hole geometry. 
Our finding is much more complicated: all physical fields can be expressed as a perturbative series with expansion parameters 
that depend not only on $\xi_I$ and $\bar{x}_{I,(2)}$ but also the zero-modes $\bar{u}^I _{(2)}$ and $\bar{g}_{m,(3)}$.

\subsubsection{Summary and discussion of perturbative solutions}
\label{subsec:combi}

The study in the subsection so far focused on technical details. This was needed because the
interplay between supersymmetry, boundary conditions, and conserved charges proved to be rather intricate. 
We now conclude the subsection with a summary of the final results and discussion of their interpretation. 

The black hole solution is parametrized primarily by the matter fields: scalar fields $f^{-1}X_I$, 
with the prefactor $f$ such that the combination $f^{-1}X_I$ is unconstrained by real special geometry,
and the magnetic potentials $u^I$. Because of supersymmetry, the electric potentials $fY^I$ can be identified with the scalar fields $fX^I$, see \eqref{eq: XY match}. Given the matter fields, $f^{-1}X_I$ and $u^I$, as well as supersymmetry, the geometry is specified by a K\"{a}hler base that depends on the function $g_m$, 
and a fibre encoding rotation through the potential $w$. All unknown 
functions $f^{-1}X_I$, $u_I$, $g_m$ and $w$ can depend only on a single radial coordinate $R^2$ and they must 
satisfy specified first order differential equations (\ref{eq: flow eq uI}-\ref{eq: flow eq fX}). 

Supersymmetry is never sufficient to specify an entire solution, because it is first order, and there is always an integrability condition that is of second order. Taking into account the Noether-Wald procedure, we find a second order constraint that satisfies a Gauss' law that was subject to detailed discussion in section \ref{section: noether-wald surface charge}. With this augmentation, the first order differential equations form a complete system. Angular momentum, with its conservation law also discussed in section~\ref{section: noether-wald surface charge}, yields nothing new, except for a formula giving the angular momentum in terms of the same parameters that define the electric charge in the near-horizon expansion. In the case of the asymptotic infinity expansion, the electric charge depends on one of the zero modes $\bar{u}^I _{(2)}$ in addition to $\bar{x}_{I,(2)}$, $\xi_I$ whereas the angular momentum depends on both the zero modes $\bar{g}_{m,(3)}$ and $\bar{u}^I _{(2)}$ as well as $\bar{x}_{I,(2)}$ and $\xi_I$.

Consistent boundary conditions for the differential equations can be specified at any radius, in principle. They must depend, at the very least, on the FI-coupling constants $\xi_I$ and the electric charges $Q_I$. We find that, when {\it starting from the black hole horizon}, this data is sufficient. Because of supersymmetry, these parameters specify the entire near horizon geometry, including the squashing of the horizon due to angular momentum. This explains why the electric charge and the angular momentum are only described with the use of the leading $x_{I,(0)}$ term in $f^{-1}X_I$, but any further subleading information about any of the fields $u^I$, $g_m$ or $w$ requires subleading contributions away from the horizon, with \textit{derivative} information of the $f^{-1}X_I$ expansion \eqref{eq:x expansion R} supplied by the $x_{I,(1)}$ coefficients.

The linchpin for establishing this claim about the near horizon expansion is the scalar field. 
In the series expansion for $R^2 f^{-1}X_I$, we have the constant at the horizon $x_{I,(0)}$ and then at ${\cal O}(R^2)$, we have $x_{I,(1)}$. For the third expansion coefficent, we find $x_{I,(2)}=0$ \eqref{eq: xI2zero}. With this starting point, the recursion relation \eqref{eq: nh x recursion} shows that 
all $x_{I,(k \geq 2)}$ actually {\it vanish}. The fact that the scalar field $f^{-1}X_I$ truncates after the first
two terms is the near horizon version of the fact that  $f^{-1}X_I$ is a harmonic function, as is familiar from ungauged supergratity. 

When analyzing the supersymmetry conditions we provisionally
considered $f^{-1}X_I$ an input that all other variables were expressed in 
terms of. The truncation $x_{I,(k \geq 2)}=0$ has the immediate effect of truncating $u^I _{(n\geq 3)}=g_{m,(n\geq 3)} = w_{(n\geq 3)} = 0$. 
The expansions at the horizon (\ref{eq:u expansion R}-\ref{eq:x expansion R}) simplify and we find
\begin{subequations}
\begin{align} \label{eq: GR uI solution}
    R^{2} u^{I} &=   \frac{\epsilon}{2}c^{IJK}\xi_{J}  x_{K,(0)}R^{2}+\frac{\epsilon\ell}{4}c^{IJK}\xi_{J}\xi_{K} R^4 \, ,
    \\ \label{eq: GR gm solution}
    R^{2}g_{m} &=  \left(1+\frac{1}{2}c^{IJK}\xi_{I}\xi_{J} x_{K,(0)}\right)R^{2}+\frac{1}{\ell^2} R^4  \, ,
    \\\label{eq: GR w solution}
    R^{2} w &=-\frac{\epsilon}{8}c^{IJK}\xi_{I}x_{J,(0)}  x_{K,(0)}   -\frac{\epsilon \ell}{4}c^{IJK}\xi_{I}\xi_{J}  x_{K,(0)}R^{2} -\frac{\epsilon}{2\ell} R^4  \, , \\
\label{eq: GR fXI solution}
    R^{2}f^{-1}X_{I} &= x_{I,(0)} + \ell \xi_I R^{2}  . 
\end{align}
\end{subequations}
These expressions exactly match the well-known Gutowski-Reall solution \cite{Gutowski:2004yv}, with the appropriate identifications of notation\footnote{$u^I$, $g_m$, $w$ and $f^{-1}X_I$ match $U^I$, $g$, $w$ and $f^{-1}X_I$  respectively in \cite{Gutowski:2004yv} via:
\begin{equation}
    q_I = \frac{1}{3}\bar{x}_{I,(2)}  \ , \ \bar{X}_I = \frac{1}{3}\ell \xi_I ~. 
\end{equation}}.
The electric charge $\widetilde{Q}_I$ \eqref{eq:Qxix} and the angular momentum $\widetilde{J}$ \eqref{eq:Jxix} computed 
in the near horizon expansion similarly agree with the familiar results. 

The analogous analysis starting from asymptotic AdS$_5$ turned out to be less straightforward. 
Recalling that coefficients starting from infinity are denoted by barred expansion coefficients, we find that, when shooting in (going from infinity towards the horizon), 
we must not only specify $\xi_I$ and $\bar{x}_{I,(2)}$, but also $\bar{u}_{I,(2)}$ and $\bar{g}_{m,(3)}$. 
The harmonic function we established at the horizon reproduces the Gutowski-Reall solution 
and has features in common with their very familiar analogues in ungauged supergravity. At infinity, it corresponds to 
\begin{equation}
\label{eq: mu matching}
    x_{I,(0)} = \bar x_{I,(2)} ~,~~\bar{u}^I _{(2)}=\bar{g}_{m,(3)}=0~.
\end{equation}
With these special values, the recursion relation \eqref{eq: x induction infty general} simplifies greatly 
\begin{align}
\label{eq: x induction infty} 
    \begin{split}
        &\frac{(n-1)^2}{n-2}\ell^{-2} \bar{x}_{I,(n+1)}
        \\&= -(n-1) \left(1+\tfrac{1}{2}(c \cdot \xi \xi \bar{x}_{(2)})\right) \bar{x}_{I,(n)} 
        - \sum_{k=4}^n \frac{n-k}{(k-2)(k-3)} (c \cdot \xi \xi \bar{x}_{(k+1)}) \bar{x}_{I,(n+1-k)}
        \\&\quad + 2c_{IJK} c^{JLM}c^{KNP} \xi_L \xi_N \sideset{}{'}\sum_{\substack{k=1}} ^{n-1}  \frac{\bar{x}_{M,(k+1)} \bar{x}_{P,(n-k+1)}}{4(k-2)(n-k-2)} +\xi_I \sideset{}{'}\sum_{\substack{k=2}} ^{n}  \frac{n+1-k}{n-1-k} \frac{(c \cdot \xi \bar{x}_{(k)}\bar{x}_{(n-k+2)})}{2n} ~. 
    \end{split}
\end{align}
Since we already know $\bar{x}_{I,(3)}=0$ from \eqref{eq: x3 zero}, and vanishing $\bar{u}^I _{(2)}$ leads to vanishing $\bar x_{I,(4)}$ as well via \eqref{eq: x4 cond}, it is not difficult to show that 
the expansion coefficients $\bar{x}_{I,(k\geq 3)}$ all vanish. Thus the 
perturbative series for $\bar x_{I,(n)}$ truncates after two terms, as expected for a harmonic function. The identification \eqref{eq: mu matching} identifies the subleading coefficient in the harmonic function at infinity with the leading one at the horizon, and {\it vice versa}. 

While \eqref{eq: mu matching} are the default, it is interesting that asymptotic boundary conditions 
with nonzero $\bar{u}_{I,(2)}$, $\bar{g}_{m,(3)}$ are consistent with supersymmetry. It has been argued that there may be missing solutions in certain supergravity theories that may not satisfy the canonical nonlinear charge constraint, see for example \cite{Bhattacharyya:2010yg, Markeviciute:2016ivy, Markeviciute:2018yal, Dias:2022eyq}. Since the value of the conserved charges do not take the canonical form, one way wonder if those parameters are somehow related to these missing solutions.

From this point of view, the possibility of $\bar{u}_{I,(2)}$, $\bar{g}_{m,(3)}$ perturbing asymptotic AdS$_5$ might be desirable. In the following, we discuss this possibility. 

First, recall that the electric charge $\widetilde{Q}_I$ and the angular momentum $\widetilde{J}$ are conserved charges, which means that they are the same whether evaluated at infinity or the horizon. Identifying \eqref{eq:Qxix} with \eqref{eq: Q infty}
we find
\begin{equation}
\label{eq: matching Q}
\begin{split}
 &\quad x_{I,(0)} -\frac{1}{2}\xi_I \left(\frac{1}{2}c^{JKL}x_{J,(0)}x_{K,(0)}\xi_L\right) + \frac{1}{2}c_{IJK}\left(\frac{1}{2}c^{JNO}\xi_N \xi_O\right) c^{KLM}x_{L,(0)}x_{M,(0)} \\ 
    &= \bar x_{I,(2)} -\frac{1}{2}\xi_I \left(\frac{1}{2}c^{JKL} \bar x_{J,(2)} \bar x_{K,(2)}\xi_L\right) + \frac{1}{2}c_{IJK}\left(\frac{1}{2}c^{JNO}\xi_N \xi_O\right) c^{KLM} \bar x_{L,(2)} \bar x_{M,(2)} \\&\quad + \epsilon \ell \left( \xi_I  \xi_{J} - c_{IJK} c^{KLM} \xi_{L} \xi_{M} \right) \bar u^J _{(2)} ~, 
\end{split} 
\end{equation} 
from matching $\widetilde{Q}_I$, and similarly \eqref{eq: J final nh} with \eqref{eq: J final inf} 
give \begin{align}
\label{eq: matching J}
\begin{split}
    &\quad \frac{\epsilon}{4}c^{IJK}
    x_{I,(0)}  x_{J,(0)} \xi_K +  \frac{1}{36}\left(c^{IJK} \xi_I \xi_J \xi_K \right) \left(c^{LMN}  x_{L,(0)} x_{M,(0)} x_{N,(0)} \right)  \\
    &= \frac{\epsilon}{4}c^{IJK}
    \bar{x}_{I,(2)}\bar{x}_{J,(2)} \xi_K +  \frac{1}{36}\left(c^{IJK} \xi_I \xi_J \xi_K \right) \left(c^{LMN} \bar{x}_{L,(2)} \bar{x}_{M,(2)} \bar{x}_{N,(2)} \right)\\
& \quad  + \frac{\epsilon}{\ell}\bar{g}_{m,(3)} + \ell \left(\frac{1}{2}\xi_I c^{JKL}\xi_J \xi_K \bar{x}_{L,(2)} + \xi_I + \frac{5}{3}\ell^{-3}\bar{x}_{I,(2)} \right)\bar{u}^I _{(2)} ~,
\end{split}
\end{align}
from matching $\widetilde{J}$. These conservation laws are consistent with a UV solution specified 
by $x_{I,(0)}$ (and the FI-couplings $\xi_I$) that flows to an IR configuration with $\bar x_{I,(2)}$ that may not even remotely agree with \eqref{eq: mu matching}. This consideration suggests that supersymmetry and charge conservation do little to constrain the IR limit of the flow.

However, there is a different source of intuition. If the perturbative series of $f^{-1}X_I$ from infinity did {\it not} truncate after exactly two terms, the third term would diverge at the horizon $R^2\to 0$, rather than approaching a constant. Other fields excited at the same order would similarly suggest a singularity. It could happen that, taking into account successive powers $R^{-2k}$ to all orders, 
there would be a finite limit $R^{2}\to 0$, after all, but determining by explicit computation 
whether this possibility is realized for any $\bar{u}_{I,(2)}$, $\bar{g}_{m,(3)}$ is technically challenging. 

From a different perspective, since the conserved charges from the near-horizon expansion do satisfy the typical charge constraint, the possibly new black hole solutions that do not seem to satisfy the typical charge constraint at infinity, would not flow to the expected near-horizon extremal AdS$_2$ geometry, which implies that these solutions may not be black holes after all.

Moreover, a change in the electric potential 
$fY^I \to fY^I + \beta^I$ with $\beta^I$ constant is trivial as it does not change the electromagnetic field strength. However, with the vielbein we have picked, such a shift must be accompanied by
$u^I \to u^I - w\beta^I$. Because $w$ includes a term $w\sim R^{-2}$ at large $R$, such a gauge transformation has the ability to remove $\bar{u}^I_{(2)}$. This mechanism shows the $\bar{u}^I_{(2)}$ are 
allowed, in principle, but also that they are not physical deformations. Indeed, these 
coefficients diverge at the horizon, so they correspond to a singular gauge which is ill-advised.

\section{Entropy Extremization} \label{section: entropy extremization}

In this section, we consider the near-horizon limit of the Legendre transform of the radial Lagrangian \eqref{eq: 1D bulk Lagrangian}, leading to a near-horizon entropy function. Extremizing this entropy function with respect to the near-horizon variables leads to an expression for the entropy in terms of the aforementioned charges.

\subsection{Near-horizon setup}

First, we consider the near-horizon of the line element $ds_2 ^2$ \eqref{eq:2d line el}, where we recall that $e^{2\rho}$ and $e^{2\sigma}$ can be expressed in terms of the variables $f$, $g_m$, and $w$ as in \eqref{eq:geometrydict}. At the horizon $R\to 0$, these variables have known near-horizon behaviors according to Table \ref{table: asymptotics}. Thus, the near-horizon limit of \eqref{eq:2d line el} becomes

\begin{equation}
\label{eq: ds2 nh}
    ds_{2,\text{nh}} ^2 = v\left(\frac{R^4}{\ell_2 ^2} dt^2 - \frac{dR^2}{R^2} \right)\, ,
\end{equation}
with $v$ and $\ell_2$ defined based on the $R\to 0$ behavior of $e^{2\rho}$ and $e^{2\sigma}$:
\begin{align}
    \left. e^{2\sigma} \right|_{\text{nh}} \equiv \frac{v}{R^2} ~, \qquad \left. e^{2\rho} \right|_{\text{nh}} \equiv  \frac{v}{\ell_2 ^2}R^4 ~.
\end{align}
Furthermore, $v^{\frac{1}{2}}$ and $\ell_2 ^{\frac{1}{3}}$ are near-horizon length scales defining the 2D $(t,R)$ part of the line element \eqref{eq: 5D metric ansatz full}
\begin{equation}
\label{eq: 5d nh metric ansatz}
    ds_{5,\text{nh}} ^2 = v\left(\frac{R^4}{\ell_2 ^2} dt^2 - \frac{dR^2}{R^2} \right) - e^{-U_1} (\sigma_1 ^2 + \sigma_2 ^2) -e^{-U_2}(\sigma_3 +a^0 )^2 ~. 
\end{equation}
The role of the variable $\ell_2$ is elucidated by noting the near-horizon limit of the K\"ahler condition \eqref{eq:Kahlercond}:
\begin{equation}
    \label{eq: Kahlercond nh}
    \ell_2 = 2ve^{-\frac{1}{2}U_2} ~. 
\end{equation}
This relation will be used to eliminate $\ell_2$ in the rest of the near-horizon 
analysis. 

Having reviewed the near-horizon 2D line element, and in anticipation of applying the entropy function formalism \cite{Sen:2005wa, Sen:2005iz, David:2006yn, Sen:2007qy, Sen:2008yk, Sen:2008vm, Sen:2009vz, Astefanesei:2006dd, Castro:2007sd} to the Lagrangian density in \eqref{eq: 1D bulk Lagrangian}, we use the following coordinate transformation
\begin{equation}
\label{eq: AdS2 nh rescale}
\begin{split}
    dt &\to \frac{1}{2}\ell_2 dt ~, \\ 
    dR &\to \frac{1}{2R}dR  \, ,
\end{split}
\end{equation}
to bring the coordinates $(t,R)$ in \eqref{eq: ds2 nh} to the canonical $\text{AdS}_2$ form 
\begin{equation}
    ds_{2,\text{nh}} ^2 = \frac{v}{4}\left(R^2 dt^2 - \frac{dR^2}{R^2}\right)  ~,
\end{equation}
where now it is clear that $v^{\frac{1}{2}}$ is related to the $\text{AdS}_2$ length scale.  

The coordinate transformation \eqref{eq: AdS2 nh rescale} will have the effect of rescaling the Lagrangian 2-form $\mathcal{L}_2 = \mathcal{L}_1 dt \wedge dR$ \eqref{eq: 2D bulk Lagrangian} by a factor
\begin{equation}
\label{eq: L2 factor}
    \mathcal{L}_2 \to \frac{\ell_2}{4R}\mathcal{L}_2 ~. 
\end{equation}
With the prescription of defining the entropy function through omitting the $dt \wedge dR$ volume form from the dimensionally-reduced action, we anticipate dividing the density $\mathcal{L}_1$ by a factor of $\frac{4R}{\ell_2}$. 

Combining the near-horizon behaviors of $e^{2\rho}$ and $e^{2\rho}$ studied above with the dictionary definitions \eqref{eq:geometrydict}, $e^{-U_1}$, $e^{-U_2}$ and $b^I$ can be shown to be constants to leading order in the near-horizon limit based on the leading-order behaviors of $f$, $g_m$ and $w$ consistent with the small $R$ asymptotics in Table~\ref{table: asymptotics}.

Concerning the matter fields, the electric fields $a^I$ and $a^0$ in \eqref{eq: 5D potential ansatz} become in the near-horizon limit
\begin{equation}
\label{eq: nh electrics}
        \left. a^I \right|_{\text{nh}} \equiv \frac{e^I R^2}{2v}e^{\frac{1}{2}U_2} dt, \quad
        \left. a^0 \right|_{\text{nh}} \equiv -\frac{e^0 R^2}{2v}e^{\frac{1}{2}U_2} dt ~.
\end{equation}
The total 1D Lagrangian density \eqref{eq: 1D bulk Lagrangian} then becomes
\begin{equation}
\label{eq: lagrangian nh}
\begin{split}
        \mathcal{L}_{1,\text{nh}} &= \frac{\pi}{2G_5} e^{-U_1 - \frac{1}{2}U_2} \frac{4R}{\ell_2} v \left[\frac{1}{v^2}e^{-U_2} (e^0)^2 - \frac{4}{v} + e^{U_1} - \frac{1}{4}e^{2U_1 - U_2} -\frac{1}{2}G_{IJ}e^{2U_1} b^I b^J\right. \\
        &\quad + \left.   \frac{2}{v^2}G_{IJ} (e^I - b^I e^0)(e^J - b^J e^0) - V \right] - \frac{\pi}{2G_5} \frac{4R}{\ell_2}\frac{1}{2}c_{IJK} b^I b^J (e^K - \frac{2}{3}e^0 b^K)  \, .
    \end{split}
\end{equation}
Apart from an overall prefactor in the integration measure, every other appearance of $\ell_2$ has been re-expressed in terms of $v$ and $U_2$ by using \eqref{eq: Kahlercond nh}. The $\frac{4R}{\ell_2}$ factor has been factored out of the volume element, and we follow the prescription made earlier to exactly cancel it out with the $\frac{\ell_2}{4R}$
factor from \eqref{eq: L2 factor} in order to obtain the Lagrangian density suitable for the entropy function. We also note the presence of the Chern-Simons boundary terms in \eqref{eq: lagrangian nh} that are crucial for calculating the near-horizon charges. 

We now obtain a near-horizon Lagrangian \eqref{eq: lagrangian nh} that is a function of the variables $v$, $U_1$, $U_2$, $e^I$, $e^0$, and $b^I$. We will next derive the charges $Q_I$ and $J$ from $\mathcal{L}_{1,\text{nh}}$, with the goal of Legendre transforming the Lagrangian into an entropy function $\mathcal{S}$ that can be ultimately extremized towards a function purely of the charges $\mathcal{S}=\mathcal{S}(Q_I, J)$. 

We note from the earlier Noether procedure \eqref{eq: 2D charges} that $Q_I$ and $J$ were obtained in terms of radially dependent variables. In terms of the near-horizon limit of the electric fields \eqref{eq: nh electrics}, this becomes
\begin{align}
    Q_I &= 
    \frac{\pi}{G_5}\left[  e^{-U_1 - \frac{1}{2}U_2}  \frac{4}{v}G_{IJ} (e^J - b^J e^0)  -  \frac{1}{2}c_{IJK} b^J b^K\right] ~, \\ 
    J&= \frac{\pi}{G_5} \left[-\frac{2}{v}e^{-U_1 - \frac{3}{2}U_2} e^0 + \frac{4}{v}e^{-U_1 - \frac{1}{2}U_2}G_{IJ}b^I (e^J - b^J e^0) -\frac{1}{3}c_{IJK} b^I b^J b^K \right] ~.
\end{align} 
This introduces the charges $Q_I$ and $J$ as conjugates to the electric fields $e^I$ and $e^0$, allowing for the inversion
\begin{align}
\label{eq:eI to charges}
    e^I - b^I e^0 &= \frac{v}{16}e^{U_1 + \frac{1}{2}U_2} G^{IJ}\left(  \widetilde{Q}_J + 2c_{JKL} b^K b^L \right) ~,\\ 
\label{eq:e0 to charges}
    e^0 &= -\frac{v}{8}e^{U_1 + \frac{3}{2}U_2} \left(\widetilde{J}- \widetilde{Q}_I b^I -\frac{2}{3}c_{IJK} b^I b^J b^K \right)~ \, ,
\end{align}
where now the rescaled $\widetilde{Q}_I$ and $\widetilde{J}$ \eqref{eq: JQ rescale} have been used. The near-horizon entropy function can now be defined as a Legendre transform of the Lagrangian density \eqref{eq: lagrangian nh} with fixed charges
\begin{equation}
\label{eq: entropy legendre}
    \mathcal{S} = 2\pi  \left(e^I \frac{\partial \mathcal{L}_{1,\text{nh}}}{\partial e^I} +  e^0 \frac{\partial \mathcal{L}_{1,\text{nh}}}{\partial e^0} - \mathcal{L}_{1,\text{nh}}\right) ~,
\end{equation}
which, after eliminating the electric fields through \eqref{eq:eI to charges} and \eqref{eq:e0 to charges}, yields 
\begin{equation}
\begin{split}
\label{eq: entropy function nh}
{\cal S} & = \frac{4\pi^2}{G_5} e^{-U_1 - \frac{1}{2} U_2} \left[  1 
+ \frac{v}{4}\left(\frac{1}{4} e^{2U_ 1 - U_2} -  e^{U_1}  + \frac{1}{64} e^{2U_1+2U_2}\left( \widetilde{J} - \widetilde{Q}_I b^I - \frac{2}{3} c_{IJK} b^I b^J b^K \right)^2
 \right. \right. \cr 
&\left. \left.
 +  V + \frac{1}{128} e^{2U_1+U_2} G^{IJ} \left( \widetilde{Q}_I + 2c_{IKL} b^K b^L \right)  \left( \widetilde{Q}_J +  2c_{JMN} b^M b^N \right) + \frac{1}{2} e^{2U_1} G_{IJ} b^I b^J \right)\right]~.
\end{split}
\end{equation}
This entropy function depends on the physical variables $v$, $U_1$, $U_2$, $b^I$, $X^I$
describing the near horizon geometry and matter fields, with the conserved charges $\widetilde{Q}_I$, $\widetilde{J}$ appearing as fixed parameters. At its extremum, it yields the physical variables and the black hole entropy as a function of the charges. 
 
 \subsection{Extremization of the Entropy Function}
 
It is exceedingly simple to extremize  with respect to $v$ which appears only as a Lagrange multiplier in front of the large round bracket that comprises nearly all of \eqref{eq: entropy function nh}. This leaves the extremized value of $\mathcal{S}$:
\begin{equation}
\label{eq: S ext value}
    \mathcal{S} = \frac{4\pi^2}{G_5} e^{-U_1 - \frac{1}{2} U_2} ~. 
\end{equation}
This is exactly the black hole entropy computed via the area law for a horizon defined by the volume 3-form $e^{-U_1 - \frac{1}{2}U_2} \sigma_1 \wedge \sigma_2 \wedge \sigma_3$ with the angular ranges specified in \eqref{eq: theta phi psi ranges}. However, the explicit dependence of $U_1$ and $U_2$ on the charges remains to be determined. For this we must
extremize with respect to the remaining variables
\begin{align}
\label{eq: S ext eqs}
    \begin{split}
         \partial_{U_{1}} \mathcal{S} = \partial_{U_{2}} \mathcal{S}  = \partial_{b^{I}} \mathcal{S} = D_I \mathcal{S}  = 0~.
    \end{split}
\end{align}
Here $D_I$ is the K\"ahler-covariantized derivative with respect to the scalars $X^I$. It
is defined such that $X^I D_I = 0$, which is the correct way to vary the scalars while also implementing the constraint \eqref{eq:Xconst}. 
The conditions \eqref{eq: S ext eqs} give
\begin{align}
\label{eq: v ext S}
    0=&V-e^{U_1} + \frac{1}{4}e^{2U_1 - U_2}   + \frac{1}{64}e^{2U_1 + 2U_2} \mathcal{M}^2 + G^{IJ} \frac{1}{128}e^{2U_1 + U_2 }\mathcal{K}_I \mathcal{K}_J  + \frac{1}{2}G_{IJ} e^{2U_1} b^I b^J  ~, \\ 
\label{eq: U1 ext S}
    0=&-4 -vV + \frac{v}{4}e^{2U_1 - U_2}  + \frac{v}{64}e^{2U_1 + 2U_2} \mathcal{M} ^2 + G^{IJ} \frac{v}{128}e^{2U_1 + U_2 } \mathcal{K}_I \mathcal{K}_J + \frac{v}{2}G_{IJ} e^{2U_1} b^I b^J  ~, \\
\nonumber 0=&-2 -\frac{v}{2}V +\frac{v}{2}e^{U_1} - \frac{3v}{8}e^{2U_1 - U_2}  + \frac{3v}{128}e^{2U_1 + 2U_2} \mathcal{M} ^2 + G^{IJ} \frac{v}{256}e^{2U_1 + U_2 } \mathcal{K}_I \mathcal{K}_J \\&\qquad - \frac{v}{4}G_{IJ} e^{2U_1} b^I b^J ~, 
 \label{eq: U2 ext S}
\\
\label{eq: X ext S}
\begin{split}
 0=&v D_I V + \frac{v}{128}e^{2U_1 + U_2} (D_I G^{JK}) \mathcal{K}_J \mathcal{K}_K + \frac{v}{2}e^{2U_1} (D_I G_{JK}) b^J b^K ~,
\end{split}
\\
\label{eq: b ext S}
0=&\frac{1}{32}e^{U_2}\left(-e^{U_2} \mathcal{K}_I \mathcal{M} \right. \left. +2 G^{JK} c_{IJN} b^N \mathcal{K}_K \right)
+ G_{IJ}b^J~,
\end{align}
where $\mathcal{M}$ and $\mathcal{K}_I$ are shorthand for
\begin{equation}
    \mathcal{M} \equiv \widetilde{J} - \widetilde{Q}_I b^I - \frac{2}{3}c_{IJK}b^I b^J b^K \ , \ \mathcal{K}_I \equiv \widetilde{Q}_I + 2c_{IJK} b^J b^K ~. 
\end{equation}
Additionally, $D_I$ acts on the scalars $X^J$ following\footnote{This can be generalized to other quantities via the product rule on $D_I$, for instance,
\begin{equation}
    \begin{split}
        D_I X_J &= \frac{1}{2}c_{JKL} D_I(X^K X^L) = c_{IJK}X^K - \frac{2}{3}X_I X_J ~.
    \end{split}
\end{equation}}
\begin{equation}
    D_I X^J = \delta_I ^J - \frac{1}{3}X_I X^J~.
\end{equation}
Ideally we seek the most general extremum that solves (\ref{eq: v ext S}-\ref{eq: b ext S}) that is consistent with our ansatz \eqref{eq: 5d nh metric ansatz}. However, the extremization equations are highly nonlinear in the variables of interest $(v,U_1,U_2,X^I,b^I)$. Therefore, in the following we specialize and find all supersymmetric solutions.

\subsubsection*{Near-horizon supersymmetric conditions}

It is straightforward to take the near-horizon limit
of the supersymmetry conditions (\ref{eq:susyrelation2}-\ref{eq:susyrelation4}), along with
the identification $X^I = Y^I$ \eqref{eq: XY match}. After inverting $e^I$ and $e^0$ in terms of $\widetilde{Q}_I$ and $\widetilde{J}$,
following \eqref{eq:eI to charges} and \eqref{eq:e0 to charges}), we obtain the following near-horizon supersymmetric relations
\begin{align}
\label{eq: nh susy rel1}
0&= \widetilde{Q}_I + 2c_{IJK}b^J b^K - 4e^{-U_2} X_I ~, \\
    \label{eq: nh susy rel2}
   0&= \widetilde{J} - \widetilde{Q}_I b^I - \frac{2}{3}c_{IJK}b^I b^J b^K  -  4e^{-U_1 - U_2} (\xi \cdot X) ~, \\ 
    \label{eq: nh susy rel3}
    0&= b^I - e^{-U_1} \left(X^I (\xi \cdot X) - G^{IJ} \xi_J \right) ~, \\
    \label{eq: nh susy rel4}
    0&= e^{U_1} - \frac{4}{v} -2V ~. 
\end{align}
We also need the near-horizon version of the K\"ahler condition \eqref{eq:Kahlercond}. We can trade the variables $\rho$ and $\sigma$ describing the 2D geometry for $f$ and $g_m$ following \eqref{eq:geometrydict} and eliminate $a^0 _t$ that results in favor of 
the charges through \eqref{eq:e0 to charges}. These steps lead to
\begin{equation}
\label{eq: nh susy rel5}
    \frac{4}{v}-(\xi \cdot X)^2 = e^{2U_1 - U_2} ~. 
\end{equation}
We have verified that when the five supersymmetric relations (\ref{eq: nh susy rel1}-\ref{eq: nh susy rel5}) 
are satisfied, then the five $\mathcal{S}$ extremization equations (\ref{eq: v ext S}-\ref{eq: b ext S}) are satisfied as well. The details of this computation are not instructive so we omit them. 
The reverse logic would be that {\it all} extremal solutions within the scope of our ansatz are supersymmetric. This we have not shown, and it is indeed not true, i.e. there are no known nonextremal supersymmetric Lorentzian black holes. Thus the specialization to supersymmetric solutions addresses a genuine subset of the extremal black holes.

In the remainder of this subsection we solve the supersymmetry relations 
(\ref{eq: nh susy rel1}-\ref{eq: nh susy rel5}) explicitly and find all variables as functions of 
the charges $Q_I$ and $J$.  

\subsubsection{Solving for the entropy and the charge constraint}

The supersymmetry conditions 
(\ref{eq: nh susy rel1}-\ref{eq: nh susy rel5}) are all algebraic, but they are far from trivial. Straightforward contractions, followed by taking simple linear combinations, give scalar identities
\begin{align}
X\cdot b & = e^{-U_1} \xi \cdot X ~,\cr
\widetilde{Q}\cdot b & = \frac{3}{2}\widetilde{J} - 8 e^{-U_1 - U_2} \xi\cdot X~,\cr
\xi\cdot b &= e^{U_1-U_2}-1~,
\label{eq: bprojections}
\end{align}
which will prove useful later. 
Our strategy will be to exploit identities like these to find
simple combinations of variables that can be expressed entirely in terms of 
the charges $Q_I$, $J$ and the couplings $\xi_I$. Combinations of those
will in turn give explicit formulae for physical variables. 

In this spirit, we expand $\widetilde{Q}_J \widetilde{Q}_K$ using the square of \eqref{eq: nh susy rel1}, and then simplify
the terms that are products of $b$'s using \eqref{eq: nh susy rel2}. We obtain
\begin{equation}
\label{eq: QQ nh susy simplified}
    \frac{1}{2}c^{IJK}\widetilde{Q}_J \widetilde{Q}_K = 32e^{-2U_1 - U_2}c^{IJK} \xi_J \xi_K + 16 e^{-U_1 - U_2} X^I  - 2\widetilde{J}b^I ~.  
\end{equation}
Contracting \eqref{eq: QQ nh susy simplified} with $\xi_I$, the first term on the right 
becomes proportional to $e^{-2U_1 - U_2}$, which is related to the black hole entropy through $\mathcal{S}$ \eqref{eq: S ext value}. There will also be a term 
$(\xi \cdot X)$ that we can eliminate with the help of 
\begin{align}
\label{eq: J xi id}
\widetilde{J} = 8 e^{-2U_1} (\xi \cdot X) + \frac{32}{3}e^{-3U_1}c^{IJK} \xi_I \xi_J \xi_K ~,
\end{align}
which is a simplification of \eqref{eq: nh susy rel2} with $\widetilde{Q}_I$ and $b^I$ eliminated using \eqref{eq: nh susy rel1} and \eqref{eq: nh susy rel3}, respectively. These steps give
\begin{equation}
\label{eq: entropy2 JQ}
  \frac{1}{6}c^{IJK}\xi_I \xi_J \xi_K \left(\frac{\mathcal{S}}{2\pi}\right)^2 = \frac{1}{2}c^{IJK}\xi_I Q_J Q_K - \frac{\pi}{4G_5}2J ~, 
\end{equation}
which amounts to an explicit formula for the black hole entropy as function of the conserved charges
\begin{equation}
\label{eq:entropybox}
    \mathcal{S} = 2\pi \sqrt{\frac{1}{2}c^{IJK}\ell^3 \xi_I Q_J Q_K - N^2 J} ~.
\end{equation}
This is in full agreement with the entropy of supersymmetric extremal AdS$_5$ black holes \cite{Cabo-Bizet:2018ehj, Benini:2018mlo, Choi:2018hmj}. We have expressed the entropy using the untilded charges $Q_I$ and $J$ \eqref{eq: JQ rescale}  and traded $G_5$ for $N^2$ using $\frac{\pi \ell ^3}{4G_5}= \frac{1}{2}N^2$ and applied $\frac{1}{6}c^{IJK}\xi_I \xi_J \xi_K = \ell^{-3}$ from \eqref{eq: xi cubed constraint} to explicitly show all the dimensionful quantities.

Continuing with the strategy of evaluating natural combinations of the conserved charges, we evaluate the cubic invariant of the charges $c^{IJK} \widetilde{Q}_I \widetilde{Q}_J \widetilde{Q}_K$ by taking the cube of $\widetilde{Q}_I$ from \eqref{eq: nh susy rel1}, with the resulting contractions of $X_{I}$ and $b^{I}$ such as $c_{IJK}X^I b^J b^K$ and $X_I b^I$ simplified using the $b^I$ relation \eqref{eq: nh susy rel3} as well as \eqref{eq: nh susy rel4} and \eqref{eq: nh susy rel5} for the terms quadratic in $\xi$. This yields
\begin{equation}
\label{eq: QQQ form 1}
    c^{IJK} \widetilde{Q}_I \widetilde{Q}_J \widetilde{Q}_K = 6e^{-U_1 - 2U_2} + \frac{1}{8}\widetilde{J} c_{IJK} b^I b^J b^K~.
\end{equation}
Alternatively, we can arrive at the cubic product of electric charges by contracting \eqref{eq: QQ nh susy simplified} 
with $Q_I$ from \eqref{eq: nh susy rel1}, giving
\begin{align}
\label{eq: QQQ form 2}
\begin{split}
    &\tfrac{1}{64}c^{IJK}\widetilde{Q}_I \widetilde{Q}_J \widetilde{Q}_K \\&= 4e^{-U_1 -2U_2} +2e^{-2U_1 - U_2} +e^{-2U_1 -U_2} c^{IJK}\widetilde{Q}_I \xi_J \xi_K -\tfrac{1}{4}e^{-U_1 - U_2} \widetilde{J}(\xi \cdot X) + \tfrac{1}{8}\widetilde{J}c_{IJK} b^{I} b^{J} b^{K} ~.
\end{split}
\end{align}
Comparing \eqref{eq: QQQ form 1} and \eqref{eq: QQQ form 2}, we find the identity
\begin{equation}
\label{eq: QJ id1}
    \frac{1}{2}c^{IJK}\widetilde{Q}_I \xi_J \xi_K + 1 = e^{U_1} \left(e^{-U_2} + \frac{1}{8}\widetilde{J}~\xi \cdot X\right)~.
\end{equation}
It is useful because it gives access to a useful combination of $U_1$, $U_2$, and $\xi\cdot X$. Indeed, we can simplify the cube of the electric charge in the form \eqref{eq: QQQ form 1} using the $b$ identity \eqref{eq: nh susy rel3}, and then \eqref{eq: J xi id} to eliminate $(\xi \cdot X)$, to give
\begin{equation}
\frac{1}{6}c^{IJK}\widetilde{Q}_I \widetilde{Q}_J \widetilde{Q}_K +\widetilde{J}^2= 64 e^{-U_1 - U_2} \left(e^{-U_2} +\frac{1}{8}\widetilde{J}~\xi \cdot X~\right) ~.
\end{equation}
The right-hand side of this equation differs from that of \eqref{eq: QJ id1} only by a factor
proportional $e^{-2U_1 - U_2}$ which is the square of the geometric measure on the black hole horizon. 
As such, it is related to the black hole entropy $\mathcal{S}$ both through the area law \eqref{eq: S ext value} and as a function of charges \eqref{eq:entropybox}. Collecting these relations, and reintroducing $Q_I$ and $J$ \eqref{eq: JQ rescale} in order to align with the conventional units for this result, we find
\begin{equation}
\label{eq: charge constraint}
    \left(\frac{1}{6}c^{IJK}Q_I Q_J Q_K + \frac{\pi}{4G_5}J^2 \right)= \ell^3 \left(\frac{1}{2}c^{IJK}\xi_I\xi_J Q_K + \frac{\pi}{4G_5} \right)\left(\frac{1}{2}c^{IJK}\xi_I Q_J Q_K - \frac{\pi}{2G_5} J\right) ~.
\end{equation}
This is the prototypical 5D nonlinear charge constraint \cite{Cabo-Bizet:2018ehj, Benini:2018mlo, Choi:2018hmj}. However, the charge constraint \eqref{eq: charge constraint} does not make progress towards solving the supersymmetry equations nor determining the near horizon solution in terms of conserved charges. Rather, it is a relation between the conserved charges that, if taken at face value, all supersymmetric black holes dual to ${\cal N}=4$ SYM must satisfy. This is extremely important and the continuing questions regarding this constraint is one of the motivations for the detailed study reported in this article. 

Even at this point of our discussion where we are deep into solving certain nonlinear equations, it is worth noting that
that the black hole entropy \eqref{eq: entropy2 JQ} and the charge constraint \eqref{eq: charge constraint} 
can be combined into one complex-valued equation
\begin{equation}
\label{eq: complex charge rel}
 ~~~\frac{1}{6} c^{IJK} \left( Q_I + i \frac{\cal S}{2\pi} \xi_I \right) \left(Q_J + i \frac{\cal S}{2\pi} \xi_J\right) \left(Q_K + i  \frac{\cal S}{2\pi}\xi_K\right) +\frac{\pi}{4G_5}
\left(- J + i \frac{\cal S}{2\pi}\right)^2 = 0 ~~~~. 
\end{equation}
The real part gives the constraint \eqref{eq: charge constraint} and the imaginary part gives the formula for the entropy \eqref{eq: entropy2 JQ}. Complexified equations are natural in problems involving supersymmetry. Also, \eqref{eq: complex charge rel} appears as the condition for a complex saddle point that
provides an accounting in ${\cal N}=4$ SYM for the entropy of black hole preserving $1/16$ of the maximal 
supersymmetry \cite{Cabo-Bizet:2018ehj,Cabo-Bizet:2019eaf, Choi:2018hmj, Benini:2018mlo}.

\subsubsection{Near-horizon variables as function of conserved charges}

Having discussed the black hole entropy and the constraint on charges, we move on to expressing all other aspects
of the near-horizon geometry and the matter content in terms of the fixed charges $Q_I$ and $J$. 

For the following computations, we make use of the expressions \eqref{eq: nh susy rel1} and \eqref{eq: nh susy rel2} for $J$ and $Q_I$ simplified using the relation for $b^I$ in \eqref{eq: nh susy rel3}. We find
\begin{align}
\label{eq: Q nh susy simplified}
    Q_I &= \frac{\pi}{G_5}e^{-2U_1}\left[X_I e^{2U_1 - U_2} + X_I (\xi \cdot X)^2 - 2\xi_I (\xi \cdot X) - 4G_{IJ}c^{JKL}\xi_K \xi_L \right] ~, \\
\label{eq: J nh susy simplified}
    J &= \frac{16\pi}{G_5}e^{-3U_1} \left[\frac{1}{8}e^{U_1} \xi \cdot X + \ell^{-3} \right] ~. 
\end{align}
Multiplying both sides of \eqref{eq: QJ id1} by $J$, using the relation \eqref{eq: J nh susy simplified}, 
we can solve for $e^{U_1}~\xi \cdot X$ in terms of the charges
\begin{equation}
\label{eq: xidotX nh JQ}
    \frac{1}{8}\xi \cdot X e^{U_1} = \frac{J\left(\frac{1}{2}c^{IJK}Q_I \xi_J \xi_K + \frac{\pi}{4G_5}\right) - \left(\frac{\mathcal{S}}{2\pi}\right)^2 \ell^{-3}}{J^2 + \left(\frac{\mathcal{S}}{2\pi}\right)^2} ~.
\end{equation}
This result ties a geometrical quantity --- as it appears in the near-horizon K\"ahler relation \eqref{eq: nh susy rel5} --- to the charges. It also has an additional immediate value as \eqref{eq: J nh susy simplified} 
relates it to $e^{-3U_1}$, giving
\begin{align}
\label{eq: U1 nh JQ}
 ~~~ e^{-3U_1} = \frac{G_5}{16\pi} \frac{J^2 + \left(\frac{\mathcal{S}}{2\pi}\right)^2}{\frac{1}{2}c^{IJK}Q_I \xi_J \xi_K + \frac{\pi}{4G_5} + J\ell^{-3}} ~.
\end{align}
This expression sets the scale of the non-deformed $S^3$ which has line element $e^{-U_1}(\sigma_1 ^2 + \sigma_2 ^2 + \sigma_3 ^2)$. 

Due to the rotation of the black hole, the horizon geometry \eqref{eq: 5D metric ansatz full} 
is deformed away from $S^3$. 
We can quantify the deformation by computing $e^{U_1- U_2}$ via $e^{3U_1}$ from \eqref{eq: U1 nh JQ} and $e^{2U_1 + U_2}$ from \eqref{eq: S ext value}: 
\begin{align}
\label{eq: U1mU2 nh JQ}
 ~~~e^{U_1 -U_2}  =
 \frac{ \frac{4G_5}{\pi} \left( \frac{1}{2} c^{IJK} Q_I \xi_J \xi_K + \frac{\pi}{4G_5} +
 J\ell^{-3}\right)   \left( \frac{\cal S}{2\pi} \right)^2 }{ J^2 + \left( \frac{\cal S}{2\pi} \right)^2 } ~.
\end{align}
The only scalar near-horizon parameter that was not yet computed is the AdS$_2$-volume $v$. Due to the alternate near-horizon K\"ahler condition \eqref{eq: nh susy rel5}, the combination $ve^{U_1}$ can be expressed in terms of $\mathcal{S}$, $(\xi \cdot X)e^{U_1}$ and $e^{-3U_1}$. These three quantities were given as functions 
of the conserved charges in \eqref{eq: S ext value}, \eqref{eq: xidotX nh JQ} and \eqref{eq: U1 nh JQ}. After simplifications, we find
\begin{align}
\label{eq: vU1 nh JQ}
  ~~~\frac{v}{4} e^{U_1} = \frac{\pi \ell^3}{4G_5}\frac{  \ell^3\left(\frac{1}{2} c^{IJK} Q_I \xi_J \xi_K+ \frac{\pi}{4G_5}\right) +
 J }  {    \ell^6 \left( \frac{1}{2} c^{IJK} Q_I \xi_J \xi_K+ \frac{\pi}{4G_5} \right)^2 +  \left( \frac{\cal S}{2\pi} \right)^2
 }.
\end{align}
This completes the explicit extremization of the entropy function for the scalar variables which at this point have all been expressed in terms of conserved charges $Q_I, J$ and FI-couplings $\xi_I$.

We must similarly determine the vectors $b^I$ and $X^I$ at the extremum which may be determined, in principle, by the input vectors $\xi_I$ and $\widetilde{Q}_I$. However, the position of the vector indices $I$ do not match 
so the full real special geometry enters. We exploit only the vectorial symmetry, and then, $X^I$ and $b^I$ must be linear combinations of {\it three} vectors: $c^{IJK}\widetilde{Q}_I Q_J$, $c^{IJK}\widetilde{Q}_I \xi_J$, and $c^{IJK}\xi_I \xi_J$. One linear relation of this kind was given in \eqref{eq: QQ nh susy simplified}. To find another, we contract \eqref{eq: QQ nh susy simplified} with $\widetilde{Q}_I$ and 
simplify using \eqref{eq: QQQ form 2}. This gives
\begin{equation}
    \widetilde{Q}\cdot X = 8 e^{-U_2} + 4 e^{-U_1}~.
\end{equation}
Combining this with \eqref{eq: bprojections} and \eqref{eq: xidotX nh JQ}, we have all four inner products of $b^I$, $X^I$, $Q_I$ and $\xi_I$. 
We already determined the scalar combinations $c^{IJK} \xi_I \xi_J \widetilde{Q}_K$ and $c^{IJK} \xi_I \widetilde{Q}_J \widetilde{Q}_K$ from \eqref{eq: entropy2 JQ} and \eqref{eq: QJ id1}, so we can establish the vectorial equation
\begin{equation}
    \frac{1}{2}c^{IJK} \widetilde{Q}_J \xi_K = b^I + \frac{1}{8}\widetilde{J} e^{U_1} X^I~.
    \label{eq: Qx nh susy simplified}
\end{equation}
Inversion of \eqref{eq: QQ nh susy simplified} and \eqref{eq: Qx nh susy simplified}
give
\begin{align}
 \label{eq: bI nh JQ}
 b^I = \frac{4G_5}{\pi} \cdot \frac{1}{2}
 \frac{  c^{IJK} \xi_J Q_K\left( \frac{\cal S}{2\pi} \right)^2 -\left( \frac{1}{2} c^{IJK} Q_J Q_K - \frac{1}{2}  c^{IJK} \xi_J \xi_K\left( \frac{\cal S}{2\pi} \right)^2\right) J  } { J^2+   \left( \frac{\cal S}{2\pi} \right)^2  }~,
 \end{align}
and
\begin{align}
\label{eq: XI nh JQ}
 X^I = 4 e^{-U_{1}}
 \frac{  J c^{IJK} \xi_J Q_K +\left(\frac{1}{2} c^{IJK} Q_J Q_K - \frac{1}{2}  c^{IJK} \xi_J \xi_K\left( \frac{\cal S}{2\pi} \right)^2 \right)} {  J^2+  \left(\frac{\cal S}{2\pi} \right)^2 } 
  ~,
\end{align}
where, once we impose the value of $e^{-U_{1}}$ given in \eqref{eq: U1 nh JQ}, $X^{I}$ is a function of the entropy \eqref{eq:entropybox} and the charges of the black hole.

In summary, we have found that the near-horizon limit of the supersymmetric equations implies that the near-horizon fields and variables of the geometry/matter ansatz are given by the charges $Q_I$ and $J$, through the relations (\ref{eq: U1 nh JQ}-\ref{eq: XI nh JQ}), which themselves parametrize a special extremum of the near-horizon entropy function \eqref{eq: entropy function nh}, for general FI coupling $\xi_I$ and $c_{IJK}$.

\subsection{Complexification of the near-horizon variables}

Each of the main results derived in the previous subsection are complicated formulae. However, they resemble one another and, in particular, it stands out that several expressions, such as \eqref{eq: bI nh JQ} and \eqref{eq: XI nh JQ}, share a common denominator. Indeed, there is an elegant way to pair them into complexified near-horizon variables
\begin{equation}
Z^I = b^I - i e^{-\frac{1}{2}U_2}X^I  =  \frac{G_5}{\pi}  \frac{c^{IJK} (Q_J + i\frac{\mathcal{S}}{2\pi}\xi_J) (Q_K + i\frac{\mathcal{S}}{2\pi}\xi_K)}{-J + i\frac{\mathcal{S}}{2\pi}} ~. 
\label{eq:ZIdefi}
\end{equation}
To the extent $X^I$ can be interpreted is an electric field it is indeed natural that its partner is a magnetic field $b^{I}$. In addition to the real part $b^I$ being given by \eqref{eq: bI nh JQ}, we recognize in the imaginary part the combination of the factor $e^{-U_1 - \frac{1}{2}U_2}$ in the entropy $\mathcal{S}$ and $X^I e^{U_1}$ given respectively by \eqref{eq:entropybox} and \eqref{eq: XI nh JQ}.

Some discussions of the AdS$_5$ black hole geometry invoke from the outset principles that are inherently complex, such as the Euclidean path integral or special geometry in four dimensions. This can give conceptual challenges so, in our discussion of entropy extremization, complex variables such as \eqref{eq:ZIdefi}  are introduced \cite{Hosseini:2017mds, Ntokos:2021duk} only for their apparent convenience. 
To make precise connections with the literature, we now to 
reintroduce the electric fields $e^I$ and $e^0$ conjugate to the conserved charges $\tilde{Q}_I$ and $\tilde{J}$. 
For $e^0$ defined in \eqref{eq:e0 to charges}, simplification using \eqref{eq: nh susy rel2}, gives an expression for $e^0$ that depends on $ve^{U_1}$ in \eqref{eq: vU1 nh JQ}, $e^{-3U_1}$ in \eqref{eq: U1 nh JQ}, $\mathcal{S}$ in \eqref{eq: S ext value}, and $(\xi \cdot X)e^{U_1}$ in \eqref{eq: xidotX nh JQ}. Collecting formulae, we then find 
\begin{align}
\label{eq: e0 nh JQ}
 e^0 = - \frac{4\pi }{\cal S} \frac{\pi \ell^3}{4G_5} \frac{ J \ell^{3} \left( \frac{1}{2} c^{IJK} Q_I \xi_J \xi_K+ \frac{\pi}{4G_5} \right) -  \left( \frac{\cal S}{2\pi} \right)^2 
 }  {    \ell^6 \left( \frac{1}{2} c^{IJK} Q_I \xi_J \xi_K+ \frac{\pi}{4G_5}\right)^2  + \left( \frac{\cal S}{2\pi} \right)^2} ~.
\end{align}
This expression for $e^0$ combines nicely with \eqref{eq: vU1 nh JQ} and gives the
complex potential
\begin{equation}
\label{eq: complex scalar nh JQ}
    \frac{1}{2}e^0 + i \frac{v}{4}e^{U_1} = \frac{\pi \ell^3}{4G_5} \left(\frac{2\pi}{\mathcal{S}}\right) \frac{-J+i\frac{\mathcal{S}}{2\pi}}{\ell^3 \left( \frac{1}{2} c^{IJK} Q_I \xi_J \xi_K+ \frac{\pi}{4G_5}\right) + i\left( \frac{\cal S}{2\pi}\right)} ~.
\end{equation}
Given $e^0$ in \eqref{eq: e0 nh JQ} as well as \eqref{eq: bI nh JQ} and \eqref{eq: XI nh JQ}, the electrical potentials dual to the vectorial charges become \eqref{eq:eI to charges}:
\begin{align}
\label{eq: eI nh JQ}
e^I = \frac{2\pi}{\mathcal{S}}  \frac{\ell^6 \left(\frac{1}{2}c^{IJK}Q_J Q_K - \frac{1}{2}c^{IJK}\xi_J \xi_K \left(\frac{\mathcal{S}}{2\pi}\right)^2\right)(\frac{1}{2} c^{IJK} Q_I \xi_J \xi_K+ \frac{\pi}{4G_5})+\ell^3 c^{IJK}Q_J \xi_K (\tfrac{\mathcal{S}}{2\pi})^2 }{ \ell^6 \left( \frac{1}{2} c^{IJK} Q_I \xi_J \xi_K+ \frac{\pi}{4G_5}\right)^2  + \left( \frac{\cal S}{2\pi} \right)^2}.
\end{align}
As preparation for the complexified version, we combine
\eqref{eq: bI nh JQ} and \eqref{eq: XI nh JQ} as
\begin{align}
\begin{split}
&\frac{v}{2}\left(b^I e^{U_1} + X^I \xi\cdot X\right) 
\\&=  \frac{\ell^3 c^{IJK}Q_J \xi_K \left(\frac{1}{2} c^{LMN} Q_L \xi_M \xi_N+ \frac{\pi}{4G_5}\right) - \ell^3 \left(\frac{1}{2}c^{IJK}Q_J Q_K -\frac{1}{2}c^{IJK}\xi_J \xi_K \left( \frac{\cal S}{2\pi}\right)^2 \right)}{\ell^6 \left( \frac{1}{2} c^{IJK} Q_I \xi_J \xi_K+ \frac{\pi}{4G_5}\right)^2 + \left( \frac{\cal S}{2\pi}\right)^2}~,
\end{split}
\end{align}
where we have imposed $\frac{v}{4}e^{U_1}$ in \eqref{eq: vU1 nh JQ}, $e^{-U_1}$ in \eqref{eq: U1 nh JQ},  
and $(\xi \cdot X)e^{U_1}$ in \eqref{eq: xidotX nh JQ}. 
We then find the complex special geometry vector
\begin{equation}
\label{eq: complex vector nh JQ}
    e^I + i\frac{v}{2}\left(b^I e^{U_1}+X^I (\xi \cdot X)\right) = \frac{2\pi}{\mathcal{S}} \frac{\frac{1}{2}c^{IJK}(Q_J+i\frac{\mathcal{S}}{2\pi}\xi_J)(Q_K+i\frac{\mathcal{S}}{2\pi}\xi_K)}{\ell^3 (\frac{1}{2}c^{LMN}Q_L \xi_M \xi_N + \frac{\pi}{4G_5}) + i\frac{\mathcal{S}}{2\pi}} ~.
\end{equation}
The complex potentials (\ref{eq: complex scalar nh JQ}-\ref{eq: complex vector nh JQ}) appear commonly in the literature, albeit with the normalization
\begin{equation}
\label{eq: omega nh}
~~~
\frac{\omega}{\pi} =     \frac{\pi\ell^3}{4G_5} \left(\frac{2\pi}{\mathcal{S}}\right) \frac{-J+i\frac{\mathcal{S}}{2\pi}}{\ell^3 \left( \frac{1}{2} c^{IJK} Q_I \xi_J \xi_K+ \frac{\pi}{4G_5}\right) + i\left( \frac{\cal S}{2\pi}\right)}  = \frac{1}{2} e^0+ 
  i  \frac{v}{4} e^{ U_1}  
  ~,
\end{equation}
and
\begin{equation}
\label{eq: DeltaI nh}
~~~\frac{\Delta^I}{\pi}  =  \frac{2\pi} {\cal S} \frac{ \frac{1}{2} c^{IJK} \ell^3 (Q_J + i \frac{\cal S}{2\pi}\xi_J)  (Q_K + i \frac{\cal S}{2\pi} \xi_K)}{ 
\ell^3 \left( \frac{1}{2} c^{IJK} Q_I \xi_J \xi_K+ \frac{\pi}{4G_5}\right) + i\left( \frac{\cal S}{2\pi}\right)} = 
  e^I  +  i  \frac{v}{2}  (  b^I e^{U_1}  + X^I \xi\cdot X)~.
\end{equation}
The real and imaginary parts of the complexified potentials 
$\omega$ and $\Delta^I$ are related to one another through 
\begin{equation}
\label{eq: eI e0 re}
    \xi_I e^I +e^0 = 0 ~.
\end{equation}
and from the identities \eqref{eq: nh susy rel3} and \eqref{eq: nh susy rel5}, we find
\begin{equation}
\label{eq: Delta omega rel}
    2\omega + \xi_I \Delta^I = 2\pi i ~.
\end{equation}
This is our version of the well-known complex constraint that is imposed on the chemical potentials conjugate to $J$ and $Q_I$ in analyses involving complex saddle points from the outset. An important example is the Hosseini-Hristov-Zaffaroni (HHZ) extremization principle 
for 5D rotating BPS black holes \cite{Hosseini:2017mds,Cabo-Bizet:2018ehj,Benini:2018mlo,Choi:2018hmj}.
The complexified potentials $\omega$ and $\Delta^I$ can be exploited to 
simplify the Lagrangian density \eqref{eq: lagrangian nh}. 
The linchpin is the identity 
\begin{equation}
    \frac{\tfrac{1}{6}c_{IJK}\Delta^I \Delta^J \Delta^K}{\omega^2} = \left(\frac{2\pi^2}{\mathcal{S}}\right) \frac{(-J+\tfrac{i\mathcal{S}}{2\pi})^2}{
\ell^3 \left( \frac{1}{2} c^{IJK} Q_I \xi_J \xi_K+ \frac{\pi}{4G_5}\right) + i\left( \frac{\cal S}{2\pi}\right)} ~.
\end{equation}
It is established using the cube of \eqref{eq: DeltaI nh}, along with \eqref{eq: cijk symmetrization} to simplify the products of $c_{IJK}$, as well as the square of \eqref{eq: omega nh} and the complexified charge relation \eqref{eq: complex charge rel}. 
The same combination of terms appears when evaluating instead
\begin{equation}
\label{eq: DeltaQ OmegaJ to S}
    \Delta^I Q_I - 2\omega J = -\left(\frac{N^2}{2}\right) \left(\frac{2\pi^2}{\mathcal{S}}\right) \frac{(-J+\tfrac{i\mathcal{S}}{2\pi})^2}{
\ell^3 \left( \frac{1}{2} c^{IJK} Q_I \xi_J \xi_K+ \frac{\pi}{4G_5}\right) + i\left( \frac{\cal S}{2\pi}\right)} + \mathcal{S} ~,
\end{equation}
with the use of the definitions of $\omega$ and $\Delta^I$ in \eqref{eq: omega nh} and \eqref{eq: DeltaI nh} respectively, as well as the complex relations \eqref{eq: complex charge rel} and \eqref{eq: Delta omega rel}, and exchanging $G_5$ for $N$ via $\tfrac{\pi \ell^3}{4G_5} = \tfrac{N^2}{2}$. This allows us to rewrite the black hole entropy $\mathcal{S}$ as
\begin{equation}
\label{eq: S to DeltaQ OmegaJ}
    \mathcal{S} = \Delta^I Q_I - 2\omega J + \frac{N^2}{2}  \frac{\tfrac{1}{6}c_{IJK}\Delta^I \Delta^J \Delta^K}{\omega^2} ~. 
\end{equation}
Referring back to the real-valued entropy functional $\mathcal{S}$ 
as the Legendre transform of the on-shell Lagrangian $\mathcal{L}_1$ \eqref{eq: entropy legendre}, and noting that $(e^I, e^0)$ constitute the real parts of $(\Delta^I, \omega)$, we obtain the greatly simplified expression 
\begin{equation}
    2\pi \mathcal{L}_{1,\text{nh}} = -\frac{N^2}{2} \text{Re} \left( \frac{\tfrac{1}{6}c_{IJK}\Delta^I \Delta^J \Delta^K}{\omega^2}\right) ~. 
\end{equation}
We have thus been able to reproduce the standard HHZ entropy function result \cite{Hosseini:2017mds}, although while remaining entirely in 5D (no reduction to 4D), with the help of the entropy function formalism. The derivation of \eqref{eq: S to DeltaQ OmegaJ} also makes the Legendre transformation between the entropy $\mathcal{S}$ and the complexified entropy function manifest.

\section{Discussion} \label{section: discussion}

We have analysed the first order attractor flow equations derived from the vanishing of the supersymmetric variations in $D=5$ $\mathcal{N}=2$ gauged supergravity with FI-couplings to $\mathcal{N}=2$ vector multiplets. We focus on solutions with electric charges $Q_{I}$ and one independent angular momentum $J$. In order to analyze the flow equations and find the conserved charges, we first assume a perturbative expansion at either the near-horizon geometry or the asymptotic boundary.  As usual, the supersymmetry conditions are not sufficient to guarantee a solution to the 
equation of motion, but we find that the conserved Noether-Wald surface charges fill this gap. 
This leads to a self-contained set of first order differential equations. 

To integrate these differential equations we need boundary conditions, or more generally integration constants. 
In the present setting, this turns out to be somewhat complicated. Generically, first order differential equations, even coupled ones, just need values at one point to compute the derivative and then, by iteration, the complete solution
follows.\footnote{Locality is among the
major caveats. In principle, first order differential equation give derivatives, and then the derivatives of the derivatives, and so on for the whole series. Generally, it is not easy to prove convergence for a series obtained this way, but this obstacle, and other mathematical fine points, do not appear significant at our level of analysis.}
We find that, whether starting from the black hole horizon or the asymptotic AdS$_5$, solving the first order equations is subtle. Supersymmetry conditions exhibit zero-modes which fail to provide a derivative, as a first order differential equation is expected to do. On the positive side, in these situations supersymmetry give relations between the first few coefficients near a boundary. 

After exploiting conserved charges extensively, the initial value problem simplifies. Indeed, at the horizon, all fields must satisfy the entropy extremization principle, discussed in detail in section~\ref{section: entropy extremization}. The relative simplicity of \textit{shooting out} from the horizon can be construed as black hole attractor behavior. The situation starting from asymptotic AdS is much more involved, as detailed in subsection~\ref{subsec:solutiontoattractoreq}. 

We are far from the first to investigate the attractor flow for rotating AdS$_5$ black holes. Some notable works are \cite{Hosseini:2017mds, Ntokos:2021duk}. In our procedure, we have remained in five dimensions, without dimensionally reducing to four dimensions, where the metric no longer contains a fibration. Our approach is complementary, in that the role of rotation is highlighted. Additionally, we have allowed for backgrounds that go beyond the omnipresent STU model. Finally, we have also considered a complexification of the near-horizon variables that elucidates some features of the theory from the near-horizon perspective. This includes the well-known complex constraint on the chemical potentials. 

Many open problems persist after our analysis of AdS$_5$ rotating black holes. For example, we derived the first order attractor flow equations from the supersymmetric variations of the $\mathcal{N}=2$ gauged theory, but it would be instructive to also derive them from the Lagrangian. After a suitable Legendre transform, the dimensionally reduced Lagrangian can be written as a sum of squares, up to a total derivative. In minimizing the Lagrangian, each square gives a condition that is equivalent to the vanishing of the supersymmetric variations. It would be interesting to extract the flow equations from this method as it can also be more directly related to the entropy extremization once the near-horizon limit is taken. We also expect this now radial entropy function to greatly simplify once the fields and variables in it are suitably complexified, such as was done at the near-horizon level. This would allow for an understanding of the underlying complex structure of the rotating $\text{AdS}_5$ black hole spacetime without the customary reduction to 4D.

Higher derivative corrections in the context of AdS$_5$ black holes have been studied by \cite{Cassani:2023vsa, Bobev:2021qxx, Cassani:2022lrk, Liu:2022sew} and references therein, and it would be interesting to understand the role of higher derivative corrections in the attractor flow. This is also interesting from the entropy extremization point of view and allows us to probe higher derivative corrections to the entropy from the near-horizon, which can be checked via holography. Finally, a similar analysis can then be completed in other dimensions, including the rotating AdS black holes in six and seven dimensions \cite{Kantor:2019lfo, Chow:2008ip}. The product of the scalar fields with one of the parameters of the metric yields a harmonic function and we would expect that one can solve the flow equations using a similar approach via a perturbative expansion. We hope to comment on these ideas in the near future.

\section*{Acknowledgements}

We thank Nikolay Bobev, Pablo Cano, Mirjam Cvetic, Alan Fukelman, Luca Illiesiu,
Sameer Murthy and Enrico Turetta for valuable discussions. 
FL thanks the Simons Foundation for support through a sabbatical fellowship. He also thanks Stanford Institute for Theoretical Physics for hospitality and support in the course of the sabbatical. MD thanks CERN for hospitality in the final stages of this work and is especially thankful for Alejandro Cabo-Bizet for letting her borrow an adaptor for her laptop charger to complete the finishing touches of the paper. MD is  supported in part by the NSF Graduate Research Fellowship Program under NSF Grant Number: DGE 1256260 and by KU Leuven C1 grant ZKD1118 C16/16/005, and by the Research Programme of The Research Foundation – Flanders (FWO) grant G0F9516N. NE is supported in part by the Leinweber Graduate Fellowship.
This work was supported in part by the U.S. Department of Energy under grant DE-SC0007859.

\begin{appendix}

\section{Conventions and notations} \label{appendix: conventions and notations}

In this Appendix, we summarize the conventions and notations used in the various expressions involving differential geometry as well as real special geometry. \\
We introduce components as
\begin{align}
   \xi = \xi^{\mu} \frac{\partial}{\partial x^{\mu}}~, \quad  \omega=\frac{1}{r!} \omega_{\mu_{1} \ldots \mu_{r}} \mathrm{~d} x^{\mu_{1}} \wedge \ldots \wedge \mathrm{d} x^{\mu_{r}}~.
\end{align}
In this notation the interior product $i_{\xi}$ of $\omega$ with respect to $\xi$ is
\begin{align}
\begin{split}
i_{\xi}  \omega &=\frac{1}{(r-1) !} \xi^{\nu} \omega_{\nu \mu_{2} \ldots \mu_{r}} \mathrm{~d} x^{\mu_{2}} \wedge \ldots \wedge \mathrm{d} x^{\mu_{r}} \\
&=\frac{1}{r !} \sum_{s=1}^{r} \xi^{\mu_{s}} \omega_{\mu_{1} \ldots \mu_{s} \ldots \mu_{r}}(-1)^{s-1} \mathrm{~d} x^{\mu_{1}} \wedge \ldots \wedge \widehat{\mathrm{d} x^{\mu_{s}}} \wedge \ldots \wedge \mathrm{d} x^{\mu_{r}}~.
\end{split}
\end{align}
The wide hat indicates that $dx^{\mu_{s}}$ is removed.

The Hodge dual is defined by
\begin{align}
\begin{aligned}
&\star_{r}\left(\mathrm{~d} x^{\mu_1} \wedge \mathrm{d} x^{\mu_2} \wedge \ldots \wedge \mathrm{d} x^{\mu_r}\right)
=\frac{\sqrt{|g|}}{(m-r) !} \varepsilon^{\mu_1 \mu_2 \ldots \mu_r}{}_{v_{r+1} \ldots v_m} \mathrm{~d} x^{v_{r+1}} \wedge \ldots \wedge \mathrm{d} x^{v_m}~,
\end{aligned}
\end{align}
where the subscript $r$ denotes the dimension of the spacetime and the totally antisymmetric tensor is
\begin{align}
    \varepsilon_{\mu_1 \mu_2 \ldots \mu_m}= \begin{cases}+1 & \text { if }\left(\mu_1 \mu_2 \ldots \mu_m\right) \text { is an even permutation of }(12 \ldots m) \\ -1 & \text { if }\left(\mu_1 \mu_2 \ldots \mu_m\right) \text { is an odd permutation of }(12 \ldots m) \\ 0 & \text { otherwise. }\end{cases}.
\end{align}
The indices on the totally antisymmetric symbol $\varepsilon_{\mu_1 \mu_2 \ldots \mu_m}$ can be raised by the metric through 
\begin{align}
    \varepsilon^{\mu_1 \mu_2 \ldots \mu_m}=g^{\mu_1 v_1} g^{\mu_2 v_2} \ldots g^{\mu_m v_m} \varepsilon_{v_1 v_2 \ldots v_m}=g^{-1} \varepsilon_{\mu_1 \mu_2 \ldots \mu_m} .
\end{align}
The Hodge dual of the identity $1$ gives the invariant volume element
\begin{align}
    \star_{r} 1=\frac{\sqrt{|g|}}{m !} \varepsilon_{\mu_1 \mu_2 \ldots \mu_m} \mathrm{~d} x^{\mu_1} \wedge \ldots \wedge \mathrm{d} x^{\mu_m}=\sqrt{|g|} \mathrm{d} x^1 \wedge \ldots \wedge \mathrm{d} x^m ~.
\end{align}
We define the $r$-forms $U$ and $V$ as
\begin{align}
    \begin{aligned}
        U &=\frac{1}{r !} U_{\mu_1 \ldots \mu_r} \mathrm{~d} x^{\mu_1} \wedge \ldots \wedge \mathrm{d} x^{\mu_r}, \qquad
        V =\frac{1}{r !} V_{\mu_1 \ldots \mu_r} \mathrm{~d} x^{\mu_1} \wedge \ldots \wedge \mathrm{d} x^{\mu_r} ~,
    \end{aligned}    
\end{align}
such that
\begin{align}
\begin{aligned}
U \wedge \star_{r} V &= V \wedge \star_{r} U = \frac{1}{r !} U_{\mu_1 \ldots \mu_r} V^{\mu_1 \ldots \mu_r} \sqrt{|g|} \mathrm{d} x^1 \wedge \ldots \wedge \mathrm{d} x^m .
\end{aligned}    
\end{align}

\section{Real special geometry}  \label{appendix subsection: real special geometry}

In this appendix we summarize the conventions and formulae needed for manipulations in real special geometry. We study $\mathcal{N}=2$ theories with $n_V$ vector multiplets and $n_H=0$ hyper-multiplets. The starting point is a collection of real 5D scalar fields $X^I$ with $I=0, 1, \ldots, n_V$. They are subject to the constraint 
\begin{equation}
 \label{eq:Xconst} 
\frac{1}{6}c_{IJK} X^I X^J X^K = 1~,
\end{equation}
where the structure constants $c_{IJK}$ are real numbers, completely symmetric in $I$, $J$, and $K$,
that satisfy the closure relation
\begin{equation}
c_{IJK}c_{J'(LM}c_{PQ)K'} \delta^{JJ'} \delta^{KK'} = \frac{4}{3} \delta_{I(L} c_{MPQ)}~.
\label{eq:cclosure down}
\end{equation}
The index $I$ takes $n_V+1$ distinct values but, because of the 
constraint \eqref{eq:Xconst}, there are $n_V$ independent scalar fields, one for each ${\cal N}=2$ vector multiplet in 5D. Round brackets $(\cdots)$ indicate symmetrization of indices with weight one so, for example, $c_{IJK} = c_{(IJK)}$. 

Using the Euclidean metric to define $c^{IJK}$ with upper indices, meaning \\
$c^{IJK} = \delta^{II'} \delta^{JJ'} \delta^{KK'} c_{I'J'K'}$, the closure relation \eqref{eq:cclosure down} can be rewritten as
\begin{equation}
\label{eq:cclosure up}
    c_{IJK}c^{J(LM}c^{PQ)K} = \frac{4}{3} \delta_I ^{(L} c^{MPQ)}~.
\end{equation}
We also note the following identities involving symmetrizations
\begin{align}
    \label{eq: cijk symmetrization}
    c_{IJK}c^{J(LM}c^{PQ)K} &= \frac{1}{3}c_{IJK}\left(c^{JLM}c^{PQK}+c^{JLP}c^{MQK} + c^{JPM}c^{LQK}\right) ~, \\ 
    \label{eq: kronecker symmetrization}
    \delta_I ^{(L} c^{MPQ)} &= \frac{1}{4}\left(\delta_I ^L c^{MPQ}+\delta_I ^M c^{LPQ} + \delta_I ^P c^{LMQ} + \delta_I ^Q c^{LMP}\right) ~.
\end{align}
Given the scalars $X^I$ and $c_{IJK}$ as inputs, we define the scalar $X_I$ (with lower index) and the metric on field space $G_{IJ}$ as
\begin{align} \label{eq: XI relation 1}
X_I  &= \frac{1}{2} c_{IJK} X^J X^K~,\cr 
G_{IJ} & = \frac{1}{2} \left(  X_I X_J - c_{IJK} X^K\right) ~.
\end{align}  
In manipulations we often use the formulae
\begin{align}
G_{IJ} X^J  &= \frac{1}{2} X_I ~,\cr 
X_I X^I & = 3 ~.
\end{align}  
The closure relation \eqref{eq:cclosure up} then requires that the inverse matrix $G^{IJ}$ satisfies
\begin{equation}
c^{IJK} X_{K}= X^I X^{J} - \frac{1}{2} G^{IJ}~.
\label{eq:cinvdef}
\end{equation}
It follows that, just as $G_{IJ}$ lowers indices on $X^J$ indices (up to a factor of $\frac{1}{2}$), the inverse $G^{IJ}$ raises indices on $X_J$
\begin{equation}
\begin{split}
        G^{IJ} X_J &= 2(X^I X^J - c^{IJK} X_K)X_J = 2X^I~.
\end{split}
\end{equation}
We also note the identity
\begin{equation}
    \begin{split}
        (c^{IJK} X_K) (c_{ILM} X^M) &= \left( X^I X^{J} - \tfrac{1}{2} G^{IJ} \right) \left(X_I X_L - 2G_{IL} \right)  = \delta^J _L + X^J X_L \, .
    \end{split}
\end{equation}
In the literature, it is common to summarize real special geometry through the
cubic polynomial
\begin{equation}
    \mathcal{V} = \frac{1}{6}c_{IJK} X^I X^J X^K~.
\end{equation}
The constraint \eqref{eq:Xconst} is simply $\mathcal{V} = 1$. Differentiating {\it first} and {\it then} imposing the constraint $\mathcal{V} = 1$, we find
\begin{align}
 \mathcal{V}_I & \equiv \frac{\partial \mathcal{V}}{\partial X^I} =  \frac{1}{2}c_{IJK}X^J X^K = X_I ~,\\ 
\mathcal{V}_{IJ}  & \equiv  \frac{\partial^2 \mathcal{V}}{\partial X^I \partial X^J}  = c_{IJK} X^K~,\\
 G_{IJ} &= - \frac{1}{2}\frac{\partial^2 \ln \mathcal{V}}{\partial X^I \partial X^J}  = \frac{1}{2}\left(\mathcal{V}_I \mathcal{V}_J - \mathcal{V}_{IJ} \right)= \frac{1}{2} \left(  X_I X_J - c_{IJK} X^K\right)~.
\end{align}
The inverse $\mathcal{V}^{IJ}$ of $\mathcal{V}_{IJ}$ (meaning it satisfies $\mathcal{V}^{IJ} \mathcal{V}_{JK} = \delta^I _K$) is given by
\begin{equation}
    \mathcal{V}^{IJ} = \frac{1}{2}(X^I X^J - G^{IJ})~.
\end{equation}
The STU-model is an important example. In this special case $n_V=2$ and we shift the labels so $I=1,2,3$ (rather than $I=0, 1, 2$). The only nonvanishing $c_{IJK}$ are $c_{123}=1$ and all its permutations. In our normalizations, the STU model has 
$$
X^1 X^2 X^3=1~, ~~X_I^{-1} = X_I ~, ~~G_{IJ} = \frac{1}{2} X_I^2 \delta_{IJ}~. 
$$
In these formulae there is no sum over $I=1, 2, 3$. We add a special note about adapting the formalism of real special geometry, this time adapted to $\xi_I$ given by the constraint
\begin{equation}
    \frac{1}{6}c^{IJK}\xi_I \xi_J \xi_K = \ell^{-3} ~.
\end{equation}
Following similar steps in terms of defining a raised version of the $\xi_I$, imposing consistency with the raised $c^{IJK}$ through the condition \eqref{eq: cijk symmetrization}, we can define the following
\begin{equation}
\begin{split}
    \xi^I &= \frac{1}{2}c^{IJK}\xi_J \xi_K ~, \\ \xi_I &= \frac{1}{2}\ell^{3} c_{IJK}\xi^J \xi^K ~. 
\end{split}
\end{equation}
We then go on defining a version of the $G_{IJ}$ and $G^{IJ}$ for the $\xi_I$
\begin{equation}
\begin{split}
\tilde{G}^{IJ} &= 2\left(\ell^3 \xi^I \xi^J - c^{IJK}\xi_K \right) ~, \\ 
    \tilde{G}_{IJ} &= \frac{1}{2}\ell^3 \left(\xi_I \xi_J - c_{IJK}\xi^K \right) ~,
\end{split}
\end{equation}
which leads to the crucial inversion identity on the $\xi_I$:
\begin{equation}
\label{eq: xi inversion}
    \frac{1}{2}\ell^3 (c_{IKM}c^{MNP}\xi_N \xi_P - \xi_I \xi_K) \left(c^{IJL}\xi_L \right) = \delta^J _K ~. 
\end{equation}

A final comment: in this article, we take 5D supergravity as the starting point. For an introduction to the geometric interpretation of the 5D fields and the formulae they satisfy in terms of Calabi-Yau compactification of 11D supergravity, we refer to \cite{Larsen:2006xm}.

 \section{Supersymmetry conditions}
\label{appendix: supersymmetry}

In this appendix we establish the conditions that our {\it ansatz} 
\eqref{eq:14ds5} must satisfy in order to preserve supersymmetry. 

\subsection{The K\"ahler condition on the base geometry}
\label{appendix subsection: Kahler condition}

We want to establish the conditions on the variables in the 4D base geometry in \eqref{eq:14ds5} that ensure that it is K\"ahler. For a given vielbein basis $e^a$ on the base $    ds_4 ^2 = \eta_{ab} e^a e^b $, such as \eqref{eq: 14 vierbein},
the K\"ahler condition is
\begin{equation}
    d(e^1 \wedge e^4 - e^2 \wedge e^3) = 0 ~. 
\end{equation}
In the $(1+4)$ split \eqref{eq:14ds5}, the base space \eqref{eq:14ds4} is automatically K\"ahler as
\begin{equation}
   \left(g_m ^{-1/2}\right) \left(\tfrac{1}{2}R g_m ^{1/2} dR \wedge \sigma_3 \right)  - \tfrac{1}{4}R^2 \sigma_1 \wedge \sigma_2 = d\left(\tfrac{1}{4}R^2 \sigma_3 \right) \, ,
\end{equation}
which is automatically closed. We look instead to the $(2+3)$ split in \eqref{eq: 5D metric ansatz full} to obtain a nontrivial K\"ahler condition. For that, we rewrite \eqref{eq: 5D metric ansatz full} in the form $ds_5 ^2 = f^2 (dt+\omega)^2 - f^{-1}ds_4 ^2$, and find the warp factor 
\begin{equation}
\label{eq:fdef}
    f = (e^{2\rho} -e^{-U_2} (a_t ^0)^2 )^{1/2}~,
\end{equation}
the 1-form
\begin{equation}
\label{eq:omdef}
\omega = - f^{-2}e^{-U_2}a^0 _t ~ \sigma_3~,
\end{equation}
and the 4D base geometry
\begin{equation}
\label{eq:ds4def}
    ds_4 ^2 = fe^{2\sigma} dR^2 + \frac{1}{4}R^2 (\sigma_1 ^2 + \sigma_2 ^2) + f^{-1}e^{2\rho - U_2} \sigma_3 ^2~.
\end{equation}
To find the condition for which \eqref{eq:ds4def} is K\"ahler, we introduce the basis 1-forms
\begin{align}
    e^1 &= f^{1/2} e^\sigma dR ~,\\ 
    e^2 &= \frac{1}{2}R\sigma_1 ~,\\ 
    e^3 &= \frac{1}{2}R\sigma_2 ~,\\ 
    e^4 &= f^{-1/2} e^{\rho - U_2/2}\sigma_3~.
\end{align}
The K\"ahler 2-form $J=e^1 \wedge e^4 - e^2\wedge e^3$ becomes
\begin{equation}
    \label{eq:Jdef} 
    J = e^{\sigma + \rho - U_2/2} dR \wedge \sigma_3  -\frac{1}{4}R^2 \sigma_1 \wedge \sigma_2~.
\end{equation}
The K\"ahler condition demands that $J$ is closed
\begin{equation}
    dJ = e^{\sigma+\rho -U_2/2} dR \wedge \sigma_1 \wedge \sigma_2   - \frac{1}{2}R dR \wedge \sigma_1 \wedge \sigma_2 = 0~. 
\end{equation}
We therefore find
\begin{equation}
    e^{\sigma + \rho - U_2/2} = \frac{1}{2}R~.    \label{eq:Kahlercondapp}
\end{equation}
This condition must be satisfied so that the general ansatz \eqref{eq: 5D metric ansatz full} can support supersymmetry.
The K\"ahler condition allow us to rewrite the base geometry \eqref{eq:ds4def} as
\begin{equation}
    ds_4 ^2 = fe^{2\sigma} dR^2 + \frac{1}{4}R^2 (\sigma_1 ^2 + \sigma_2 ^2 + f^{-1}e^{-2\sigma} \sigma_3 ^2)~. 
    \label{eq:basebetter}
\end{equation}
This form of the base geometry depends on a single function $fe^{2\sigma}$. 

\subsection{K\"{a}hler potential}
The K\"{a}hler condition \eqref{eq:Kahlercondapp} relates the 1-forms $e^1$ and $e^4$ in \eqref{eq: 14 vierbein}. If we define
a radial coordinate $r$ such that 
\begin{equation}
\label{eq:Rvsrdef}
    \partial_r R = f^{-\frac{1}{2}}e^{-\sigma}~,
\end{equation}
the tetrad simplifies so 
\begin{align}
    e^1 &= dr ~,\\ 
    e^4 &= \partial_r \left( \tfrac{1}{4}R^2 \right) \sigma_3~.
\end{align}
with $e^2$ and $e^3$ unchanged. In these coordinates, the unique spin connections solving Cartan's equations $de^a + \omega^a_{~b} e^b=0$ are
\begin{align}
\label{eq: omega21 4d}
    ^4\omega^2_{~1} = ^4\omega^4_{~3} &= \frac{\partial_r R}{R} e^2~, \\ 
   ^4\omega^3_{~1} = ^4\omega^2_{~4} &= \frac{\partial_r R}{R} e^3~,
    \\ 
    ^4\omega^4_{~1}  &= \left( \frac{\partial_r R}{R}+
    \frac{\partial^2_r R}{\partial_r R}\right)e^4 ~, \\ 
\label{eq: omega23 4d}
     ^4\omega^2_{~3}  &= \left( \frac{\partial_r R}{R} - \frac{2}{R\partial_r R} \right) e^4~,
\end{align}
where the $^4$ superscript distinguishes these 4D spin connections from the 5D spin connections that will appear in later computations. The resulting curvature 2-forms $R^a_{~b}= d\omega^a_{~b} + \omega^a_{~c}\omega^c_{~b}$ on the 4D base become
\begin{align}
    R^2_{~1}  &= R^4_{~3}  = \frac{\partial^2_r R}{R} (e^1 e^2 +e^3 e^4)~, \\ 
     R^3_{~1}  &= R^2_{~4} = \frac{\partial^2_r R}{R} (e^1 e^3 - e^2 e^4) ~,
    \\ 
      R^4_{~1}  &= \left( \frac{\partial^3_r R}{\partial_r R} + 
      3\frac{\partial^2_r R}{R}  \right) e^1 e^4
    + 2\frac{\partial^2_r R}{R} e^3 e^2~,
    \\   R^2_{~3}  &= 2\frac{\partial^2_r R}{R} e^1 e^4
    +\frac{4}{R^2} ((\partial_r R)^2 - 1) e^3 e^2 ~.
\end{align}
The components of the Riemann curvature are read off from $R^a_{~b}= \frac{1}{2} {\rm Riem}^a_{~bcd} e^c e^d$. For a complex manifold they are collected succinctly in the K\"{a}hler curvature 2-form 
with components ${\cal R}_{ab}= \frac{1}{2} {\rm Riem}_{abcd}J^{cd}$. In the context of our {\it ansatz} \eqref{eq:14ds4}, we have
\begin{align}
{\cal R}_{14} &= \epsilon ({\rm Riem}_{1423} - {\rm Riem}_{1414})
= \epsilon \left(\frac{\partial^3_r R}{\partial_r R} +  5 \frac{\partial^2_r R}{R} \right)~, \cr
{\cal R}_{23} &= \epsilon ({\rm Riem}_{2323} - {\rm Riem}_{1423})
= - \epsilon \left(2 \frac{\partial^2_r R}{R} +  \frac{4}{R^2} ((\partial_r R)^2 - 1 ) \right)~,
\end{align}
and so the K\"{a}hler curvature 2-form becomes
\begin{align}
{\cal R} & = \frac{1}{2} \epsilon\left( R \partial^3_r R
+ 5 \partial_r R\partial^2_r R\right) dr \sigma_3
- \frac{1}{2}\epsilon \left( R \partial^2_r R + 2 ((\partial_r R)^2 - 1 \right) \sigma_1 \sigma_2 \cr
& =  \epsilon d \left( (\frac{1}{2} R \partial^2_r R  + 
(\partial_r R)^2 - 1 )\sigma_3\right)~.
\end{align}
It is manifestly of the form ${\cal R} = dP$ where $P = p\sigma_3$ with 
\begin{equation}
    p = \epsilon \left( \frac{1}{2} R \partial^2_r R  + 
(\partial_r R)^2 - 1  \right)=
\epsilon \left( \frac{1}{4} R \partial_R (\frac{1}{f}e^{-2\sigma}) + \frac{1}{f}e^{-2\sigma} - 1 \right)~.
\label{eq:ponbase}
\end{equation}
The second equation follows by repeated use of \eqref{eq:Rvsrdef}. 
Since ${\cal R}$ is the exterior derivative of something, it is clearly closed. Thus the base manifold is K\"{a}hler.

The final expression \eqref{eq:ponbase} depends on the single scalar function $f e^{2\sigma}$ that determines the base geometry \eqref{eq:basebetter}. It encapsulates everything about the curvature of the 4D base. 

\subsection{Supersymmetry conditions}
\label{subsection:susyconds}

The ${\cal N}=2$ supergravity theory we consider is, in particular, invariant under the fermionic transformations of the gaugino and the gravitino
\begin{align}
    \label{eq:gaugvar}
 \delta\lambda & =     \left[ G_{IJ}  \left( \frac{1}{2} \gamma^{\mu\nu} F^J_{~\mu\nu}   - \gamma^\mu \nabla_\mu X^J \right) \epsilon^\alpha - \xi_I \epsilon^{\alpha \beta} \epsilon^\beta \right] 
\partial_i X^I  ~, \\ 
 \label{eq:gravvar}
\delta \psi^\alpha _\mu  & =  \left[ (\partial_\mu - \frac{1}{4} \omega_\mu^{\nu\rho} \gamma_{\nu\rho}) + \frac{1}{24} (\gamma_\mu^{~\nu\rho} - 4\delta^{~\nu}_\mu\gamma^\rho) X_I F^I_{\nu\rho} \right] 
 \epsilon^\alpha  +\frac{1}{6}  \xi_I (3A^I_\mu - X^I\gamma_\mu) \epsilon^{\alpha \beta} \epsilon^\beta ~,
\end{align}
where 
$\epsilon^\alpha \ (\alpha=1,2)$ are symplectic Majorana spinors. 
For bosonic solutions to the theory that respect at least some supersymmetry these variations vanish for the spinors $\epsilon^\alpha$ that generate the preserved supersymmetry. 
Supersymmetric black holes in AdS$_5$ with finite horizon area preserve the supersymmetry 
generated by the spinors $\epsilon^\alpha$ 
that satisfy the projections
\begin{align}
\label{eq: proj g0 app}
    \gamma^0 \epsilon^\alpha &= \epsilon^\alpha ~, \\
\label{eq: proj gmn app}
    \frac{1}{4}J^{(1)}_{mn} \gamma^{mn} \epsilon^\alpha &= -\epsilon^{\alpha \beta} \epsilon^\beta ~.
\end{align}
Each of these equations impose two projections on the spinor $\epsilon^\alpha$. All these projections commute, so the resulting black holes preserve $2^{-4}= 1/16$ of the maximal supersymmetry.   

We seek to work out the conditions that set the supersymmetric variations \eqref{eq:gaugvar} and \eqref{eq:gravvar} to zero,
 satisfying the projections \eqref{eq: proj g0 app} and \eqref{eq: proj gmn app}
imposed on purely bosonic solutions. We use the  matter ansatz and geometry in \eqref{eq: 5D potential ansatz} and \eqref{eq:14ds5}, respectively.

The gamma matrices are defined with respect to a flat 5D space and satisfy the Clifford algebra
\begin{equation}
    \{\gamma^\mu, \gamma^\nu \} = 2\eta^{\mu \nu} ~,
\end{equation}
with the flat 5D space defined in \eqref{eq:14ds5} via the following veilbein
\begin{align}
\label{eq:funfbE0}
    E^0 &= f(dt + w \sigma_3) ~,\\ 
\label{eq:funfbEi}
    E^i &= f^{-\frac{1}{2}}e^i~,
\end{align}
where $e^i$ with spatial indices refers to the 4D veilbein introduced in \eqref{eq: 14 vierbein}. Furthermore, the gamma matrices $\gamma^\mu$ following the projection \eqref{eq:projJ} satisfy
\begin{equation}
\label{eq:projJ2}
    -\frac{\epsilon}{2}(\gamma^{23} - \gamma^{14}) = \epsilon \epsilon^{\alpha \beta} \epsilon^\beta ~,
\end{equation}
where $\gamma^{\mu \nu}$ is the antisymmetrized product for $a\neq b$, which means that after squaring \eqref{eq:projJ2} we obtain
\begin{equation}
    \gamma^{1234} \epsilon^\alpha = \epsilon^\alpha ~,
\end{equation}
and thus
\begin{equation}
\label{eq:1423gammaproj}
    \gamma^{14}\epsilon^\alpha = -\gamma^{23} \epsilon^\alpha ~. 
\end{equation}
This becomes relevant for evaluating inner products of components of 2-forms and $\gamma^{ab}$ as well as their decomposition into self-dual and anti-self-dual terms.

\subsubsection{The gaugino equation}

Recall that the gaugino equation is given by \eqref{eq:gaugvar}, where the 5D 2-form $F^{I}=dA^{I}$ can be computed from \eqref{eq:AIdict}
\begin{equation}
\label{eq: FI ansatz app}
    F^I = \partial_R (fY^I) e^{-\sigma} f^{-1} E^1 \wedge E^0 + 4f \left(fY^I \partial_{R^2} w + \partial_{R^2} u^I\right) E^1 \wedge E^4 - \frac{4f}{R^2}\left(fY^I w + u^I\right) E^2 \wedge E^3 ~.
\end{equation} 
The spatial $F_{mn}^I$ components can be rearranged into self-dual and anti-self-dual terms
\begin{equation}
\begin{split}
    F^I &= \partial_R (fY^I) e^{-\sigma} f^{-1} E^1 \wedge E^0 \\ &\quad +2f \left(fY^I \left(\partial_{R^2} - \frac{1}{R^2} \right)w + \left(\partial_{R^2} - \frac{1}{R^2} \right)u^I\right) (E^1 \wedge E^4 + E^2 \wedge E^3)\\
    &\quad+ 2f \left(fY^I \left(\partial_{R^2} + \frac{1}{R^2} \right)w + \left(\partial_{R^2} + \frac{1}{R^2} \right)u^I\right) (E^1 \wedge E^4 - E^2 \wedge E^3) ~.
\end{split}
\end{equation}
Since $(\gamma^{14} + \gamma^{23})\epsilon^\alpha =0$ per \eqref{eq:1423gammaproj}, only the anti-self-dual components of $F^{\mu \nu}$ via $F_{\mu \nu} ^J \gamma^{\mu \nu}$ contributes to the gaugino variation. We thus simplify $G_{IJ}\frac{1}{2}\gamma^{\mu \nu} F_{\mu \nu}^J$ to find
\begin{align}
\begin{split}
    &G_{IJ}\frac{1}{2}\gamma^{\mu \nu} F_{\mu \nu}^J \epsilon^\alpha \\&= G_{IJ} \left[ \partial_R(fY^I) e^{-\sigma} f^{-1} \gamma^{10} +2f  \left(fY^I \left(\partial_{R^2} + \tfrac{1}{R^2} \right)w + \left(\partial_{R^2} + \tfrac{1}{R^2} \right)u^I\right)(\gamma^{14}- \gamma^{23}) \right] \epsilon^\alpha ~.
\end{split}
\end{align}
We then move on to the second term of \eqref{eq:gaugvar}, noting that $X^I$ is only a function of $R$
\begin{equation}
    G_{IJ} (-\gamma^\mu \nabla_\mu X^J )\epsilon^\alpha = G_{IJ} \left( -\gamma^1 e^{-\sigma} \partial_R X^J\right)\epsilon^\alpha ~. 
\end{equation}
Lastly, the third term of \eqref{eq:gaugvar} becomes
\begin{equation}
    -\xi_I \epsilon^{\alpha \beta}\epsilon^\beta = +\epsilon \xi_I \gamma^{23} \epsilon^\alpha ~.
\end{equation}
Combining all three contributions, we obtain the following equations
\begin{align}
    G_{IJ} \left[\partial_R (fY^I) - f \partial_R X^I \right] \partial_i X^I &= 0 ~, \\ 
\label{eq:gaugvar2}
    \left[4G_{IJ} f \left(\left(\partial_{R^2} + \frac{1}{R^2}\right)u^J + fY^J  \left(\partial_{R^2} + \frac{1}{R^2}\right)w\right) + \epsilon \xi_I \right] \partial_i X^I &= 0 ~. 
\end{align}
Since $X_I \partial_i X^I =\frac{1}{2}\partial_i (X_I X^I) = 0$, the $f\partial_R X^J$ term can be rewritten as $\partial_R (fX^J)$ and thus we obtain
\begin{equation}
\label{eq:susyrel1}
    \boxed{G_{IJ}\left[\partial_R (fY^J) - \partial_R (fX^J) \right]\partial_i X^I = 0 ~,} 
\end{equation}

This can be reexpressed by defining a vector $\delta^I = fY^I - fX^I$, to imply that $\partial_R \delta^I$ is orthogonal to $\partial_i X^I$, and thus proportional to $X^I$:
\begin{equation}
    \partial_R \delta^I = k X^I \, ,
\end{equation}
for some constant $k$. We will focus on the special solution where $\delta^I$ vanishes, meaning
\begin{equation}
\label{eq: XY match}
    X^I = Y^I ~. 
\end{equation}
Using this relation, we now move on to \eqref{eq:gaugvar2}, the second gaugino variation result. It is a projection of the vector quantity in the square brackets, along the direction of $\partial_i X^I$. The immediate consequence of it is that this quantity is proportional to $X_I$. Rearranging terms, we obtain the ambiguous result

\begin{equation}
\label{eq:susyrel2lambda}
    \left(\partial_{R^2} + \frac{1}{R^2}\right) u^I = \frac{1}{2}\epsilon f^{-1} c^{IJK} X_J \xi_K + \frac{1}{2}f^{-1}\lambda  X^I ~,
\end{equation}
with $\lambda$ a scalar coefficient that arises from the ambiguity in defining the quantity in square brackets in \eqref{eq:gaugvar2} as orthogonal to $\partial_i X^I$. Determining this quantity requires resorting to further supersymmetry relations, which leads us to the vanishing of the gravitino variation \eqref{eq:gravvar}.

\subsubsection{The gravitino equation}

In order to simplify the vanishing of the gravitino equation \eqref{eq:gravvar}, we need to establish the components of the 5D spin connection that appears in the term $-\frac{1}{4}\ \omega_\mu ^{\nu \rho} \gamma_{\nu \rho}\epsilon^\alpha$. Based on the vielbein \eqref{eq:funfbE0} and \eqref{eq:funfbEi}, we have
\begin{equation}
\begin{aligned}
     \omega^0 _{\ 1} &= f^{-1} e^{-\sigma} \partial_R f E^0 + 2f^2 \partial_{R^2}w E^4 ~ , & \quad
    \omega^0 _{\ 2} &= -\frac{2f^2 w}{R^2} E^3 ~ ,  \\ 
    \omega^0 _{\ 3} &= \frac{2f^2 w}{R^2}E^2 ~ , & \quad 
    \omega^0 _{\ 4} &= -2f^2 \partial_{R^2} w E^1 ~ , \\ 
    \omega^2 _{\ 1} &=\ ^4 \omega^2 _1 - \frac{1}{2}f^{-1} e^{-\sigma} \partial_R f E^2~ ,  & \quad 
    \omega^3 _{\ 1} &=\ ^4 \omega^3 _{\ 1} - \frac{1}{2}f^{-1} e^{-\sigma} \partial_R f  E^3 ~ , \\ 
     \omega^4 _{\ 1} &=\ ^4 \omega^4 _{\ 1} - \frac{1}{2}f^{-1}e^{-\sigma} \partial_R f E^4 - 2f^2 \partial_{R^2}w E^0 ~ , & \quad 
    \omega^2 _{\ 3} &=\ ^4\omega^2 _{\ 3} - \frac{2f^2 w}{R^2}E^0 ~ , \\ 
     \omega^3 _{\ 4} &=\ ^4\omega^3 _{\ 4} ~ , & \quad
     \omega^4 _{\ 2} &=\ ^4\omega^4 _{\ 2} ~ , 
\end{aligned}
\end{equation}
where $^4\omega^m _{\ n}$ represents the 4D spin connections (\ref{eq: omega21 4d}-\ref{eq: omega23 4d}), and $e^m$ are the 4D tetrad 1-forms, which are related to the $E^\mu$ \eqref{eq:funfbE0} and \eqref{eq:funfbEi} via $e^m = f^{1/2} E^m$. We now proceed to evaluate the components of the gravitino variation \eqref{eq:gravvar}, starting with $\mu = 0$: 
\begin{align}
\label{eq: gravitino three contributions from mu=0 part 1}
    \left(\partial_0 - \frac{1}{4}  \omega_0 ^{\nu \rho} \gamma_{\nu \rho}\right)\epsilon^\alpha &= \left(\partial_0 + \gamma^{23} f^2 \left(\partial_{R^2} + \frac{1}{R^2}\right)w - \frac{1}{2}f^{-1} e^{-\sigma} \partial_R f \gamma^1 \right) \epsilon^\alpha ~,
    \\ 
    \label{eq: gravitino three contributions from mu=0 part 2}
    \begin{split}
        \frac{1}{24}(\gamma_0 ^{\nu \rho} - 4\delta^\nu _0 \gamma^\rho)X_I F^I _{\nu \rho} \epsilon^\alpha &= \left(-\gamma^{23}f^2 \left[\left(\partial_{R^2} + \frac{1}{R^2}\right)w+\frac{1}{3}f^{-1}X_I \left(\partial_{R^2} + \frac{1}{R^2}\right)u^I \right] \right. \\& \left. \qquad + \frac{1}{2}f^{-1}e^{-\sigma} \partial_R f \gamma^1 \right)\epsilon^\alpha ~,
    \end{split}
    \\ \label{eq: gravitino three contributions from mu=0 part 3}
    \frac{1}{6}\xi_I (3A^I _0 - X^I \gamma_0)\epsilon^{\alpha \beta}\epsilon^\beta &= \epsilon \frac{1}{3}\xi_I X^I \gamma^{23}\epsilon^\alpha ~,
\end{align}
where $A^I _0 $ stands for the component of the $A^I$ 1-form along the $E^0$ flat veilbein, which amounts to $A^I _0 = f^{-1} A^I _t = f^{-1} (fX^I) = X^I$. Thus, adding up the three contributions in \eqref{eq: gravitino three contributions from mu=0 part 1}, \eqref{eq: gravitino three contributions from mu=0 part 2} and \eqref{eq: gravitino three contributions from mu=0 part 3}, we note that the terms proportional to $\gamma^1$ cancel out identically. What is left is terms proportional to the identity and to $\gamma^{23}$, which when made to vanish separately, lead to two results
\begin{equation}
\partial_0 \epsilon = 0 ~, 
\end{equation}
\begin{equation}
\label{eq:susyrel3}
    fX_I \left(\partial_{R^2} + \frac{1}{R^2}\right)u^I - \epsilon \xi_I X^I =  0~.
\end{equation}
This equation is another expression involving a projection of the quantity $\left(\partial_{R^2} + \frac{1}{R^2}\right)u^I$. Rather than a redundant relation, it can in fact be used to further constrain the ambiguity in $u^I$ that arose from the projection in the gaugino variation \eqref{eq:gaugvar2}. In fact, \eqref{eq:susyrel2lambda} and \eqref{eq:susyrel3} imply that
\begin{equation}
    fX_I \left(\frac{1}{2}\epsilon f^{-1} c^{IJK} X_J \xi_K + \frac{1}{2}f^{-1}\lambda  X^I\right) - \epsilon \xi_I X^I = 0 ~,
\end{equation}
which immediately means that $\lambda = 0$. The final result is given by
\begin{equation}
    \boxed{\left(\partial_{R^2} + \frac{1}{R^2}\right)u^I = \frac{1}{2}\epsilon f^{-1}c^{IJK} X_J \xi_K ~.}
\end{equation}
We now move on to the spatial components of \eqref{eq:gravvar}. For $\mu = 1$: 
\begin{align} \label{eq: gravitino three contributions from mu=1 part 1}
    \left(\partial_1 - \frac{1}{4}  \omega_1 ^{\nu \rho} \gamma_{\nu \rho}\right)\epsilon^\alpha &= \left(\partial_1 + \gamma^4 f^2 \partial_{R^2} w \right) \epsilon^\alpha ~, \\
    \begin{split} \label{eq: gravitino three contributions from mu=1 part 2}
    \frac{1}{24}(\gamma_1 ^{\nu \rho} - 4\delta^\nu _1 \gamma^\rho)X_I F^I _{\nu \rho} \epsilon^\alpha &= \left(-\gamma^{4}f^2 \left[\left(2\partial_{R^2} - \frac{1}{R^2}\right)w+\frac{1}{3}f^{-1}X_I \left(2\partial_{R^2} - \frac{1}{R^2}\right)u^I \right] \right. \\ & \left. \qquad  - \frac{1}{2}f^{-1}e^{-\sigma} \partial_R f \gamma^1 \right)\epsilon^\alpha 
    \end{split} ~,
    \\ \label{eq: gravitino three contributions from mu=1 part 3}
    \frac{1}{6}\xi_I (3A^I _1 - X^I \gamma_1)\epsilon^{\alpha \beta}\epsilon^\beta &= \frac{1}{6}\xi_I X^I \gamma^{4}\epsilon^\alpha = \frac{1}{6}fX_I \left(\partial_{R^2} + \frac{1}{R^2}\right)u^I \gamma^4 \epsilon^\alpha ~.
\end{align}
Again, adding the contributions \eqref{eq: gravitino three contributions from mu=1 part 1}, \eqref{eq: gravitino three contributions from mu=1 part 2} and \eqref{eq: gravitino three contributions from mu=1 part 3}, and separating out the terms proportional to the identity, $\gamma^1$ and $\gamma^4$, we obtain
\begin{equation}
\label{eq:susyrel4}
    \boxed{\left(\partial_{R^2} - \frac{1}{R^2}\right)w + \frac{1}{2}f^{-1}X_I \left(\partial_{R^2}-\frac{1}{R^2}\right)u^I = 0 ~,} 
\end{equation}
as well as the spatial dependence of the spinor $\epsilon$: $\partial_R \epsilon = \frac{1}{2}f^{-1}(\partial_R f)\epsilon$, which leads to $\epsilon = \epsilon_0 f^{1/2}$ for some constant $\epsilon_0$. The $\mu=2$ and $\mu=3$ components of \eqref{eq:gravvar} yield the same condition \eqref{eq:susyrel4}, which leaves us with $\mu=4$ that introduces an additional term due to the appearance of the 4D spin connection terms: 
\begin{align}
    \left(\partial_4 - \frac{1}{4}  \omega_4 ^{\nu \rho} \gamma_{\nu \rho}\right)\epsilon^\alpha &= \left(\partial_4 - \gamma^1 f^2 \partial_{R^2} w - \frac{1}{4}f^{-1}e^{-\sigma} \partial_R f \gamma^{14} + \frac{1}{4}\ ^4\omega_{mn} \gamma^{mn} \right) \epsilon^\alpha ~, \\
    \begin{split}
    \frac{1}{24}(\gamma_4 ^{\nu \rho} - 4\delta^\nu _4 \gamma^\rho)X_I F^I _{\nu \rho} \epsilon^\alpha &= \left(\gamma^{1}f^2 \left[(2\partial_{R^2} - \frac{1}{R^2})w+\frac{1}{3}f^{-1}X_I (2\partial_{R^2} - \frac{1}{R^2})u^I \right] \right.  \\ & \left. \qquad +\frac{1}{4}f^{-1}e^{-\sigma} \partial_R f \gamma^{14} \right)\epsilon^\alpha ~,
    \end{split}
    \\ 
    \frac{1}{6}\xi_I (3A^I _4 - X^I \gamma_4)\epsilon^{\alpha \beta}\epsilon^\beta &= \left( \frac{1}{2}\epsilon \xi_I u^I \gamma^{23} - \frac{1}{6}fX_I (\partial_{R^2} + \frac{1}{R^2})u^I \gamma^1 \right) \epsilon^\alpha ~.
\end{align}
Combining these terms leads to the condition \eqref{eq:susyrel4} as well as the 4D relation
\begin{equation}
    \left(\frac{1}{4}\ ^4\omega_{mn} \gamma^{mn} + \frac{1}{2}\xi_I u^I \gamma^{23}\right)\epsilon^\alpha = 0 ~.
\end{equation}
Using the 4D spin connections (\ref{eq: omega21 4d}-\ref{eq: omega23 4d}), we find that $^4\omega_{mn} \gamma^{mn} \epsilon^\alpha = -2p \gamma^{23} \epsilon^\alpha$ with $p$ from \eqref{eq:ponbase}. We relate $fe^{2\sigma}$ to $g_m$ based on \eqref{eq:geometrydict}, and find that
\begin{equation}
\label{eq:susyrel5}
    \boxed{p = \epsilon\left(\frac{1}{2}R^2 (\partial_{R^2} g_m) + g_m - 1)\right) = \xi_I u^I ~.} 
\end{equation}
We can now gather the four main supersymmetry relations that were derived:
\begin{align}
\label{eq:susyrel1f}
    0 &= G_{IJ}\left(\partial_R(fY^I) -\partial_R (fX^I)\right)\partial_i X^J ~,
    \\
\label{eq:susyrel2f}
    0 &=\left(\partial_{R^2} + \frac{1}{R^2}\right)u^I - \frac{1}{2}\epsilon f^{-1}c^{IJK} X_J \xi_K ~,
    \\
\label{eq:susyrel3f}
    0 &= \left(\partial_{R^2} - \frac{1}{R^2}\right)w + \frac{1}{2}f^{-1}X_I \left(\partial_{R^2}-\frac{1}{R^2}\right)u^I ~,
    \\
\label{eq:susyrel4f}
    0&= - \epsilon R^2 (\partial_{R^2} g_m) + 2\epsilon(1-g_m) + 2\xi_I u^I~. 
\end{align}

 \end{appendix}

\bibliographystyle{utphys}
\bibliography{references}

\end{document}